\documentclass[11pt]{article}

\usepackage[letterpaper,margin=1in]{geometry}
\usepackage{amsmath,amssymb,amsthm,mathtools,bm}
\usepackage[authoryear,round]{natbib}
\usepackage{booktabs,longtable,array,threeparttable}
\usepackage{graphicx}
\usepackage{pdflscape}
\usepackage[section]{placeins}
\usepackage{caption}
\usepackage{float,needspace}
\usepackage{enumitem}
\usepackage{xcolor}
\usepackage{microtype}
\usepackage[colorlinks=true,linkcolor=blue,citecolor=blue,urlcolor=blue]{hyperref}
\usepackage{orcidlink}

\newtheorem{theorem}{Theorem}
\newtheorem{lemma}{Lemma}
\newtheorem{corollary}{Corollary}
\newtheorem{proposition}{Proposition}

\newtheorem{condition}{Condition}

\newcommand{\R}{\mathbb R}
\newcommand{\1}{\mathbf{1}}
\newcommand{\dd}{\,\mathrm d}
\newcommand{\MCD}{\operatorname{MCD}}

\newcommand{\supp}{\operatorname{supp}}
\newcommand{\tr}{\operatorname{tr}}
\newcommand{\argmax}{\operatorname*{arg\,max}}
\newcommand{\argmin}{\operatorname*{arg\,min}}
\newcommand{\Ee}{\mathbb E}
\newcommand{\Pp}{\mathbb P}

\newcommand{\method}{\textsc{FDB-AH}}
\newcommand{\fs}{\textsc{FDB-FS}}
\newcommand{\fsm}{\textsc{FSDA}}
\newcommand{\hf}{\textsc{FDB-HF}}
\newcommand{\hseven}{\textsc{FDB-75}}

\newlist{conditions}{enumerate}{1}
\setlist[conditions]{
  label=\textbf{Condition \arabic*.},
  leftmargin=0pt,
  labelsep=0.8em,
  itemindent=!,
  align=left,
  itemsep=1.0ex,
  topsep=1.0ex,
  parsep=0pt
}

\graphicspath{{figures/}}


\title{Initial Subset-Size Selection for Minimum Covariance Determinant Estimation}
\author{%
Chikun Li\textsuperscript{a}\quad
Baisuo Jin\textsuperscript{b,*}\,\orcidlink{0000-0001-5302-053X}\\[0.55em]
\footnotesize \textsuperscript{a}Institute of Statistics and Applied Mathematics,\\[-0.1em]
\footnotesize Anhui University of Finance and Economics, Bengbu, China\\[-0.1em]
\footnotesize \href{mailto:120240009@aufe.edu.cn}{120240009@aufe.edu.cn}\\[0.15em]
\footnotesize \textsuperscript{b}Department of Statistics and Finance,\\[-0.1em]
\footnotesize University of Science and Technology of China, Hefei, China\\[0.3em]
\footnotesize \textsuperscript{*}Corresponding author: \href{mailto:jbs@ustc.edu.cn}{jbs@ustc.edu.cn}}
\date{}
\begin{document}

\maketitle

\begin{abstract}
	The minimum covariance determinant (MCD) method estimates multivariate location and scatter from a subset of the observations.
	For a given size \(h\), it chooses a subset whose sample covariance matrix has the smallest determinant.
	The choice of \(h\) affects both robustness and efficiency.
	We propose a nonparametric rule for selecting the initial subset size \(h\).
	We study the rule under a smooth unimodal elliptical model.
	With a small amount of contamination at a single point, we identify the largest candidate proportion for which an MCD solution can still exclude that point.
	Under suitable conditions, the rule selects this proportion at the population level.
	Without contamination, we prove that the proportion of observations in the selected subset converges almost surely to one as the sample size grows.
	For computation, we use a fast depth-based algorithm (FDB) to approximate the subset at each candidate size.
	We call the resulting method \method{} (AH for adaptive \(h\)).
	Simulations and real-data examples illustrate the benefits of choosing \(h\) adaptively for scatter estimation and outlier detection.
\end{abstract}


\section{Introduction}
\label{sec:introduction}

The minimum covariance determinant (MCD) method is well known in robust statistics
\citep{rousseeuw1985,BDJ93,CH99}.
For observations \(X_1,\ldots,X_n\in\R^p\), it selects a subset of size \(h\)
with the smallest empirical covariance determinant and estimates location
and scatter from that subset.
The robustness and efficiency of MCD estimation depend directly on \(h\).
An overly large subset can include outliers, displace the estimated centre,
and distort scatter. An unnecessarily small subset uses less information from regular
observations and can reduce estimation efficiency.

Two common choices for the initial subset size are
\begin{equation}
	h_{\min}=\left\lfloor\frac{n+p+1}{2}\right\rfloor,\qquad
	h_{0.75}=\lfloor0.75n\rfloor.
	\label{eq:fixed-h}
\end{equation}
With \(h_{\min}\), the MCD estimator can withstand a high proportion of
contamination.
When little contamination is expected, \(h_{0.75}\) is often used to include
more regular observations and improve estimation efficiency
\citep{hubert2012detmcd,hubert2018mcd,zhang2023fdb}.
For a new dataset, the contamination proportion and the distribution of the
outliers are often unknown, making it difficult to choose \(h\) in advance.

One-step reweighting can improve the efficiency of MCD estimates by including
regular observations outside the initial subset \citep{CH99,hubert2018mcd}.
In high dimensions, the cutoff for retaining observations can be based on the
asymptotic distribution of robust distances \citep{ro2015outlier,li2024ridge}.
These distances depend on the initial subset, so changing \(h\) may affect
which observations are retained even under the same cutoff rule.

A question related to the initial subset size is how many observations to
exclude when identifying outliers.
\citet{riani2019monitoring} use reweighted MCD estimates to plot each
observation's Mahalanobis distance against the subset size \(h\).
The plots show whether the same observations are flagged as outliers at
different trimming proportions.
\citet{heng2026stability} proposed Stable MCD, which chooses \(h\) based on the
stability of both parameter estimates and outlier classifications under resampling.

Our nonparametric rule for choosing \(h\) draws on a pattern that
\citet{riani2009unknown} observed in the Swiss banknote data.
The smallest Mahalanobis distance among observations outside the subset
peaked just before a group of outliers began to enter, then fell.
For each candidate size \(h\), we use the mean and scale-corrected covariance
matrix of the corresponding subset to compute Mahalanobis distances for all
observations and record the \((h+1)\)st smallest value.
For an exact MCD subset with a positive-definite covariance matrix, this value
is the smallest Mahalanobis distance among the observations outside the subset.
For each candidate size \(h\), we form a comparison window containing \(h\) and
nearby candidate sizes. We check the candidate sizes in
increasing order and select the first \(h\) whose recorded distance is a maximum
within its window.

For computation, we use the fast depth-based method (FDB) of
\citet{zhang2023fdb}.
We denote the FDB procedure with our window rule by \method{}. The two FDB
procedures with fixed sizes \(h_{\min}\) and \(h_{0.75}\) are denoted by \hf{}
and \hseven{}, respectively.

\begin{figure}[!htbp]
	\centering
	\includegraphics[width=0.98\textwidth]{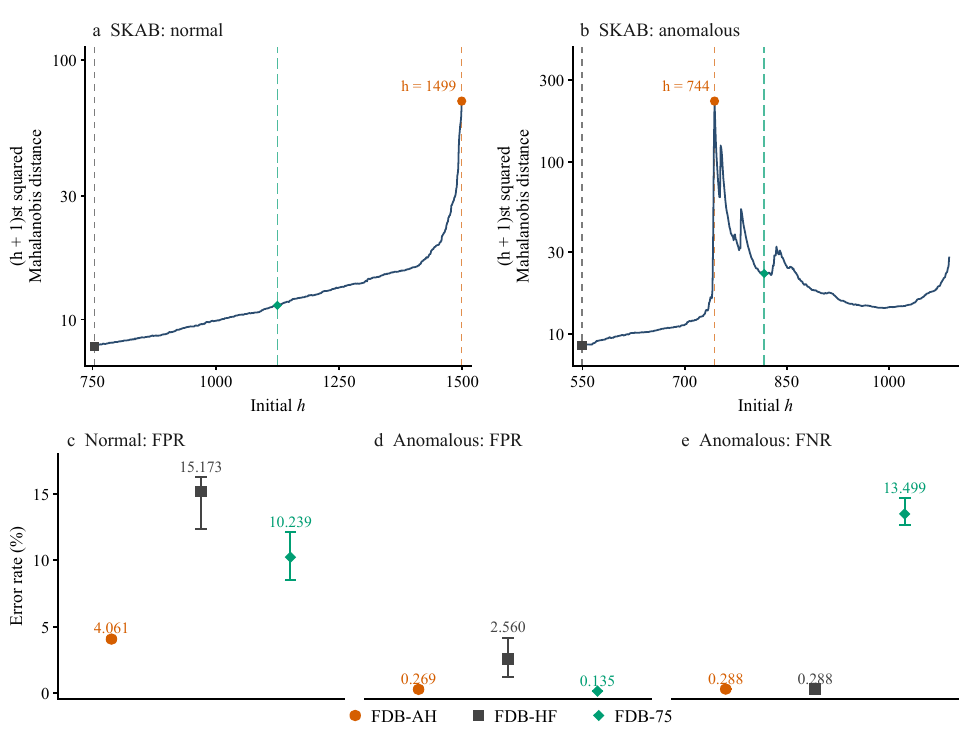}
	\caption{Results for two SKAB
	records.
		(a,b) Curves of the \((h+1)\)st smallest squared Mahalanobis distance for the
		normal-operation and rotor-imbalance records. The vertical axes are on a
		logarithmic scale.
		Points show average error rates over \(100\) runs for the normal-operation
		record (c) and the rotor-imbalance record (d,e). Error bars show the range.}
	\label{fig:motivation}
\end{figure}

We illustrate the rule using two records from a water-circulation testbed in
the Skoltech Anomaly Benchmark (SKAB, \citealp{katser2020skab}).
In the normal-operation record, the first window peak in the distance curve
occurs only when the initial subset contains nearly all the observations. In
the rotor-imbalance record, the first window peak occurs at a smaller retained
proportion (Figure~\ref{fig:motivation}a,b).
Figure~\ref{fig:motivation}c--e compares the error rates of the three methods
after one-step reweighting. The false positive rate (FPR) is the proportion of normal
observations excluded, and the false negative rate (FNR) is the proportion of anomalous
observations retained.

The forward search of \citet{riani2009unknown} uses a parametric rule for
outlier detection. At each step, it compares the smallest Mahalanobis distance
outside the current subset with a threshold based on a multivariate
normal model.
Even in the absence of outliers, a distance may cross the threshold by chance.
The authors use different signalling rules at different stages of the search,
choosing both the quantile levels of the thresholds and the number of crossings
needed to trigger a signal empirically.
For example, in the middle of the search, crossings in several consecutive
steps or a single crossing of a higher threshold can trigger a signal, with
further confirmation if needed.
Our rule selects \(h\) without a distance threshold and thus requires
weaker distributional assumptions.

Section~\ref{sec:formulation} defines the MCD subsets and the corresponding
Mahalanobis distances, Section~\ref{sec:selector} introduces the window rule
and algorithm, and Section~\ref{sec:theory-summary} gives the theoretical
results.
Section~\ref{sec:simulation} reports the simulation results,
Section~\ref{sec:real-data} presents the real data analysis, and
Section~\ref{sec:discussion} discusses our findings. Proofs and additional
results are relegated to the Supplementary Material.

\section{MCD subsets and Mahalanobis distances}
\label{sec:formulation}

To extend subset selection from samples to probability distributions, we define
the MCD location and covariance functionals through trimming functions
\citep{BDJ93,CL12}.
Let \(P\) be a probability measure on \(\R^p\) and let \(\gamma\in(0,1)\) be
the minimum retained proportion.
A trimming function is a measurable function \(\phi:\R^p\to[0,1]\), where
\(\phi(z)\) is the trimming weight assigned to an observation \(z\).
For example, \(\phi=\1_B\) retains only the probability mass in a measurable
set \(B\).
Allowing partial retention also makes this formulation applicable to
distributions that assign positive mass to a single point.

We require \(\int\phi\dd P\ge\gamma\) and
\(\int\|z\|^2\phi(z)\dd P(z)<\infty\), and denote the class of trimming
functions satisfying these conditions by \(K_P(\gamma)\). Throughout,
\(\|\cdot\|\) denotes the Euclidean norm for vectors and the spectral norm for
matrices. For \(\phi\in K_P(\gamma)\), normalizing by the mass actually
retained gives the trimmed mean and covariance
\begin{equation}
 \begin{split}
 T_P(\phi)&=\frac{1}{\int\phi\dd P}\int z\phi(z)\dd P(z),\\
 C_P(\phi)&=\frac{1}{\int\phi\dd P}\int
 (z-T_P(\phi))(z-T_P(\phi))^\top\phi(z)\dd P(z).
 \end{split}
 \label{eq:population-trimming}
\end{equation}
When \(\phi\equiv1\) and \(P\) has finite second moments, these are the usual
mean and covariance.
If \(\phi_P\in K_P(\gamma)\) minimizes \(\det C_P(\phi)\) over this class, the
corresponding pair \((T_P(\phi_P),C_P(\phi_P))\) is called an MCD functional at
\(P\).
We denote its two components by \(T_\gamma(P)\) and \(C_\gamma(P)\) and call
them the MCD location and covariance functionals, respectively. Neither the
minimizing trimming function \(\phi_P\) nor the value of the MCD functional at
\(P\) is necessarily unique.

\citet[Lemma~2.1]{CL12} show that for \(\phi\in K_P(\gamma)\), whenever
\(0<\lambda\le1\) and \(\lambda\phi\in K_P(\gamma)\), we have
\[
 T_P(\lambda\phi)=T_P(\phi),\qquad
 C_P(\lambda\phi)=C_P(\phi).
\]
Taking \(\lambda=\gamma/\int\phi\dd P\) gives a trimming function with retained
mass exactly \(\gamma\) and the same covariance determinant.
Thus, in computing the MCD, we may restrict attention to
\(\int\phi\dd P=\gamma\) without loss of generality.

To correct for the change in scale due to trimming, we set
\(\Sigma_\gamma(P)=c_\gamma C_\gamma(P)\), where \(c_\gamma>0\) is a population
scale correction factor. For example, let \(P\) be the standard multivariate
normal distribution \(N_p(0,I_p)\), whose covariance matrix is the identity
matrix \(I_p\) of order \(p\).
In this case, we choose \(c_\gamma\) so that the corrected scatter equals this
covariance matrix, giving \(\Sigma_\gamma(P)=I_p\). For a given \(\gamma\), the
factor does not depend on the trimming function. Thus the scale correction
does not change which trimming functions minimize the determinant.

For observations \(X_1,\ldots,X_n\in\R^p\), with \(n\ge p+2\), consider
\begin{equation}
 \mathcal H_n=\{h_{\min},\ldots,n-1\}.
 \label{eq:h-grid}
\end{equation}
For \(h\in\mathcal H_n\), put \(\gamma=h/n\) and let \(\mathcal I_{n,\gamma}\)
be an index set of size \(h\) that minimizes the determinant of the empirical
covariance over all such index sets.
We call the observations indexed by \(\mathcal I_{n,\gamma}\) an MCD subset.
The corresponding sample MCD estimators are
\begin{align}
 \widehat T_{n,\gamma}
 &=\frac1h\sum_{i\in\mathcal I_{n,\gamma}}X_i,\nonumber\\
 \widehat C_{n,\gamma}
 &=\frac1h\sum_{i\in\mathcal I_{n,\gamma}}
 (X_i-\widehat T_{n,\gamma})(X_i-\widehat T_{n,\gamma})^\top,\nonumber\\
 \widehat\Sigma_{n,\gamma}&=c_\gamma\widehat C_{n,\gamma}.
 \label{eq:sample-mcd}
\end{align}
For \(z\in\R^p\), define its squared Mahalanobis distance by
\begin{equation}
 \widehat D_{n,\gamma}(z)
 =(z-\widehat T_{n,\gamma})^\top
 \widehat\Sigma_{n,\gamma}^{-1}(z-\widehat T_{n,\gamma}).
 \label{eq:sample-distance}
\end{equation}
We consider the following sequence of sample distances outside the subset,
\begin{equation}
 \widehat d_{{\rm out},n}^{\,2}(h/n)
 =\min_{i\notin\mathcal I_{n,h/n}}\widehat D_{n,h/n}(X_i),
 \qquad h\in\mathcal H_n.
 \label{eq:mcd-path}
\end{equation}
For brevity, write \(Y_{n,h}=\widehat d_{{\rm out},n}^{\,2}(h/n)\).
Let \(\widehat D_{n,h/n,(j)}\) denote the \(j\)th smallest of the \(n\)
distances in \eqref{eq:sample-distance}.
For an MCD subset, the Mahalanobis distance of any observation in the subset is
no greater than that of any observation outside it. Otherwise, replacing an
observation in the subset by a closer observation outside it would reduce the
covariance determinant, contrary to the definition of an MCD subset.
Lemma~\ref{lem:sample-separation} proves this result, including ties, using the
concentration-step principle \citep{rousseeuw1999fast,CL12}.
Consequently,
\begin{equation}
 Y_{n,h}=\widehat D_{n,h/n,(h+1)}.
 \label{eq:mcd-order-identity}
\end{equation}

For a numerical approximation to the MCD, let
\(\widetilde{\mathcal I}_{n,h/n}\) denote the index set of the \(h\)
observations selected by the algorithm. We use tildes for numerical
approximations below.
Compute \(\widetilde T_{n,h/n}\) and \(\widetilde C_{n,h/n}\) as in
\eqref{eq:sample-mcd}, and set
\(\widetilde\Sigma_{n,h/n}=a_{n,p,h}\widetilde C_{n,h/n}\), where
\(a_{n,p,h}>0\) is the scale correction factor used by the numerical method.
The corresponding squared Mahalanobis distance is
\begin{equation}
	\widetilde D_{n,h/n}(z)
	=(z-\widetilde T_{n,h/n})^\top
	\widetilde\Sigma_{n,h/n}^{-1}(z-\widetilde T_{n,h/n}),
	\label{eq:computed-distance}
\end{equation}
Using the same notation for the ordered distances, we write
\begin{equation}
	\widetilde Y_{n,h}=\widetilde D_{n,h/n,(h+1)}.
	\label{eq:implemented-path}
\end{equation}

\section{Adaptive selection of the initial subset size}
\label{sec:selector}

We select the initial subset size adaptively by comparing
\(\widetilde Y_{n,h}\) at nearby subset sizes.
Fix \(c\in(0,1)\) and let
\begin{equation}
	g_n=\max\{1,\lfloor cn\rfloor\}.
	\label{eq:window-floor}
\end{equation}
For each \(h\), we restrict the comparison to eligible subset sizes that differ
from it by at most \(g_n\), so the comparison window is
\begin{equation}
	\mathcal W_n(h)=\{k\in\mathcal H_n:|k-h|\le g_n\}.
	\label{eq:window-set}
\end{equation}
We include \(h\) in the candidate set if \(\widetilde Y_{n,h}\) is at least as
large as the value at every other subset size in the window, giving
\begin{equation}
	\mathcal A_n=
	\{h\in\mathcal H_n:
	\widetilde Y_{n,h}\ge\widetilde Y_{n,k},
	\ \forall k\in\mathcal W_n(h)\}.
	\label{eq:candidate-set}
\end{equation}
This condition allows ties for the maximum within a window.
Any \(h\) at which \(\widetilde Y_{n,h}\) attains its global maximum on
\(\mathcal H_n\) belongs to \(\mathcal A_n\), so the set is nonempty.
From this set we select the smallest subset size,
\begin{equation}
	\widehat h=\min\mathcal A_n.
	\label{eq:hhat}
\end{equation}
Reweighting then determines which observations enter the final estimates.

Algorithm~1 gives the computational steps.
The inputs are the observations, the set of candidate subset sizes
\(\mathcal H_n\), the window proportion \(c\), a method for computing or
approximating MCD subsets, and a scale correction rule.

\begin{center}
\fbox{\begin{minipage}{0.94\linewidth}
\textbf{Algorithm 1: Adaptive Subset-Size Selection for MCD Estimation}
\begin{enumerate}[leftmargin=1.5em,itemsep=0.3em,topsep=0.4em]
\item For each \(h\in\mathcal H_n\), obtain \(\widetilde{\mathcal I}_{n,h/n}\)
and compute the mean and scale-corrected covariance of the corresponding
subset.
\item For each candidate subset, compute the squared distances of all \(n\)
observations using \eqref{eq:computed-distance} and record the \((h+1)\)st
smallest value, \(\widetilde Y_{n,h}\).
\item Select \(\widehat h\) using \eqref{eq:window-floor}--\eqref{eq:hhat} and
take the observations indexed by \(\widetilde{\mathcal I}_{n,\widehat h/n}\) as
the initial subset.
\item Use this subset to compute weights for all observations,
\[
w_i=\1\{\widetilde D_{n,\widehat h/n}(X_i)\le\chi^2_{p,0.975}\},
\qquad i=1,\ldots,n,
\]
where \(\chi^2_{p,0.975}\) is the \(0.975\) quantile of a chi-squared
distribution with \(p\) degrees of freedom.
\item Use the sample mean and covariance of the observations with \(w_i=1\) as
the final estimates.
\end{enumerate}
\end{minipage}}
\end{center}

We use FDB to approximate the MCD subset at each candidate size
\citep{zhang2023fdb}.
Statistical depth measures how central a point is relative to a distribution or
sample, with larger depth indicating greater centrality.
We use projection depth, which measures how far a point lies from the sample
median when both the point and the sample are projected onto a line in each
direction.
For a unit vector \(u\), let \(m_n(u)\) be the median of
\(u^\top X_1,\ldots,u^\top X_n\) and let \(s_n(u)\) be their median absolute
deviation from \(m_n(u)\).
The sample projection depth of \(z\) is
\[
 \operatorname{PD}_n(z)=
 \left\{1+\sup_{\|u\|=1}
 \frac{|u^\top z-m_n(u)|}{s_n(u)}\right\}^{-1}.
\]
The ratio standardizes the deviation by the median absolute deviation. The
smaller the supremum of the standardized deviations over all directions, the
greater the projection depth. When the denominator is zero, we define the ratio
as zero if the numerator is also zero and as infinity otherwise.
Consider the unimodal elliptical model in Lemmas~1--2 of \citet{zhang2023fdb}.
Suppose that the retained proportions of the subset selected by projection depth
and the MCD subset converge to the same fixed positive value as the sample size increases.
Under these conditions, the two corresponding regions converge to the same
central ellipsoid.
Their Theorem~1 shows that the probability mass of the symmetric difference
between the two regions tends to zero.
This result supports the use of projection depth to approximate MCD subsets.

For computation, we approximate the supremum by the maximum over \(K\) random
directions.
Let \(\pi=(\pi_1,\ldots,\pi_n)\) be the indices of the observations in
decreasing order of depth.
For each \(h\in\mathcal H_n\), take
\begin{equation}
	\widetilde{\mathcal I}_{n,h/n}=\{\pi_1,\ldots,\pi_h\}.
	\label{eq:fdb-subset}
\end{equation}
A single ordering thus gives the subsets for all candidate sizes
\(h\in\mathcal H_n\).
In Step~1 of Algorithm~1, we use this ordering to select the subset at each
candidate size.
We call the resulting adaptive method \method{}.

\section{Theoretical results}
\label{sec:theory-summary}

Under one-point contamination, we first examine how the radius of the
population MCD ellipsoid varies with the retained proportion and give a
first-order approximation to the location of its local peak.
When a sample MCD subset is used at each candidate size, we examine how the
retained proportion selected by the window rule relates to this peak location.
For sampling without contamination, we also show the role of a comparison
window of width proportional to \(n\) in selecting the initial subset size.

Let \(Z\sim P\). When \(\Sigma_\gamma(P)\) is positive definite, define
\begin{align}
	D_\gamma(z;P)
	&=\{z-T_\gamma(P)\}^\top\Sigma_\gamma(P)^{-1}
	\{z-T_\gamma(P)\},\label{eq:main-Dgamma}\\
	\rho_\gamma^2(P)
	&=\inf\{t>0:P(D_\gamma(Z;P)\le t)\ge\gamma\}.
	\label{eq:main-rho}
\end{align}
The corresponding MCD ellipsoid is
\(E_\gamma(P)=\{z:D_\gamma(z;P)\le\rho_\gamma^2(P)\}\),
where \(\rho_\gamma(P)\) denotes its radius.
Writing \(\supp(P)\) for the support of \(P\), we define the infimum of the
squared Mahalanobis distances of support points outside the ellipsoid by
\[
d_{\rm out}^2(\gamma;P)
=\inf_{z\in\supp(P)\setminus E_\gamma(P)}D_\gamma(z;P).
\]
If every open shell immediately outside the ellipsoid has positive probability,
this infimum equals \(\rho_\gamma^2(P)\).
\emph{If the MCD location and covariance are not unique, we choose a pair with
the largest \(\rho_\gamma^2(P)\).}

Consider the one-point contamination model
\begin{equation}
	P_{\varepsilon,x}=(1-\varepsilon)F+\varepsilon\delta_x,
	\qquad 0<\varepsilon<1/2,
	\label{eq:main-contamination}
\end{equation}
where \(F\) is the uncontaminated distribution, \(x\in\R^p\) is a fixed
contaminating point, and \(\delta_x\) is unit point mass at \(x\).
This model is often used to study the effect of one-point contamination on the
MCD functional \citep{CH99,CL12}.
\begin{condition}
\label{cond:elliptical}
The distribution \(F\) has density
\begin{equation}
 f_F(z)=\{\det(S_0)\}^{-1/2}
 g\!\left((z-\mu_0)^\top S_0^{-1}(z-\mu_0)\right),\qquad z\in\R^p,
 \label{eq:elliptical-density}
\end{equation}
where \(\mu_0\in\R^p\), \(S_0\) is positive definite, and
\(g:[0,\infty)\to(0,\infty)\) is twice continuously differentiable with
\(g'<0\).
\end{condition}
Write \(x_{\mathrm{orig}}\) for the contaminating point in the original
coordinates.
We standardize the coordinates using the affine transformation
\(z\mapsto S_0^{-1/2}(z-\mu_0)\).
The contaminating point in the new coordinates is
\[
x_{\mathrm{std}}=S_0^{-1/2}(x_{\mathrm{orig}}-\mu_0).
\]

By the affine equivariance of the MCD location and scatter functionals
\citep[Section~2]{CL12}, it suffices to work in standardized coordinates (see
Supplementary Section~A.1 for the derivation).
In these coordinates, the uncontaminated distribution has centre \(\mu_0=0\)
and scatter matrix \(S_0=I_p\).
We continue to use \(F\) and \(P_{\varepsilon,x}\) for the standardized
uncontaminated and contaminated distributions, respectively, and write
\(x=x_{\mathrm{std}}\).

Let \(Z\sim F\) and denote the squared Euclidean distance of \(Z\) from the
origin by \(R^2=Z^\top Z\).
Write \(F_{R^2}\) and \(f_{R^2}\) for the distribution function and density of
\(R^2\), respectively. Define
\begin{equation}
	\begin{split}
		q_\gamma&=F_{R^2}^{-1}(\gamma),\qquad
		f_\gamma=f_{R^2}(q_\gamma),\\
		M(\gamma)&=\Ee\{R^2\1(R^2\le q_\gamma)\},\qquad
		c_\gamma=\frac{p\gamma}{M(\gamma)}.
	\end{split}
	\label{eq:A-radial}
\end{equation}
\citet[Theorem~3.2]{CL12} show that when \(F\) has a density, an optimal
trimming function equals the indicator of the corresponding MCD ellipsoid
\(E_\gamma(F)\) almost everywhere under \(F\).
Under Condition~\ref{cond:elliptical}, this ellipsoid is unique for a given
retained proportion \(\gamma\) \citep[Theorem~1]{BDJ93}.
In the standardized coordinates above, it is the ball centred at the origin
with probability \(\gamma\), namely \(E_\gamma(F)=\{z:\|z\|^2\le q_\gamma\}\).

By spherical symmetry, \(T_\gamma(F)=0\), and after trimming all coordinates
have the same variance and distinct coordinates have zero covariance.
The sum of these \(p\) variances is the mean of \(R^2\) within the ball,
\(M(\gamma)/\gamma\), so \(C_\gamma(F)=M(\gamma)I_p/(p\gamma)\) (see
Supplementary Section~A.1 for details).
After scale correction by \(c_\gamma\) in \eqref{eq:A-radial}, we have
\(\Sigma_\gamma(F)=I_p\) and hence \(D_\gamma(z;F)=\|z\|^2\).
Since \(F\) has a positive density everywhere, support points outside the ball
can lie arbitrarily close to its boundary, so
\(d_{\rm out}^2(\gamma;F)=q_\gamma\).
We also use the \(C^2\) smoothness of \(g\) below in our analysis of one-point
contamination.

For a contaminating point \(x\ne0\), let \(q_x=\|x\|^2\) denote its squared
distance from the origin.
Write \(\eta_x=F_{R^2}(q_x)\) for the probability that an uncontaminated
observation \(Z\sim F\) satisfies \(\|Z\|\le\|x\|\).
In model \eqref{eq:main-contamination}, all the probability mass
\(\varepsilon\) of the contaminating component is concentrated at \(x\).
Let \(w\in[0,1]\) be the value of the trimming function at \(x\), that is, the
proportion of this mass retained.
Thus the contaminating point \(x\) contributes \(\varepsilon w\) to the total
retained mass. For a given retained proportion \(\gamma\), if a trimming function
minimizes the determinant of the trimmed covariance, we call its value \(w\) at
\(x\) an optimal weight.

We now consider sufficiently small \(\varepsilon\) and a retained proportion
\(\gamma\) close to \(\eta_x\).
How the radius \(\rho_\gamma(P_{\varepsilon,x})\) of the MCD ellipsoid varies
with \(\gamma\) depends on the probability mass \(\varepsilon w\) retained at
\(x\).
To describe how the optimal weight \(w\) varies, we define the following
coefficient \(\beta_x\).
\begin{align}
	L_x&=\eta_x-\frac{2q_xf_{\eta_x}}p,&
	H_x&=M(\eta_x)-\frac{2q_x^2f_{\eta_x}}{p+2},\label{eq:A-LH}\\
	V_x&=\frac{2q_x}{L_x}+\frac{(p-1)q_x^2}{H_x},&
	\beta_x&=1-f_{\eta_x}V_x.\label{eq:A-beta}
\end{align}
The condition \(g'<0\) implies that both \(L_x\) and \(H_x\) are positive (see
Supplementary Section~A.1 for the proof).
As \(\gamma\) increases, the optimal weight \(w\) increases continuously from
\(0\) to \(1\) when \(\beta_x>0\), and jumps directly from \(0\) to \(1\) when
\(\beta_x<0\).
A first-order approximation to the retained proportion \(\gamma\) at which the
radius reaches a local peak is
\begin{equation}
	\theta_\varepsilon=(1-\varepsilon)\eta_x
	+\frac{\varepsilon}{2}\min\{\beta_x,0\}.
	\label{eq:A-theta}
\end{equation}

Let \(J\subset(0,1)\) be a sufficiently small compact neighbourhood of
\(\eta_x\).
Suppress the distribution argument by writing
\(\rho_\gamma^2=\rho_\gamma^2(P_{\varepsilon,x})\).
For positive constants \(\delta_-,\delta_+\), define
\[
 I_\varepsilon=[\theta_\varepsilon-\delta_-,\,
                         \theta_\varepsilon+\delta_+\varepsilon],
 \qquad
 \rho_{\max}^2=\max_{\gamma\in I_\varepsilon}\rho_\gamma^2.
\]
The following theorem bounds the error in approximating the peak location on
\(I_\varepsilon\) by \(\theta_\varepsilon\).

\begin{theorem}
	\label{thm:population-peak}
	Under Condition~\ref{cond:elliptical}, fix \(p\ge1\) and \(x\ne0\).
	Under model \eqref{eq:main-contamination}, there are \(\delta_-,\delta_+,C>0\)
	such that \(I_\varepsilon\subset J\) for all sufficiently small
	\(\varepsilon\).
	For any retained proportion \(\gamma_\varepsilon\in I_\varepsilon\) at which the
	squared radius attains its maximum \(\rho_{\max}^2\) on this interval, we then
	have
	\begin{equation}
		|\gamma_\varepsilon-\theta_\varepsilon|\le C\varepsilon^2.
		\label{eq:A-peak-location}
	\end{equation}
	For each fixed \(0<\tau<\delta_+\), there is \(c_\tau>0\) such that, for all
	sufficiently small \(\varepsilon\),
	\begin{equation}
		\rho_{\max}^2-
		\sup_{\substack{\gamma\in I_\varepsilon\\
				|\gamma-\theta_\varepsilon|\ge\tau\varepsilon}}
		\rho_\gamma^2
		\ge c_\tau\varepsilon
		\label{eq:A-peak-gap}
	\end{equation}
	If \(\beta_x>0\), the radius has a unique maximizer on \(I_\varepsilon\),
	namely \(\gamma_\varepsilon=(1-\varepsilon)\eta_x\).
\end{theorem}

Equation~\eqref{eq:A-peak-location} gives the first-order approximation
\(\theta_\varepsilon\) to the peak location on \(I_\varepsilon\).
Equation~\eqref{eq:A-peak-gap} shows that, within this interval, a positive gap
separates the squared radius from its maximum \(\rho_{\max}^2\) when the
retained proportion \(\gamma\) is sufficiently far from \(\theta_\varepsilon\).

For \(\beta_x>0\), as \(\gamma\) increases within \(I_\varepsilon\), the radius
first increases while \(w=0\) is optimal and then decreases while the optimal
weight satisfies \(0<w<1\).
When \(\beta_x<0\), the optimal weight \(w\) jumps directly from \(0\) to
\(1\), and the radius drops at this point.
When \(\beta_x=0\), there is an interval centred at \(\theta_\varepsilon\) of
width \(O(\varepsilon^2)\) such that the optimal weight is \(0\) to the left of
the interval and \(1\) to its right.

Now suppose \(\eta_x>1/2\). Fix \(b\in(\eta_x,1)\) and consider any total
retained proportion \(\gamma\) in the candidate interval \(\Gamma_b=[1/2,b]\).
Given the trimming weight \(w\) at the contaminating point \(x\), choose a
trimming function that satisfies these two conditions and minimizes the
determinant of the trimmed covariance, and denote the resulting trimmed
covariance matrix before scale correction by \(C_{\gamma,w}\). Define
\[
\ell_\gamma(w)=\log\det C_{\gamma,w},\qquad
W_\gamma=\argmin_{0\le w\le1}\ell_\gamma(w).
\]
Thus \(W_\gamma\) is the set of optimal weights at the contaminating point
\(x\) for retained proportion \(\gamma\).
In particular, \(0\in W_\gamma\) means that at least one optimal trimming
function excludes all the contaminating mass at \(x\).

For \(c>0\), define
\begin{equation}
	\mathcal A_{\varepsilon,c}^{(b)}
	=\{\gamma\in\Gamma_b:\rho_\gamma^2\ge\rho_u^2
	\text{ for all }u\in\Gamma_b\cap[\gamma-c,\gamma+c]\}.
	\label{eq:population-window-set}
\end{equation}

\begin{proposition}
\label{prop:population-first-window-entry}
Under Condition~\ref{cond:elliptical}, fix \(p\ge1\) and \(x\in\R^p\) such that
\(\eta_x>1/2\), and consider model \eqref{eq:main-contamination}.
Take \(I_\varepsilon\) and \(\delta_+\) from Theorem~\ref{thm:population-peak}.
Fix \(b\in(\eta_x,1)\) and \(0<\kappa<\delta_+\).
For sufficiently small \(\varepsilon>0\), \(\rho_\gamma^2\) has a unique
maximizer \(\gamma_\varepsilon\) on \(I_\varepsilon\).
For every \(0<c\le\kappa\varepsilon\), \(\mathcal A_{\varepsilon,c}^{(b)}\) has
a smallest element and
\begin{equation}
 \min\mathcal A_{\varepsilon,c}^{(b)}
 =\gamma_\varepsilon
 =\max\{\gamma\in\Gamma_b:0\in W_\gamma\}.
 \label{eq:pfw-identification}
\end{equation}
Moreover, \(W_\gamma=\{0\}\) for \(\gamma\in[1/2,\gamma_\varepsilon)\).
At \(\gamma_\varepsilon\), the largest-radius convention stated before
\eqref{eq:main-contamination} selects a solution with \(w=0\).
\end{proposition}

Within \(\Gamma_b\), the retained proportions for which excluding the
contaminating point entirely (\(w=0\)) is optimal form the interval
\([1/2,\gamma_\varepsilon]\).
The radius is strictly increasing on this interval.
The radius attains its maximum over \(I_\varepsilon\) at
\(\gamma_\varepsilon\), and the comparison window at this point lies within
\(I_\varepsilon\).
Thus \(\gamma_\varepsilon\) is the smallest retained proportion satisfying the
window condition in \eqref{eq:population-window-set}.

For each candidate size \(h\in\mathcal H_n\), we choose a sample MCD subset.
If several subsets minimize the determinant at a given size, we list the
observation indices in each subset in increasing order. To choose between
any two of these subsets, we take the one with the smaller index at the
first position where their lists differ.
When the covariance matrices at all candidate sizes are positive definite,
we define
\begin{equation}
	\begin{split}
		\mathcal A_n^{\MCD}
		&=\{h\in\mathcal H_n:Y_{n,h}\ge Y_{n,k}
		\text{ for all }k\in\mathcal W_n(h)\},\\
		\widehat h_{\MCD}&=\min\mathcal A_n^{\MCD}.
	\end{split}
	\label{eq:mcd-window-selector}
\end{equation}
Any \(h\) at which \(Y_{n,h}\) attains its maximum on \(\mathcal H_n\) belongs
to \(\mathcal A_n^{\MCD}\), so this set is nonempty.
If the covariance matrix at any candidate size is singular, set
\(\widehat h_{\MCD}=n-1\).
Under the contaminated and uncontaminated models considered below, with
probability \(1\) the covariance matrices at all candidate sizes are positive
definite for all sufficiently large \(n\) (see Supplementary Section~A.6).
The following theorem bounds the difference between the retained proportion
\(\widehat h_{\MCD}/n\) and
\(\theta_\varepsilon\).

\begin{theorem}
\label{thm:contaminated-window}
Under Condition~\ref{cond:elliptical}, fix \(p\ge1\) and \(x\in\R^p\) such that
\(\eta_x>1/2\), and take \(\delta_+\) from Theorem~\ref{thm:population-peak}.
For every fixed \(0<\tau<\delta_+\), there is \(\varepsilon_\tau>0\) such that
the following holds.
Fix \(0<\varepsilon<\varepsilon_\tau\) and \(0<c<(\delta_+-\tau)\varepsilon\).
For independent observations with distribution \(P_{\varepsilon,x}\),
\begin{equation}
 \left|\widehat h_{\MCD}/n-\theta_\varepsilon\right|
       <\tau\varepsilon
 \qquad\text{eventually almost surely}.
 \label{eq:contaminated-window-location}
\end{equation}
\end{theorem}

By \eqref{eq:A-theta} and \eqref{eq:A-peak-location}, the retained proportion
\(\gamma_\varepsilon\) at the population peak tends to \(\eta_x\) as
\(\varepsilon\to0\).
Since \(\eta_x>1/2\), we have \(\gamma_\varepsilon>1/2\) for sufficiently small
\(\varepsilon\).
With such an \(\varepsilon\) fixed, the smallest candidate proportion
\(h_{\min}/n\) tends to \(1/2\) as \(n\) increases and hence eventually lies
below \(\gamma_\varepsilon\).
Now consider a candidate proportion \(h/n\) that differs from
\(\theta_\varepsilon\) by less than \(\tau\varepsilon\).
Since \(c<(\delta_+-\tau)\varepsilon\), all retained proportions compared to the
right of \(h/n\) lie within the interval \(I_\varepsilon\) considered in
\eqref{eq:A-peak-gap}.

Combining \eqref{eq:contaminated-window-location} and
\eqref{eq:A-peak-location} shows that, with probability \(1\), for all
sufficiently large \(n\) the proportion \(\widehat h_{\MCD}/n\) selected from the
sample differs from the population choice \(\gamma_\varepsilon\) by at most
\(\tau\varepsilon+C\varepsilon^2\).

We next consider samples from the uncontaminated distribution \(F\).
\citet[Propositions~3.1--3.2 and Theorem~4.2]{CL12} proved consistency of the
sample MCD location and scatter estimates for a fixed retained proportion.
Under the uncontaminated model, we show that this result holds uniformly over
all candidate retained proportions \(h/n\) in any fixed interval
\([a,b]\subset[1/2,1)\).
The maximum of \(\lvert Y_{n,h}-q_{h/n}\rvert\) over these proportions
also converges almost surely to zero.

\begin{lemma}
\label{thm:clean-uniform}
Under Condition~\ref{cond:elliptical}, let \(X_1,X_2,\ldots\) be independent
observations from \(F\), with \(p\) fixed.
For every \(\mathcal K=[a,b]\) with \(1/2\le a<b<1\), almost surely
\begin{equation}
 \sup_{\substack{h\in\mathcal H_n\\h/n\in\mathcal K}}
 \left\{\|\widehat T_{n,h/n}\|
       +\|\widehat\Sigma_{n,h/n}-I_p\|
       +|Y_{n,h}-q_{h/n}|\right\}\longrightarrow0.
 \label{eq:B-uniform-path}
\end{equation}
\end{lemma}

To rule out a candidate size \(h\), it suffices to find a larger size \(k\)
in its window with \(Y_{n,k}>Y_{n,h}\).
Under the uncontaminated model, the population quantiles satisfy
\(q_{k/n}>q_{h/n}\) whenever \(k>h\).
We compare this increase with the uniform approximation error in
Lemma~\ref{thm:clean-uniform}.

Fix \(c\in(0,1/2)\) and \(1/2<\gamma_L<\gamma_U<1-c\), and consider candidate
sizes \(h\) such that \(h/n\in[\gamma_L,\gamma_U]\).
To cover the right side of each window, extend the interval to
\(\Gamma_c=[\gamma_L,\gamma_U+c]\). The maximum absolute error over \(\Gamma_c\) is
\begin{equation}
	r_n=\sup_{\substack{h\in\mathcal H_n\\h/n\in\Gamma_c}}
	|Y_{n,h}-q_{h/n}|.
	\label{eq:B-uniform-condition}
\end{equation}
Define the event
\(\mathcal F_n
=\left\{\exists\,h\in\mathcal H_n:\ 
h/n\in[\gamma_L,\gamma_U]
\ \text{and}\ 
Y_{n,h}\ge Y_{n,k}\text{ for all }k\in\mathcal W_n(h)
\right\}\). As the retained proportion increases from \(\gamma\) to
\(\gamma+c/2\), the smallest increase in the population quantile is
\[
\Delta_c=\inf_{\gamma\in[\gamma_L,\gamma_U]}
(q_{\gamma+c/2}-q_\gamma)>0,
\]
where \(\Delta_c>0\) because \(q_\gamma\) is continuous and strictly increasing
and \([\gamma_L,\gamma_U]\) is a closed interval.
Write \(\Pp_F\) for probability under sampling from \(F\).

\begin{proposition}
\label{prop:B-floor-control}
Under Condition~\ref{cond:elliptical}, let \(X_1,X_2,\ldots\) be independent
observations from \(F\), with \(p\) fixed.
Fix \(c\in(0,1/2)\) and \(1/2<\gamma_L<\gamma_U<1-c\), with
\(r_n,\Delta_c,\mathcal F_n\) as defined above.
Almost surely, \(\mathcal F_n\) does not occur for any sufficiently large
\(n\).
Moreover, for all sufficiently large \(n\),
\[
 \Pp_F(\mathcal F_n)\le
     \Pp_F(r_n\ge\Delta_c/3)\longrightarrow0.
\]
\end{proposition}

\begin{corollary}
\label{cor:clean-retention}
Under Condition~\ref{cond:elliptical}, let \(X_1,X_2,\ldots\) be independent
observations from \(F\), with \(p\) fixed.
For every fixed \(c\in(0,1)\),
\[
 \widehat h_{\MCD}/n\longrightarrow1
 \qquad\text{almost surely}.
\]
\end{corollary}

\section{Simulation results}
\label{sec:simulation}
The simulations use three settings for sample size and dimension, denoted by
\(A=(n,p)=(100,10)\), \(B=(500,50)\), and \(C=(2000,200)\).
We run each simulation in this section \(1000\) times. We generate the latent data matrix \(Y\) and compute the observed data matrix
\(X=YG\). The matrix \(G\) has ones on the diagonal and \(0.75\) elsewhere.

We run simulations without contamination in Settings A, B, and C.
In each repetition, we draw \(n\) independent observations from \(N_p(0,I_p)\)
to form the latent data matrix \(Y\) and obtain the uncontaminated sample as
\(X=YG\).

In the contamination experiments, we first fix the signal parameter at \(r=15\)
and consider two contamination proportions \(\varepsilon\in\{0.10,0.30\}\).
In Setting B, we also use \(r=6,6.5,\ldots,10\) to examine how signal strength
affects the performance of each method.
We use the following four contamination designs.
\begin{enumerate}[leftmargin=1.6em,itemsep=0.5em]
	\item \emph{Point contamination:} The uncontaminated observations follow a
	standardized multivariate \(t_3\) distribution with \(3\) degrees of freedom
	and covariance matrix \(I_p\). The outliers follow
	\(N_p(r\sqrt p\,u,0.1^2I_p)\), where \(u\) is a unit vector orthogonal to the
	all-ones vector.
	\item \emph{Point-mass contamination:} The uncontaminated observations have
	the same distribution as in the point contamination design. All outliers equal
	\(r\sqrt p\,b\), where \(b\) is a random unit vector.
	\item \emph{Random-direction contamination:} The uncontaminated observations
	follow \(N_p(0,I_p)\).
	The outliers are generated independently. For the \(i\)th outlier, we first
	draw a random unit vector \(u_i\) and then draw one observation from
	\(N_p(rp^{1/4}u_i,I_p)\).
	\item \emph{Cluster contamination:} The uncontaminated observations follow
	\(N_p(0,I_p)\), and the outliers follow \(N_p(m,I_p)\), where
	\(m=(rp^{1/4}/\sqrt p)\bm1\) and \(\bm1\in\R^p\) is the all-ones vector.
\end{enumerate}

The random-direction and cluster contamination designs are from
\citet{zhang2023fdb}. We generate correlated data using an affine
transformation, as in \citet{hubert2012detmcd} and \citet{zhang2023fdb}.

For comparison with \method{}, we introduce a parametric method for selecting
the initial subset size, denoted by \fs{}.
This method compares \(\widetilde Y_{n,h}\)
at each candidate size \(h\) with a threshold \(b_{n,p,h}\).
Let \(\mathsf F_{\nu_1,\nu_2}^{-1}(u)\) denote the \(u\)-quantile of an \(F\)
distribution with degrees of freedom \(\nu_1,\nu_2\).
We first compute \(u_h\) and then use it to determine the
threshold \(b_{n,p,h}\),
\begin{align}
	u_h&=\frac{h+1}{h+1+(n-h)
		\mathsf F_{2(n-h),\,2(h+1)}^{-1}(0.001)},
	\label{eq:fs-beta-quantile}\\
	b_{n,p,h}&=\frac{p(h+1)(h-1)}{h(h-p)}
	\mathsf F_{p,h-p}^{-1}(u_h).
	\label{eq:fs-envelope}
\end{align}
Here \(u_h\) is the \(0.999\)-quantile of \(\operatorname{Beta}(h+1,n-h)\).
\fs{} selects the first \(h\) for which \(\widetilde Y_{n,h}>b_{n,p,h}\). If no
crossing occurs, it selects \(n-1\).
In our comparisons, \fsm{} denotes the forward search of \citet{riani2009unknown}
as implemented in the FSDA toolbox \citep{riani2012fsda}.
\fsm{} updates its subset at each step using Mahalanobis distances.
\fs{} uses one ordering by projection depth for all subset sizes.
The formula for \(u_h\) follows equation~(8) of \citet{riani2009unknown}.
The threshold \(b_{n,p,h}\) is the \(u_h\)-quantile of the scaled \(F\) distribution
in \citet{atkinson2010forward}.
Supplementary Section~C.2 gives the derivation and calibration of the threshold.

We compare \method{}, \fs{}, \hf{}, and \hseven{} after reweighting.
\hf{} and \hseven{} use fixed initial subset sizes \(h_{\min}\) and
\(h_{0.75}\), respectively. In each repetition, we compute the projection
depths of the simulated observations using \(K=\max(1000,10p)\) random
directions \citep{pokotylo2019ddalpha}.
All four methods use the same ordering, from highest to lowest projection depth.
For a given \(h\), each selects the first \(h\) observations, whose indices
form the set \(\widetilde{\mathcal I}_{n,h/n}\) in \eqref{eq:fdb-subset}.
All four methods use the same rules for the Gaussian MCD consistency correction
and the finite-sample correction.
After choosing the initial subset, each method reweights all observations using
the same cutoff and reweighting rule.
Supplementary Section~C.1 gives details of the scale correction and
reweighting.

Additional comparisons with \fsm{} and Stable MCD appear in Supplementary
Sections~\ref{sec:supp-fs-comparison} and~\ref{sec:supp-stable-comparison},
respectively.

We recommend \(c=0.08\) as the default window parameter for \method{} and use
this value in all simulations and real data analyses in this paper.
When the window is too narrow, it compares only subsets of similar sizes, and
sampling fluctuations may give a window maximum at a small subset size, leading
to an initial subset that is too small (see
Proposition~\ref{prop:B-floor-control} for the sample MCD analysis without
contamination).
If a local peak due to contamination occurs at subset size \(h\), a window wide
enough to include a candidate size \(k\) far from \(h\) with
\(\widetilde Y_{n,k}>\widetilde Y_{n,h}\) will rule out \(h\).

For each method, let \(\widehat T_X\) and \(\widehat\Sigma_X\) denote the
reweighted location and scatter estimates, respectively, based on the observed
data \(X\).
Since \(G\) is symmetric, the estimates in latent coordinates are
\(\widehat T=G^{-1}\widehat T_X\) and
\(\widehat\Sigma=G^{-1}\widehat\Sigma_X G^{-1}\), respectively.
Following the simulation study of \citet{zhang2023fdb}, we use the mean \(0\)
and covariance matrix \(I_p\) of the uncontaminated distribution as targets.
In each repetition, we compute the location error \(e_\mu\), shape error
\(e_\Sigma\), covariance mean squared error \(\mathrm{MSE}_\Sigma\), and
Stein loss \(L_{\mathrm{Stein}}\) as follows.
{\setlength{\jot}{10pt}
	\begin{align*}
		e_\mu&=\|\widehat T\|,\\
		e_\Sigma&=\log_{10}\kappa(\widehat\Sigma),\\
		\mathrm{MSE}_\Sigma&=p^{-2}\|\widehat\Sigma-I_p\|_F^2,\\
		L_{\mathrm{Stein}}&=\tr(\widehat\Sigma)-\log\det(\widehat\Sigma)-p.
	\end{align*}
}
Here \(\|\cdot\|_F\) is the Frobenius norm and \(\kappa\) is the spectral
condition number. \emph{The tables in this section report Monte Carlo means
with standard deviations in parentheses.}

\Needspace{26\baselineskip}
Under uncontaminated Gaussian sampling, the mean losses for \method{} and \fs{}
are very similar (Table~\ref{tab:clean}).
Across all four measures and three settings, both methods have mean losses
no higher than those of \hf{} and \hseven{}.

\begin{table}[H]
\centering
\caption{Performance of the four methods under uncontaminated Gaussian
sampling.}
\label{tab:clean}
\normalsize
\renewcommand{\arraystretch}{1.55}
\setlength{\tabcolsep}{3.2pt}
\begin{tabular}{llrrrr}
\toprule
Setting & Method & \(e_\mu\) & \(e_\Sigma\) & \(\mathrm{MSE}_\Sigma\) & \(L_{\mathrm{Stein}}\)\\
\midrule
A & \method & 0.314 (0.068) & 0.504 (0.060) & 0.012 (0.002) & 0.633 (0.122)\\
 & \fs     & 0.314 (0.067) & 0.502 (0.060) & 0.012 (0.002) & 0.629 (0.123)\\
 & \hf     & 0.329 (0.071) & 0.533 (0.068) & 0.012 (0.002) & 0.715 (0.155)\\
 & \hseven & 0.324 (0.070) & 0.521 (0.062) & 0.012 (0.002) & 0.683 (0.138)\\
\addlinespace
B & \method & 0.321 (0.032) & 0.552 (0.020) & 0.002 (\(9.480\mathrm{e}{-5}\)) & 2.760 (0.117)\\
 & \fs     & 0.321 (0.032) & 0.552 (0.020) & 0.002 (\(9.474\mathrm{e}{-5}\)) & 2.760 (0.117)\\
 & \hf     & 0.327 (0.033) & 0.567 (0.022) & 0.002 (\(1.005\mathrm{e}{-4}\)) & 2.902 (0.135)\\
 & \hseven & 0.323 (0.032) & 0.557 (0.021) & 0.002 (\(9.685\mathrm{e}{-5}\)) & 2.806 (0.122)\\
\addlinespace
C & \method & 0.320 (0.016) & 0.567 (0.008) & \(5.130\mathrm{e}{-4}\) (\(5.693\mathrm{e}{-6}\)) & 10.736 (0.116)\\
 & \fs     & 0.320 (0.016) & 0.567 (0.008) & \(5.130\mathrm{e}{-4}\) (\(5.691\mathrm{e}{-6}\)) & 10.736 (0.116)\\
 & \hf     & 0.330 (0.016) & 0.588 (0.008) & \(5.409\mathrm{e}{-4}\) (\(6.867\mathrm{e}{-6}\)) & 11.522 (0.156)\\
 & \hseven & 0.324 (0.016) & 0.575 (0.008) & \(5.241\mathrm{e}{-4}\) (\(6.148\mathrm{e}{-6}\)) & 11.037 (0.132)\\
\bottomrule
\end{tabular}
\end{table}

\FloatBarrier

At \(r=15\), each contamination proportion gives 12 combinations of the three
settings and four contamination types.
For each combination, we compute the mean initial subset proportion
\(\widehat h/n\) selected by \method{}.
At \(\varepsilon=0.10\), these means range from \(0.867\) to \(0.901\).
At \(\varepsilon=0.30\), they range from \(0.698\) to \(0.701\) (see Supplementary
Table~\ref{tab:supp-hfrac}).

Tables~\ref{tab:contam010} and~\ref{tab:contam030} report the four losses at
\(r=15\) for the two contamination proportions, respectively.
Both tables include \method{} and \fs{}, along with one method that uses a
fixed initial subset size.
At \(\varepsilon=0.10\), \(h_{0.75}\) is smaller than the number of regular
observations and a larger initial subset can improve estimation efficiency, so
we use \hseven{} as the benchmark.
At \(\varepsilon=0.30\), \(h_{0.75}\) exceeds the number of regular
observations, so we use \hf{}.
In practice, the contamination proportion is usually unknown.
\begin{table}[H]
\centering
\caption{Performance of three methods at \(r=15\) and \(\varepsilon=0.10\).}
\label{tab:contam010}
\fontsize{9}{10.4}\selectfont
\setlength{\tabcolsep}{1.0pt}
\renewcommand{\arraystretch}{1.1}
\begin{tabular}{@{}ll*{9}{c}@{}}
\toprule
& & \multicolumn{3}{c}{A} & \multicolumn{3}{c}{B} & \multicolumn{3}{c}{C}\\
\cmidrule(lr){3-5}\cmidrule(lr){6-8}\cmidrule(lr){9-11}
& & \method & \fs & \hseven & \method & \fs & \hseven & \method & \fs & \hseven\\
\midrule
Point & \(e_\mu\) & \(0.243\) & \(0.232\) & \(0.236\) & \(0.234\) & \(0.233\) & \(0.230\) & \(0.232\) & \(0.235\) & \(0.230\)\\
 & & \((0.058)\) & \((0.055)\) & \((0.055)\) & \((0.024)\) & \((0.023)\) & \((0.023)\) & \((0.012)\) & \((0.012)\) & \((0.011)\)\\
\addlinespace[2pt]
 & \(e_\Sigma\) & \(0.668\) & \(0.686\) & \(0.671\) & \(0.736\) & \(0.831\) & \(0.769\) & \(0.770\) & \(0.911\) & \(0.828\)\\
 & & \((0.082)\) & \((0.089)\) & \((0.081)\) & \((0.028)\) & \((0.032)\) & \((0.029)\) & \((0.011)\) & \((0.015)\) & \((0.013)\)\\
\addlinespace[2pt]
 & \(\mathrm{MSE}_\Sigma\) & \(0.032\) & \(0.042\) & \(0.038\) & \(0.008\) & \(0.010\) & \(0.009\) & \(0.002\) & \(0.003\) & \(0.003\)\\
 & & \((0.006)\) & \((0.006)\) & \((0.005)\) & \((4.098\mathrm{e}{-4})\) & \((3.433\mathrm{e}{-4})\) & \((3.594\mathrm{e}{-4})\) & \((4.888\mathrm{e}{-5})\) & \((3.735\mathrm{e}{-5})\) & \((4.115\mathrm{e}{-5})\)\\
\addlinespace[2pt]
 & \(L_{\mathrm{Stein}}\) & \(3.296\) & \(4.781\) & \(4.150\) & \(19.756\) & \(31.935\) & \(25.948\) & \(91.148\) & \(148.870\) & \(122.794\)\\
 & & \((0.857)\) & \((0.982)\) & \((0.725)\) & \((1.430)\) & \((1.656)\) & \((1.477)\) & \((2.968)\) & \((3.230)\) & \((3.048)\)\\
\addlinespace[5pt]
Point mass & \(e_\mu\) & \(0.243\) & \(0.234\) & \(0.236\) & \(0.237\) & \(0.235\) & \(0.232\) & \(0.232\) & \(0.235\) & \(0.230\)\\
 & & \((0.059)\) & \((0.057)\) & \((0.057)\) & \((0.024)\) & \((0.023)\) & \((0.024)\) & \((0.012)\) & \((0.012)\) & \((0.011)\)\\
\addlinespace[2pt]
 & \(e_\Sigma\) & \(0.673\) & \(0.686\) & \(0.673\) & \(0.739\) & \(0.833\) & \(0.770\) & \(0.770\) & \(0.912\) & \(0.828\)\\
 & & \((0.086)\) & \((0.087)\) & \((0.083)\) & \((0.028)\) & \((0.034)\) & \((0.030)\) & \((0.011)\) & \((0.015)\) & \((0.012)\)\\
\addlinespace[2pt]
 & \(\mathrm{MSE}_\Sigma\) & \(0.033\) & \(0.043\) & \(0.038\) & \(0.008\) & \(0.010\) & \(0.009\) & \(0.002\) & \(0.003\) & \(0.003\)\\
 & & \((0.007)\) & \((0.006)\) & \((0.005)\) & \((4.035\mathrm{e}{-4})\) & \((3.468\mathrm{e}{-4})\) & \((3.635\mathrm{e}{-4})\) & \((4.800\mathrm{e}{-5})\) & \((3.865\mathrm{e}{-5})\) & \((4.113\mathrm{e}{-5})\)\\
\addlinespace[2pt]
 & \(L_{\mathrm{Stein}}\) & \(3.382\) & \(4.848\) & \(4.140\) & \(19.654\) & \(32.034\) & \(25.977\) & \(91.170\) & \(149.218\) & \(122.950\)\\
 & & \((0.949)\) & \((0.957)\) & \((0.706)\) & \((1.404)\) & \((1.663)\) & \((1.492)\) & \((2.904)\) & \((3.354)\) & \((3.053)\)\\
\addlinespace[5pt]
Random & \(e_\mu\) & \(0.323\) & \(0.324\) & \(0.332\) & \(0.331\) & \(0.331\) & \(0.335\) & \(0.334\) & \(0.334\) & \(0.339\)\\
 & & \((0.071)\) & \((0.071)\) & \((0.074)\) & \((0.033)\) & \((0.033)\) & \((0.033)\) & \((0.016)\) & \((0.016)\) & \((0.016)\)\\
\addlinespace[2pt]
 & \(e_\Sigma\) & \(0.518\) & \(0.520\) & \(0.542\) & \(0.576\) & \(0.577\) & \(0.585\) & \(0.593\) & \(0.593\) & \(0.603\)\\
 & & \((0.065)\) & \((0.064)\) & \((0.070)\) & \((0.021)\) & \((0.021)\) & \((0.022)\) & \((0.008)\) & \((0.008)\) & \((0.008)\)\\
\addlinespace[2pt]
 & \(\mathrm{MSE}_\Sigma\) & \(0.013\) & \(0.013\) & \(0.013\) & \(0.002\) & \(0.002\) & \(0.002\) & \(5.609\mathrm{e}{-4}\) & \(5.612\mathrm{e}{-4}\) & \(5.741\mathrm{e}{-4}\)\\
 & & \((0.003)\) & \((0.003)\) & \((0.003)\) & \((9.917\mathrm{e}{-5})\) & \((9.956\mathrm{e}{-5})\) & \((1.017\mathrm{e}{-4})\) & \((6.390\mathrm{e}{-6})\) & \((6.385\mathrm{e}{-6})\) & \((6.893\mathrm{e}{-6})\)\\
\addlinespace[2pt]
 & \(L_{\mathrm{Stein}}\) & \(0.665\) & \(0.671\) & \(0.730\) & \(2.989\) & \(2.994\) & \(3.080\) & \(11.697\) & \(11.704\) & \(12.083\)\\
 & & \((0.136)\) & \((0.137)\) & \((0.154)\) & \((0.119)\) & \((0.120)\) & \((0.129)\) & \((0.122)\) & \((0.121)\) & \((0.141)\)\\
\addlinespace[5pt]
Cluster & \(e_\mu\) & \(0.321\) & \(0.322\) & \(0.329\) & \(0.333\) & \(0.333\) & \(0.337\) & \(0.334\) & \(0.334\) & \(0.339\)\\
 & & \((0.076)\) & \((0.076)\) & \((0.078)\) & \((0.033)\) & \((0.033)\) & \((0.033)\) & \((0.017)\) & \((0.017)\) & \((0.017)\)\\
\addlinespace[2pt]
 & \(e_\Sigma\) & \(0.514\) & \(0.517\) & \(0.539\) & \(0.577\) & \(0.578\) & \(0.586\) & \(0.593\) & \(0.593\) & \(0.602\)\\
 & & \((0.062)\) & \((0.063)\) & \((0.066)\) & \((0.022)\) & \((0.022)\) & \((0.023)\) & \((0.008)\) & \((0.008)\) & \((0.008)\)\\
\addlinespace[2pt]
 & \(\mathrm{MSE}_\Sigma\) & \(0.013\) & \(0.013\) & \(0.013\) & \(0.002\) & \(0.002\) & \(0.002\) & \(5.611\mathrm{e}{-4}\) & \(5.612\mathrm{e}{-4}\) & \(5.738\mathrm{e}{-4}\)\\
 & & \((0.003)\) & \((0.003)\) & \((0.003)\) & \((1.035\mathrm{e}{-4})\) & \((1.039\mathrm{e}{-4})\) & \((1.064\mathrm{e}{-4})\) & \((6.003\mathrm{e}{-6})\) & \((6.019\mathrm{e}{-6})\) & \((6.484\mathrm{e}{-6})\)\\
\addlinespace[2pt]
 & \(L_{\mathrm{Stein}}\) & \(0.661\) & \(0.667\) & \(0.727\) & \(2.992\) & \(2.995\) & \(3.080\) & \(11.701\) & \(11.702\) & \(12.073\)\\
 & & \((0.128)\) & \((0.131)\) & \((0.145)\) & \((0.124)\) & \((0.125)\) & \((0.133)\) & \((0.118)\) & \((0.118)\) & \((0.137)\)\\
\bottomrule
\end{tabular}
\end{table}

Table~\ref{tab:contam010} shows that at \(\varepsilon=0.10\), the mean
\(L_{\mathrm{Stein}}\) for \method{} is lower than for the other two
methods in all 12 combinations.
Under point and point-mass contamination, \method{} also has a lower mean
\(\mathrm{MSE}_\Sigma\) than the other two methods.
The three methods have similar means for \(e_\mu\).

Under point contamination in Setting A, the mean values of \(e_\Sigma\) for
\method{} and \hseven{} are \(0.668\) and \(0.671\), respectively, and their
mean values of \(L_{\mathrm{Stein}}\) are \(3.296\) and \(4.150\).

\begin{table}[H]
\centering
\centering
\caption{Performance of three methods at \(r=15\) and \(\varepsilon=0.30\).}
\label{tab:contam030}
\fontsize{9}{10.4}\selectfont
\setlength{\tabcolsep}{1.0pt}
\renewcommand{\arraystretch}{1.1}
\begin{tabular}{@{}ll*{9}{c}@{}}
\toprule
& & \multicolumn{3}{c}{A} & \multicolumn{3}{c}{B} & \multicolumn{3}{c}{C}\\
\cmidrule(lr){3-5}\cmidrule(lr){6-8}\cmidrule(lr){9-11}
& & \method & \fs & \hf & \method & \fs & \hf & \method & \fs & \hf\\
\midrule
Point & \(e_\mu\) & \(0.287\) & \(0.276\) & \(0.275\) & \(0.270\) & \(0.264\) & \(0.264\) & \(0.264\) & \(0.262\) & \(0.262\)\\
 & & \((0.067)\) & \((0.066)\) & \((0.065)\) & \((0.028)\) & \((0.027)\) & \((0.027)\) & \((0.014)\) & \((0.013)\) & \((0.013)\)\\
\addlinespace[2pt]
 & \(e_\Sigma\) & \(0.784\) & \(0.794\) & \(0.798\) & \(0.847\) & \(0.886\) & \(0.886\) & \(0.883\) & \(0.947\) & \(0.947\)\\
 & & \((0.096)\) & \((0.099)\) & \((0.100)\) & \((0.033)\) & \((0.035)\) & \((0.035)\) & \((0.014)\) & \((0.016)\) & \((0.016)\)\\
\addlinespace[2pt]
 & \(\mathrm{MSE}_\Sigma\) & \(0.029\) & \(0.035\) & \(0.036\) & \(0.007\) & \(0.009\) & \(0.009\) & \(0.002\) & \(0.003\) & \(0.003\)\\
 & & \((0.005)\) & \((0.005)\) & \((0.005)\) & \((4.197\mathrm{e}{-4})\) & \((4.169\mathrm{e}{-4})\) & \((4.169\mathrm{e}{-4})\) & \((5.264\mathrm{e}{-5})\) & \((4.779\mathrm{e}{-5})\) & \((4.779\mathrm{e}{-5})\)\\
\addlinespace[2pt]
 & \(L_{\mathrm{Stein}}\) & \(2.958\) & \(3.906\) & \(4.049\) & \(20.017\) & \(26.010\) & \(26.010\) & \(95.017\) & \(124.430\) & \(124.430\)\\
 & & \((0.662)\) & \((0.825)\) & \((0.812)\) & \((1.499)\) & \((1.722)\) & \((1.722)\) & \((3.249)\) & \((3.571)\) & \((3.571)\)\\
\addlinespace[5pt]
Point mass & \(e_\mu\) & \(0.288\) & \(0.278\) & \(0.277\) & \(0.270\) & \(0.266\) & \(0.266\) & \(0.264\) & \(0.263\) & \(0.263\)\\
 & & \((0.070)\) & \((0.065)\) & \((0.065)\) & \((0.028)\) & \((0.028)\) & \((0.028)\) & \((0.014)\) & \((0.013)\) & \((0.013)\)\\
\addlinespace[2pt]
 & \(e_\Sigma\) & \(0.786\) & \(0.789\) & \(0.791\) & \(0.847\) & \(0.883\) & \(0.883\) & \(0.883\) & \(0.946\) & \(0.946\)\\
 & & \((0.098)\) & \((0.102)\) & \((0.101)\) & \((0.035)\) & \((0.035)\) & \((0.035)\) & \((0.014)\) & \((0.016)\) & \((0.016)\)\\
\addlinespace[2pt]
 & \(\mathrm{MSE}_\Sigma\) & \(0.030\) & \(0.035\) & \(0.036\) & \(0.007\) & \(0.009\) & \(0.009\) & \(0.002\) & \(0.002\) & \(0.002\)\\
 & & \((0.005)\) & \((0.005)\) & \((0.005)\) & \((4.416\mathrm{e}{-4})\) & \((4.449\mathrm{e}{-4})\) & \((4.449\mathrm{e}{-4})\) & \((5.346\mathrm{e}{-5})\) & \((5.619\mathrm{e}{-5})\) & \((5.619\mathrm{e}{-5})\)\\
\addlinespace[2pt]
 & \(L_{\mathrm{Stein}}\) & \(2.983\) & \(3.857\) & \(4.035\) & \(19.902\) & \(25.812\) & \(25.812\) & \(94.551\) & \(123.953\) & \(123.953\)\\
 & & \((0.680)\) & \((0.813)\) & \((0.820)\) & \((1.566)\) & \((1.832)\) & \((1.832)\) & \((3.282)\) & \((4.254)\) & \((4.254)\)\\
\addlinespace[5pt]
Random & \(e_\mu\) & \(0.367\) & \(0.370\) & \(0.377\) & \(0.374\) & \(0.375\) & \(0.380\) & \(0.377\) & \(0.377\) & \(0.384\)\\
 & & \((0.080)\) & \((0.080)\) & \((0.081)\) & \((0.039)\) & \((0.039)\) & \((0.039)\) & \((0.020)\) & \((0.020)\) & \((0.020)\)\\
\addlinespace[2pt]
 & \(e_\Sigma\) & \(0.586\) & \(0.595\) & \(0.614\) & \(0.657\) & \(0.658\) & \(0.672\) & \(0.678\) & \(0.678\) & \(0.694\)\\
 & & \((0.072)\) & \((0.074)\) & \((0.081)\) & \((0.025)\) & \((0.025)\) & \((0.027)\) & \((0.010)\) & \((0.010)\) & \((0.010)\)\\
\addlinespace[2pt]
 & \(\mathrm{MSE}_\Sigma\) & \(0.016\) & \(0.016\) & \(0.016\) & \(0.003\) & \(0.003\) & \(0.003\) & \(7.185\mathrm{e}{-4}\) & \(7.186\mathrm{e}{-4}\) & \(7.414\mathrm{e}{-4}\)\\
 & & \((0.004)\) & \((0.003)\) & \((0.004)\) & \((1.331\mathrm{e}{-4})\) & \((1.332\mathrm{e}{-4})\) & \((1.351\mathrm{e}{-4})\) & \((8.316\mathrm{e}{-6})\) & \((8.318\mathrm{e}{-6})\) & \((9.339\mathrm{e}{-6})\)\\
\addlinespace[2pt]
 & \(L_{\mathrm{Stein}}\) & \(0.847\) & \(0.872\) & \(0.925\) & \(3.852\) & \(3.860\) & \(4.011\) & \(15.112\) & \(15.116\) & \(15.818\)\\
 & & \((0.164)\) & \((0.173)\) & \((0.197)\) & \((0.153)\) & \((0.153)\) & \((0.170)\) & \((0.159)\) & \((0.159)\) & \((0.201)\)\\
\addlinespace[5pt]
Cluster & \(e_\mu\) & \(0.369\) & \(0.370\) & \(0.376\) & \(0.376\) & \(0.376\) & \(0.380\) & \(0.377\) & \(0.377\) & \(0.385\)\\
 & & \((0.083)\) & \((0.083)\) & \((0.085)\) & \((0.036)\) & \((0.036)\) & \((0.036)\) & \((0.018)\) & \((0.018)\) & \((0.019)\)\\
\addlinespace[2pt]
 & \(e_\Sigma\) & \(0.584\) & \(0.589\) & \(0.608\) & \(0.656\) & \(0.656\) & \(0.668\) & \(0.678\) & \(0.678\) & \(0.693\)\\
 & & \((0.071)\) & \((0.072)\) & \((0.077)\) & \((0.024)\) & \((0.024)\) & \((0.026)\) & \((0.010)\) & \((0.010)\) & \((0.010)\)\\
\addlinespace[2pt]
 & \(\mathrm{MSE}_\Sigma\) & \(0.016\) & \(0.016\) & \(0.016\) & \(0.003\) & \(0.003\) & \(0.003\) & \(7.184\mathrm{e}{-4}\) & \(7.185\mathrm{e}{-4}\) & \(7.396\mathrm{e}{-4}\)\\
 & & \((0.003)\) & \((0.003)\) & \((0.003)\) & \((1.278\mathrm{e}{-4})\) & \((1.278\mathrm{e}{-4})\) & \((1.291\mathrm{e}{-4})\) & \((8.112\mathrm{e}{-6})\) & \((8.108\mathrm{e}{-6})\) & \((9.266\mathrm{e}{-6})\)\\
\addlinespace[2pt]
 & \(L_{\mathrm{Stein}}\) & \(0.839\) & \(0.852\) & \(0.905\) & \(3.843\) & \(3.848\) & \(3.972\) & \(15.111\) & \(15.113\) & \(15.757\)\\
 & & \((0.159)\) & \((0.161)\) & \((0.181)\) & \((0.147)\) & \((0.147)\) & \((0.159)\) & \((0.152)\) & \((0.152)\) & \((0.196)\)\\
\bottomrule
\end{tabular}
\end{table}

When the contamination proportion increases to \(\varepsilon=0.30\), \method{}
has lower mean \(L_{\mathrm{Stein}}\) than both \fs{} and \hf{} in all 12 combinations
(Table~\ref{tab:contam030}).
Here the initial subset size \(h_{0.75}=0.75n\) used by \hseven{} exceeds the
number of regular observations, \(0.70n\), so the subset contains at least
\(0.05n\) outliers.

Under point contamination in Setting C, the means of \(e_\mu\),
\(\mathrm{MSE}_\Sigma\), and \(L_{\mathrm{Stein}}\) for \hseven{} are \(76.912\),
\(2.707\mathrm{e}{3}\), and \(1.056\mathrm{e}{4}\), respectively (see
Supplementary Table~\ref{tab:supp-omitted030}).
The corresponding values for \method{} are \(0.264\), \(0.002\), and
\(95.017\).
Under random-direction contamination, \hseven{} has slightly lower mean
\(L_{\mathrm{Stein}}\) than \method{} in all three settings.

\FloatBarrier

\begin{figure}[!htbp]
\centering
\includegraphics[width=0.98\textwidth]{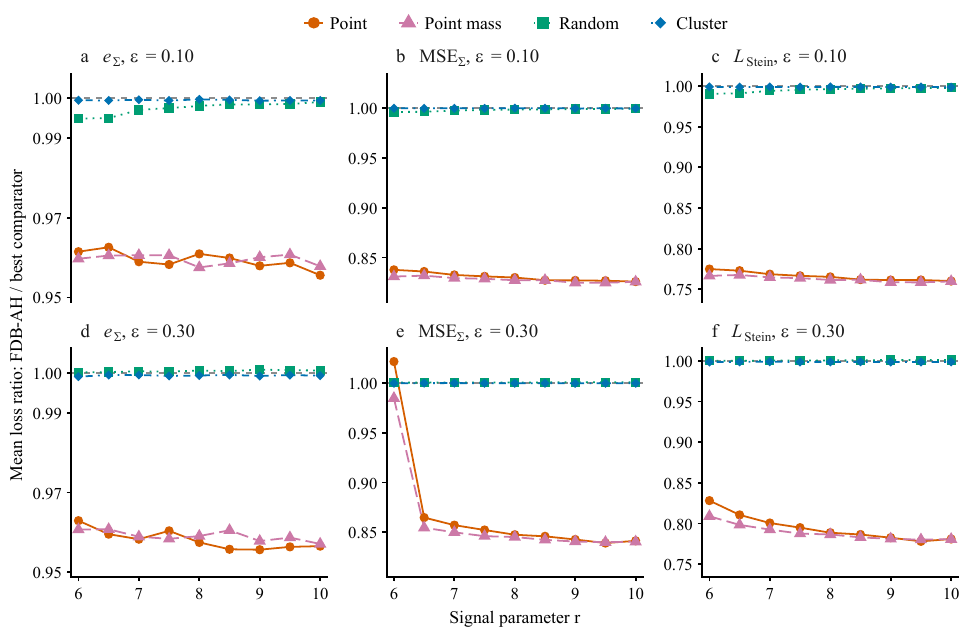}
\caption{Relative scatter losses for \method{} across signal strengths \(r\) in
Setting B.
	The top and bottom rows correspond to \(\varepsilon=0.10\) and
	\(\varepsilon=0.30\), respectively.
	From left to right, the columns show the ratios for \(e_\Sigma\),
	\(\mathrm{MSE}_\Sigma\), and \(L_{\mathrm{Stein}}\).}
\label{fig:relative-losses}
\end{figure}

\Needspace{17\baselineskip}
For each of the three scatter measures, Figure~\ref{fig:relative-losses} plots
the loss ratio against signal strength \(r\) in Setting B.
The numerator is the mean loss for \method{},
and the denominator is the smallest mean loss among \fs{}, \hf{}, and
\hseven{}.
A ratio below \(1\) indicates that \method{} has a lower mean loss than all
three comparison methods.
The relative reductions in \(\mathrm{MSE}_\Sigma\) and \(L_{\mathrm{Stein}}\) are
especially large under point and point-mass contamination.

At \(\varepsilon=0.30\), the ratios for all three losses are slightly above
\(1\) under random-direction contamination, and \hseven{} has the lowest mean
losses among the three methods.
At the same contamination proportion, the ratio for \(\mathrm{MSE}_\Sigma\)
is also above \(1\) under point contamination at \(r=6\), where \hf{} performs best.

\begin{figure}[!htbp]
\centering
\includegraphics[width=0.98\textwidth]{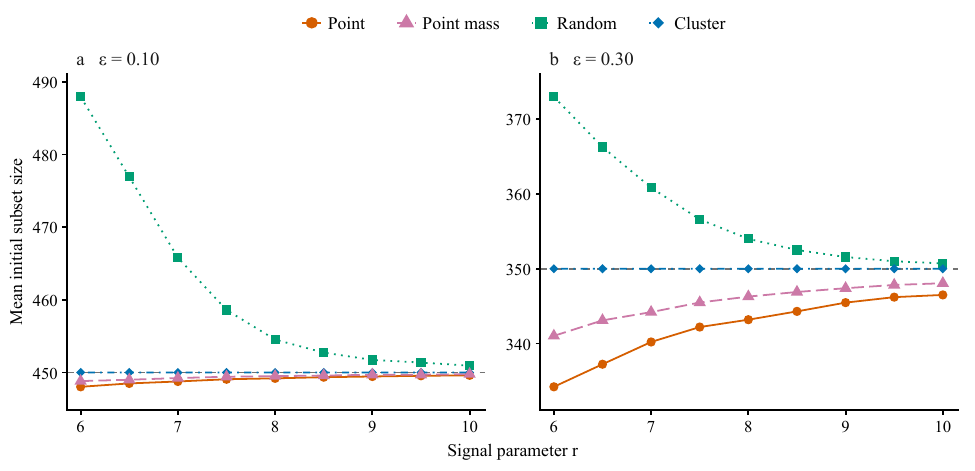}
\caption{Mean initial subset size for \method{} across signal strengths in
Setting B. The dashed grey horizontal lines indicate the number of
uncontaminated observations.}
\label{fig:supp-radius-h}
\end{figure}

Figure~\ref{fig:supp-radius-h} plots the mean \(\widehat h\) for \method{}
against \(r\) in Setting B.
Under cluster contamination, the mean \(\widehat h\) stays close to the number
of regular observations for all nine values of \(r\). It ranges from \(450.000\) to \(450.006\) at
\(\varepsilon=0.10\) and from \(350.000\) to \(350.004\) at \(\varepsilon=0.30\).

As \(r\) increases from \(6\) to \(10\), the mean subset size increases
gradually under both point and point-mass contamination.
At \(\varepsilon=0.10\), the mean increases from \(448.035\) to \(449.610\)
under point contamination and from \(448.822\) to \(449.775\) under point-mass
contamination.
At \(\varepsilon=0.30\), the mean increases from \(334.214\) to \(346.502\)
under point contamination and from \(341.027\) to \(348.066\) under point-mass
contamination.

Under random-direction contamination, the mean subset size exceeds the number
of regular observations, and the gap narrows as \(r\) increases.
At \(\varepsilon=0.10\), the mean decreases from \(487.935\) to \(450.942\). At
\(\varepsilon=0.30\), it decreases from \(373.005\) to \(350.713\).

\Needspace{23\baselineskip}
\section{Real data analysis}
\label{sec:real-data}

We analyse the Skin Segmentation dataset \citep{bhatt2009skin} from the
ADBench anomaly detection benchmark \citep{han2022adbench}.
Each observation is a pixel with three colour components, blue, green, and red
(\(B,G,R\)), so \(p=3\).
We use the colour data and class labels supplied by the benchmark, treating
non-skin pixels as the reference class and skin pixels as anomalies.

We select \(2500\) pixels using only the original row indices and retain their
original order.
The selected sample contains \(2009\) non-skin pixels and \(491\) skin pixels,
the latter accounting for \(19.640\%\) of the sample.
For each colour component, we subtract its median in the selected sample and
divide by \(Q_n\), the robust scale estimator of \citet{rousseeuw1993qn}.

We compare the four FDB methods in Section~\ref{sec:simulation}, Stable MCD,
and \fsm{}.
Supplementary Sections~\ref{sec:supp-stable-comparison}
and~\ref{sec:supp-fs-comparison} give the settings for Stable MCD and \fsm{},
respectively.
The FPR is the proportion of the \(2009\) non-skin pixels
classified as anomalous, and the FNR is the proportion of
the \(491\) skin pixels classified as normal.

Table~\ref{tab:skin-procedures} reports means over \(100\) runs of each
method on the same fixed sample.
For the four FDB methods, the initial error rates treat pixels inside the
initial subset as normal and those outside it as anomalous. The final error
rates are computed after reweighting.
In this example, Stable MCD does not reweight after selecting its subset, so
its two sets of error rates in the table are identical.
\fsm{} gives only a final list of anomalous observations, so dashes appear in
all three columns under \textquotedblleft{}Initial subset\textquotedblright{}
for this method.

\begin{table}[H]
	\centering
	\begin{threeparttable}
		\caption{Subset sizes, initial and final error rates, and running times for
		the Skin dataset.}
		\label{tab:skin-procedures}
		\small
		\setlength{\tabcolsep}{3pt}
		\renewcommand{\arraystretch}{1.15}
		\begin{tabular*}{\linewidth}{@{\extracolsep{\fill}}lrrrrrr@{}}
			\toprule
			& \multicolumn{3}{c}{Initial subset} & \multicolumn{3}{c}{Final results and running time} \\
			\cmidrule(lr){2-4}\cmidrule(lr){5-7}
			Method & \multicolumn{1}{c}{\(h\)} & FPR (\%) & FNR (\%) & FPR (\%) & FNR (\%) & Time (s)\\
			\midrule
			\method & 1770.880 & 13.692 & 7.527 & 15.306 & 2.411 & 0.466\\
			\fs & 1252.000 & 37.680 & 0.000 & 31.459 & 0.000 & 0.102\\
			\hf & 1252.000 & 37.680 & 0.000 & 31.459 & 0.000 & 0.077\\
			\hseven & 1875.000 & 12.135 & 22.360 & 14.213 & 20.415 & 0.077\\
			\fsm & --- & --- & --- & 24.938 & 0.000 & 0.157\\
			Stable MCD & 2308.880 & 9.269 & 99.000 & 9.269 & 99.000 & 7.289\\
			\bottomrule
		\end{tabular*}
	\end{threeparttable}
\end{table}

In the final detection results, \method{} has a lower mean FPR
(\(15.306\%\)) than \fsm{} (\(24.938\%\)).
Their mean FNRs are \(2.411\%\) and \(0\%\), respectively.
Of the six methods, Stable MCD has the lowest mean FPR
(\(9.269\%\)) and the highest mean FNR (\(99.000\%\)).

\fs{} selects \(h_{\min}=1252\) in every run, and after reweighting it retains
exactly the same pixels as \hf{} in that run.
The initial subsets of both methods contain only non-skin pixels. After
reweighting, their mean FPRs both decrease from \(37.680\%\) to
\(31.459\%\), and their FNRs remain at \(0\%\).
Compared with \method{}, \hseven{} has a larger initial subset on average and
includes more skin pixels.
After reweighting, the mean FPR for \hseven{} is only about
\(1.1\) percentage points lower, while its mean FNR is about
\(18\) percentage points higher.

\section{Discussion}
\label{sec:discussion}

In minimum covariance determinant estimation, the initial subset size \(h\)
determines how many observations contribute to the initial estimates of
location and scatter.
To balance robustness and estimation efficiency when the extent of
contamination is unknown, we compare the squared Mahalanobis distances of rank
\(h+1\) at different values of \(h\) and select the \(h\) at the first window
peak.
For a unimodal elliptical distribution with a small amount of one-point
contamination, Proposition~\ref{prop:population-first-window-entry} shows that
the first window peak of the population MCD ellipsoid radius occurs at retained
proportion \(\gamma_\varepsilon\).
At this proportion, the probability mass retained by at least one MCD solution
comes entirely from the uncontaminated distribution. At any larger candidate
proportion, every MCD solution assigns positive weight to the contaminating
point.
Thus \(\gamma_\varepsilon\) is the largest candidate retained proportion at which
an MCD solution can exclude the contaminating point entirely.
Equation~\eqref{eq:A-theta} gives its first-order approximation
\(\theta_\varepsilon\).

For the sample results, suppose the dimension \(p\) is fixed and a sample MCD
subset is used at every candidate size.
Under the window conditions of Theorem~\ref{thm:contaminated-window}, the
proportion \(\widehat h_{\MCD}/n\) of the full sample in the initial subset
eventually lies near \(\theta_\varepsilon\) almost surely.
If the observations are drawn independently from an uncontaminated distribution
\(F\) satisfying Condition~\ref{cond:elliptical}, then for any fixed window
parameter \(c\in(0,1)\) this proportion converges almost surely to \(1\) as
\(n\to\infty\).

\method{} uses FDB to approximate the MCD subset at each candidate size
\citep{zhang2023fdb}.
Simulations with contamination show that the mean \(e_\mu\) for
\method{} is usually close to the lowest among the four FDB methods.
The reductions in \(\mathrm{MSE}_\Sigma\) and \(L_{\mathrm{Stein}}\) for \method{}
are especially large under point and point-mass contamination.
The real data analyses show that even with the same reweighting rule, the
initial subset size affects the final FPR and FNR.

When \(p\) is close to \(h\), the subset covariance matrix may be difficult to
invert reliably. If the covariance matrix or its inverse has a sparse
structure, the initial robust estimates can be regularized
\citep{avella2018robust}.
The choice of the initial subset size \(h\) in such high-dimensional problems
warrants further study.

\section*{Data availability}
The SKAB data are available at \url{https://github.com/waico/SKAB}.
The Skin data are available at
\url{https://github.com/Minqi824/ADBench/tree/main/adbench/datasets}.
The R package for \method{} is available at \url{https://github.com/chikunli/FDB-AH}.
Code for \fsm{} and Stable MCD is available at
\url{https://github.com/UniprJRC/FSDA-MATLAB_Coder} and
\url{https://github.com/qhengncsu/StableMCD}, respectively.

\section*{Declaration of the use of generative AI and AI-assisted technologies}
During the preparation of this work, the authors used ChatGPT (OpenAI) to
assist with mathematical exposition.
After using this tool or service, the authors reviewed and edited the content
as necessary and take full responsibility for the content of the publication.

\section*{Acknowledgement}
This work is supported by the University of Science and Technology of
China--Xinjiang Normal University Counterpart Cooperation and Development Joint
Fund (KY2040002501) and the National Natural Science Foundation of China
(12231017, 72293573 and 12401373).

\section*{Supplementary material}
The Supplementary Material contains technical details including proofs, as well
as additional simulation results and real data comparisons.

\FloatBarrier
\clearpage
\section*{Supplementary material}
\setcounter{section}{0}
\setcounter{equation}{0}
\setcounter{table}{0}
\setcounter{figure}{0}
\setcounter{lemma}{0}
\renewcommand{\thesection}{\Alph{section}}
\renewcommand{\thesubsection}{\thesection.\arabic{subsection}}
\renewcommand{\theequation}{S\arabic{equation}}
\renewcommand{\thetable}{S\arabic{table}}
\renewcommand{\thefigure}{S\arabic{figure}}
\renewcommand{\thelemma}{S\arabic{lemma}}
\renewcommand{\theHsection}{supp.\Alph{section}}
\renewcommand{\theHsubsection}{supp.\Alph{section}.\arabic{subsection}}
\renewcommand{\theHequation}{supp.\arabic{equation}}
\renewcommand{\theHtable}{supp.\arabic{table}}
\renewcommand{\theHfigure}{supp.\arabic{figure}}
\renewcommand{\theHlemma}{supp.\arabic{lemma}}

\section{Proofs for the contaminated model}
\label{sec:supp-theory}

\subsection{Properties of the MCD functionals}
\label{sec:supp-model}

We use the functionals and assumptions in
Sections~\ref{sec:formulation} and \ref{sec:theory-summary}.
Throughout Section~A, \(p\ge1\) and \(x\ne0\) are fixed, and
all small-contamination statements concern \(\varepsilon\downarrow0\).

For clarity, affine equivariance refers to the following identities.
If \(A\) is nonsingular, \(b\in\R^p\), and \(P^{A,b}\) is the probability
measure of \(AZ+b\) for \(Z\sim P\), then corresponding MCD solutions satisfy
\[
 T_\gamma(P^{A,b})=AT_\gamma(P)+b,\qquad
 C_\gamma(P^{A,b})=AC_\gamma(P)A^\top
\]
\citep[Section~2]{CL12}. To verify these identities, associate a trimming
\(\phi\) for \(P\) with
\(\phi^{A,b}(y)=\phi\{A^{-1}(y-b)\}\) for \(P^{A,b}\).
Changing variables gives
\[
 \int\phi^{A,b}\dd P^{A,b}=\int\phi\dd P=\gamma,
 \qquad T_{P^{A,b}}(\phi^{A,b})=AT_P(\phi)+b.
\]
Subtracting this mean inside the covariance integral gives
\(C_{P^{A,b}}(\phi^{A,b})=AC_P(\phi)A^\top\).
The determinant of this covariance is \((\det A)^2\det C_P(\phi)\).
The correspondence between trimmings is bijective, and this positive
factor is the same for every candidate. It therefore preserves the
solutions of the determinant minimization problem.
Using the same \(c_\gamma\) also preserves squared Mahalanobis distances.

For an elliptical distribution with center \(\mu_0\) and positive-definite
scatter matrix \(S_0\), taking \(A=S_0^{-1/2}\) and
\(b=-S_0^{-1/2}\mu_0\) gives the standardization in
Section~\ref{sec:theory-summary}.
If the distribution has a finite, nonsingular covariance matrix, we
may choose \(S_0\) to be that covariance matrix and standardize to
covariance \(I_p\). This argument remains valid even if the covariance does not exist, in which
case \(S_0\) represents the scatter matrix in the elliptical representation.
The calibration in \eqref{eq:A-radial} follows from spherical symmetry.
Write \(Z=RU\), where \(U\) is uniformly distributed on the unit sphere and
independent of \(R\).
Then \(\Ee U=0\), \(\Ee(UU^\top)=I_p/p\), and
\[
 \int_{\{\|z\|^2\le q_\gamma\}}zz^\top\dd F(z)
   =\frac{M(\gamma)}p I_p.
\]
Dividing by the retained mass \(\gamma\) gives
\(C_\gamma(F)=M(\gamma)I_p/(p\gamma)\).
Multiplication by \(c_\gamma=p\gamma/M(\gamma)\) gives
\(\Sigma_\gamma(F)=I_p\).
To differentiate these quantities, first use
\(F_{R^2}(q_\gamma)=\gamma\) and \(f_\gamma>0\) to obtain
\(q_\gamma'=1/f_\gamma\). Next write
\(M(\gamma)=\int_0^{q_\gamma}u f_{R^2}(u)\dd u\).
The chain rule then gives
\(M'(\gamma)=q_\gamma f_\gamma q_\gamma'=q_\gamma\).

When the MCD solution is not unique, we select a solution with the
largest squared quantile radius. This convention concerns only
solutions with the same minimum determinant. Its existence near the
contamination transition is proved below.
Throughout Section~A, \(T_\gamma,C_\gamma,\Sigma_\gamma,D_\gamma\), and
\(\rho_\gamma^2\) refer to \(P_{\varepsilon,x}\).
Because the clean density is positive, every open shell just outside a
MCD ellipsoid has positive probability. Thus
\begin{equation}
 d_{\rm out}^2(\gamma;P_{\varepsilon,x})=\rho_\gamma^2.
 \label{eq:A-shell}
\end{equation}
Indeed, the outside infimum is at least the boundary radius, while each
shell of positive probability contains a support point arbitrarily close
to that boundary from outside.

The denominators \(L_x\) and \(H_x\) in \(V_x\) in \eqref{eq:A-beta} are positive.
To see this, write
\(A_p=\pi^{p/2}/\Gamma(p/2)\), where \(\Gamma\) is the gamma function.
Polar coordinates give \(f_{R^2}(u)=A_pu^{p/2-1}g(u)\). Thus
\[
 \eta_x=A_p\int_0^{q_x}u^{p/2-1}g(u)\dd u,\qquad
 M(\eta_x)=A_p\int_0^{q_x}u^{p/2}g(u)\dd u.
\]
The boundary terms can be written as integrals:
\[
 \frac{2q_xf_{\eta_x}}p
 =A_pg(q_x)\int_0^{q_x}u^{p/2-1}\dd u,\qquad
 \frac{2q_x^2f_{\eta_x}}{p+2}
 =A_pg(q_x)\int_0^{q_x}u^{p/2}\dd u.
\]
Subtracting the corresponding expressions gives
\begin{equation}
 \begin{split}
 L_x&=A_p\int_0^{q_x}u^{p/2-1}\{g(u)-g(q_x)\}\dd u>0,\\
 H_x&=A_p\int_0^{q_x}u^{p/2}\{g(u)-g(q_x)\}\dd u>0.
 \end{split}
 \label{eq:A-LH-integrals}
\end{equation}
Both integrands are strictly positive for \(0<u<q_x\), because
\(g'<0\) and \(q_x>0\). Their integrals are therefore positive.
Consequently \(V_x>0\) and \(\beta_x<1\). When \(p=1\), the second term
in \(V_x\) is zero. The value of \(\beta_x\) can be positive, zero, or negative.

\subsection{Conditional trimming}
\label{sec:conditional}

Let \(w\in[0,1]\) be the fraction of the contamination atom retained.
With \(w\) fixed, minimize the covariance determinant of
\begin{equation}
 \nu=(1-\varepsilon)\psi F+\varepsilon w\delta_x,\qquad
 (1-\varepsilon)\int\psi\dd F+\varepsilon w=\gamma,
 \quad 0\le\psi\le1.
 \label{eq:A-retained}
\end{equation}
The retained measure has location \(\gamma^{-1}\int z\dd\nu\) and
covariance defined by its centred second moment divided by \(\gamma\).
Write \(T_{\gamma,w}\), \(C_{\gamma,w}\), and
\(\Sigma_{\gamma,w}=c_\gamma C_{\gamma,w}\) for the conditional solution.
Its trimming ellipsoid, squared radius, and Mahalanobis distance are
denoted by \(E_{\gamma,w}\), \(\rho_{\gamma,w}^2\), and
\[
 D_{\gamma,w}(z)
 =(z-T_{\gamma,w})^\top\Sigma_{\gamma,w}^{-1}(z-T_{\gamma,w}),\qquad
 E_{\gamma,w}=\{z:D_{\gamma,w}(z)\le\rho_{\gamma,w}^2\}.
\]
These objects depend on \(\varepsilon\), which is suppressed in their
subscripts. The subscript \(w\) denotes a conditional solution.
Define its objective and signed atom-to-boundary gap by
\begin{equation}
 \ell_\gamma(w)=\log\det C_{\gamma,w},\qquad
 B_\gamma(w)=D_{\gamma,w}(x)-\rho_{\gamma,w}^2.
 \label{eq:A-objective-gap}
\end{equation}

\begin{lemma}
\label{lem:conditional}
There are a compact neighborhood \(J\subset(0,1)\) of \(\eta_x\) and
\(\varepsilon_0>0\) such that, for \(\gamma\in J\),
\(0\le\varepsilon<\varepsilon_0\), and \(w\in[0,1]\), the conditional
solution exists and is unique up to clean null sets. Its location, scatter,
and squared radius are restrictions of a \(C^2\) function of
\((\gamma,\varepsilon,\zeta)\), evaluated at \(\zeta=\varepsilon w\).
The full MCD solutions are obtained by minimizing \(\ell_\gamma(w)\) over
\([0,1]\).
\end{lemma}

\begin{proof}
We first establish compactness of the minimizers, then identify them
through their moment equations.

\emph{Existence and bounded support.}
For \(R>0\), write \(B_R=\{z\in\R^p:\|z\|\le R\}\).
Restrict \(\gamma\) to
\([\gamma_-,\gamma_+]\subset(0,1)\), with
\(\gamma_-<\eta_x<\gamma_+\). Choose
\(\varepsilon_0<\min(\gamma_-,1-\gamma_+)\).
For \(0\le\varepsilon<\varepsilon_0\), the clean retained fraction
\((\gamma-\varepsilon w)/(1-\varepsilon)\) is uniformly separated
from zero and one, for all \(w\in[0,1]\).
Indeed, after fixing a smaller \(\varepsilon_0\) if necessary, it lies
between the constants
\[
 a_-:=\frac{\gamma_--\varepsilon_0}{1-\varepsilon_0}>0,
 \qquad a_+:=\frac{\gamma_+}{1-\varepsilon_0}<1.
\]
These bounds make every conditional constraint feasible: a centred
clean ball can supply any probability in \([a_-,a_+]\).
A one-dimensional projection of \(F\) has a continuous distribution,
independent of its unit direction. Uniform continuity of its distribution
function gives
\[
 \sup_{\|u\|=1,\,b\in\R}
 F\{|u^\top Z-b|\le d\}\longrightarrow0\qquad(d\downarrow0).
\]
Choose \(d>0\), and reduce \(\varepsilon_0\) if necessary, so that each
such slab has \(P_{\varepsilon,x}\)-probability below \(\gamma_-/2\).
Every feasible retained measure has at least \(\gamma_-/2\) mass
outside the slab centred at its own projected mean. Thus its covariance
satisfies
\[
 \lambda_{\min}(C)\ge d^2\gamma_-/(2\gamma_+)>0.
\]
To obtain this bound, use \(\nu\le P_{\varepsilon,x}\) and
\(\nu(\R^p)=\gamma\). For every unit \(u\),
\[
 u^\top Cu=\gamma^{-1}\int\{u^\top(z-T)\}^2\dd\nu(z)
 \ge\frac{d^2}{\gamma}
       \nu\{|u^\top(z-T)|>d\}
 \ge\frac{d^2\gamma_-}{2\gamma_+}.
\]
Denote this lower bound by \(\lambda_0\).
A centred clean ball of the required probability, together with the
prescribed atomic mass, supplies a competitor of uniformly bounded
determinant. We may restrict a minimizing sequence to that determinant
bound. The eigenvalue lower bound then bounds \(\lambda_{\max}(C)\).
More explicitly, the competitor is supported in a ball containing
both \(x\) and the clean ball with probability \(a_+\).
Its second moments, and hence its determinant, are uniformly bounded.
Choose a constant \(D_0\) larger than that bound. If \(\det C\le D_0\),
then
\[
 \lambda_{\max}(C)\le D_0\lambda_0^{-(p-1)}=:\Lambda_0,
 \qquad \tr C\le p\Lambda_0.
\]
Choose a fixed ball whose \(P_{\varepsilon,x}\)-mass outside is less
than \(\gamma_-/2\). Every retained measure puts at least
\(\gamma_-/2\) mass in that ball. Integrating \(\|z-T\|^2\) there
shows that the covariance trace bound also bounds \(T\).
In detail, if that ball has radius \(R_0\), then
\[
 \gamma\tr C=\int\|z-T\|^2\dd\nu(z)
 \ge\frac{\gamma_-}{2}\bigl(\max\{\|T\|-R_0,0\}\bigr)^2.
\]
In particular,
\(\|T\|\le R_0+(2\gamma_+p\Lambda_0/\gamma_-)^{1/2}\).
Write \(T_0\) for this upper bound.

Choose an inner ball with uncontaminated probability exceeding every possible
clean retained fraction. The preceding bounds allow an outer ball,
also containing \(x\), such that every point outside it has a larger
Mahalanobis distance than every point in the inner ball.
For example, let the inner ball have radius \(R_1\) and clean
probability greater than \(a_+\). Choose \(R_2>\max(R_1,\|x\|,T_0)\)
so that
\[
 \frac{(R_2-T_0)^2}{\Lambda_0}
 >\frac{(R_1+T_0)^2}{\lambda_0}.
\]
The left side bounds raw squared distances from below outside the
outer ball. The right side bounds them from above inside the inner ball.
Let \(\xi_-\) be the selected clean mass outside the outer ball,
and choose an equal amount \(\xi_+\) of unselected clean mass in the
inner ball. This is possible because \(F(B_{R_1})>a_+\), whereas
the total retained clean fraction is at most \(a_+\).
Define the replacement measure and its moments before comparing them:
\[
 \nu'=\nu-\xi_-+\xi_+,\qquad
 T'=\gamma^{-1}\int z\dd\nu',\qquad
 C'=\gamma^{-1}\int(z-T')(z-T')^\top\dd\nu'.
\]
If \(\xi_-(\R^p)>0\), the strict distance ordering gives
\[
 \int(z-T)^\top C^{-1}(z-T)\dd(\xi_+-\xi_-)<0.
\]
The identity
\[
 C'=\gamma^{-1}\int(z-T)(z-T)^\top\dd\nu'
                  -(T'-T)(T'-T)^\top
\]
therefore implies
\begin{equation}
 \tr(C^{-1}C')
 \le\gamma^{-1}\int(z-T)^\top C^{-1}(z-T)\dd\nu'
 <\gamma^{-1}\int(z-T)^\top C^{-1}(z-T)\dd\nu=p.
 \label{eq:A-replacement-trace}
\end{equation}
Writing \(\lambda'_1,\ldots,\lambda'_p\) for the eigenvalues of
\(C^{-1/2}C'C^{-1/2}\), we obtain
\[
 \frac{\det C'}{\det C}
 =\prod_{j=1}^p\lambda'_j
 \le\left\{p^{-1}\tr(C^{-1}C')\right\}^p<1.
\]
Replacement preserves the prescribed atomic weight.
It confines a minimizing sequence to a common bounded ball and shows
that every minimizer has this support bound.
On that ball, take a weak-* convergent subsequence of the clean
trimmings in \(L^\infty(F)\).
Its limit remains between zero and one and preserves mass.
The coordinate functions and their products are bounded on the ball,
so all retained first and second moments converge.
The determinant consequently converges to the infimum.
This proves existence without assuming a global second moment.

\emph{Moment equations and the clean limit.}
Fix a conditional minimizer and put
\(Q(z)=(z-T)^\top C^{-1}(z-T)\).
Suppose positive selected clean mass has larger \(Q\) than positive
unselected clean mass. Choose two such portions whose \(Q\) values
are separated by a positive amount, and normalize them to equal mass.
Transfer an amount \(s>0\) from the first portion to the second.
The derivative of \(\log\det C\) at \(s=0\) is
\(\gamma^{-1}\int Q\dd(\xi_+-\xi_-)<0\).
For small \(s\), the transfer is feasible and decreases the objective,
a contradiction. Thus the essential upper bound of selected
\(Q\) values cannot exceed the essential lower bound of unselected
\(Q\) values. A threshold separates them.
Since \(C\) is positive definite and \(F\) has a density, every
level set of \(Q\) has \(F\)-probability zero.
The mass constraint lies strictly between zero and the full clean mass,
so this threshold is finite and positive. It follows that
\(\psi=\1_{E_{\gamma,w}}\), up to a clean null set.
For this proof, write \(T=T_{\gamma,w}\),
\(S=\Sigma_{\gamma,w}=c_\gamma C_{\gamma,w}\), and
\(r=\rho_{\gamma,w}^2\).
Thus \(S\) is the scaled covariance, not the raw covariance, and
\(r\) is the squared radius in its metric.
Put \(E=\{z:(z-T)^\top S^{-1}(z-T)\le r\}\) and
\(\zeta=\varepsilon w\).
The three moment integrals are
\begin{equation}
 J_0(T,S,r)=\int_E1\dd F,\quad
 J_1(T,S,r)=\int_Ez\dd F,\quad
 J_2(T,S,r)=\int_Ezz^\top\dd F.
 \label{eq:A-moment-integrals}
\end{equation}
They satisfy
\begin{subequations}\label{eq:A-moments}
\begin{align}
 (1-\varepsilon)J_0+\zeta&=\gamma,
      \label{eq:A-mass}\\
 (1-\varepsilon)J_1+\zeta x&=\gamma T,
      \label{eq:A-first-moment}\\
 (1-\varepsilon)J_2+\zeta xx^\top
      &=\frac{\gamma}{c_\gamma}S+\gamma TT^\top.
      \label{eq:A-second-moment}
\end{align}
\end{subequations}

Consider any sequence
\((\gamma,\varepsilon,\zeta)\to(\eta_x,0,0)\) with
\(0\le\zeta\le\varepsilon\).
The support bound gives a subsequential retained-measure limit
\(\nu_*\le F\) of mass \(\eta_x\), and convergence of its moments.
The minimum covariance determinant for \(F\) at that mass is
\[
 d_*=\det\!\left\{\frac{M(\eta_x)}{p\eta_x}I_p\right\}
     =\left\{\frac{M(\eta_x)}{p\eta_x}\right\}^{p}.
\]
Feasibility gives \(\det C(\nu_*)\ge d_*\).
For the reverse bound, use the centred clean ball of probability
\((\gamma-\zeta)/(1-\varepsilon)\) together with atomic mass
\(\zeta\). Its determinant tends to \(d_*\), so optimality gives
\(\det C(\nu_*)\le d_*\).
Uniqueness under Condition~\ref{cond:elliptical} now yields
\(T\to0\) and \(S\to I_p\).

For any \(0<s<q_x\), the ellipsoid probabilities at squared
thresholds \(q_x-s\) and \(q_x+s\) tend respectively to
\[
 F_{R^2}(q_x-s)<\eta_x<F_{R^2}(q_x+s).
\]
By \eqref{eq:A-mass}, the required clean probability tends to
\(\eta_x\). Hence \(q_x-s<r<q_x+s\) eventually, proving
\(r\to q_x\).
The argument applies to every parameter sequence, including varying
\(w\in[0,1]\), and therefore gives uniform convergence in \(w\).

\emph{The Jacobian.}
Transform the integrals by \(z=T+\sqrt r\,S^{1/2}y\).
Their domains become the fixed unit ball, and their integrands and
Jacobian determinants are \(C^2\) near \((T,S,r)=(0,I_p,q_x)\).
The moment equations are therefore \(C^2\).
The transformed density is evaluated on a common compact set.
Its derivatives through order two are bounded there, because \(g\)
is \(C^2\). The square-root map on positive definite matrices and
\(\sqrt r\) are smooth in this neighborhood.
These bounds justify differentiation under the integral sign for
each of \(J_0,J_1,J_2\) in \eqref{eq:A-moment-integrals}. To differentiate with respect to the estimates, let
\[
 T_v=vu,\qquad S_v=I_p+vA,\qquad r_v=q_x+vs,
\]
where \(A\) is symmetric. For a unit direction \(U\), write the
Euclidean radial boundary as \(R_v(U)U\). Expanding
\((R_vU-T_v)^\top S_v^{-1}(R_vU-T_v)=r_v\) gives
\[
 R_v^2(U)=q_x+v\Delta(U)+O(v^2),\qquad
 \Delta(U)=s+2\sqrt{q_x}\,U^\top u+q_xU^\top AU.
\]
This representation is valid for all directions near the clean ball,
since the origin remains in the ellipsoid's interior.
Let \(J_0(v),J_1(v)\) and \(J_2(v)\) denote its clean mass, first and second moment. Radial integration and differentiation
at the moving upper endpoint give
\begin{align}
 J_0'(0)&=f_{\eta_x}\Ee\Delta(U)
   =f_{\eta_x}\left(s+\frac{q_x}{p}\tr A\right),\nonumber\\
 J_1'(0)&=\sqrt{q_x}f_{\eta_x}\Ee\{U\Delta(U)\}
   =\frac{2q_xf_{\eta_x}}p u,\nonumber\\
 J_2'(0)&=q_xf_{\eta_x}\Ee\{UU^\top\Delta(U)\}\nonumber\\
   &=\frac{q_xf_{\eta_x}s}{p}I_p+
     \frac{q_x^2f_{\eta_x}}{p(p+2)}
                 \{2A+(\tr A)I_p\}.
 \label{eq:A-moment-derivatives}
\end{align}
For example,
\(J_2(v)=\Ee_U[UU^\top\int_0^{R_v^2(U)}u f_{R^2}(u)\dd u]\).
The factor \(q_x\) arises from the boundary value of the integrand.
The spherical identities used in \eqref{eq:A-moment-derivatives} are
\[
 \Ee U=0,\quad \Ee(UU^\top)=I_p/p,\quad
 \Ee\{UU^\top(U^\top AU)\}
       =\frac{2A+(\tr A)I_p}{p(p+2)}.
\]
The last identity follows by expanding \(U^\top AU\) and using
\(\Ee(U_iU_jU_kU_l)
=(\delta_{ij}\delta_{kl}+\delta_{ik}\delta_{jl}
+\delta_{il}\delta_{jk})/\{p(p+2)\}\), where
\(\delta_{ij}\) is the Kronecker delta. Odd spherical moments vanish.
One can verify the fourth-moment constants directly.
For \(p\ge2\), rotation invariance gives
\(\Ee U_1^4=3\Ee(U_1^2U_2^2)\): expand
\(\Ee\{(U_1+U_2)/\sqrt2\}^4=\Ee U_1^4\) and use the vanishing
odd terms. Expanding \(\Ee(\sum_iU_i^2)^2=1\) then gives
\(\Ee(U_1^2U_2^2)=1/\{p(p+2)\}\).
Sign symmetry ensures that every fourth moment with an odd multiplicity is zero.
These observations give the fourth-moment identity. For \(p=1\),
\(U\) takes the values \(-1\) and \(1\) with equal probabilities,
and the same formula holds directly.

Hold \((\gamma,\varepsilon,\zeta)=(\eta_x,0,0)\) fixed.
Differentiating \eqref{eq:A-mass}--\eqref{eq:A-second-moment}
and using \eqref{eq:A-moment-derivatives}, a kernel increment satisfies
\begin{subequations}\label{eq:A-kernel}
\begin{align}
 f_{\eta_x}(s+q_x\tr A/p)&=0,\label{eq:A-kernel-mass}\\
 L_xu&=0,\label{eq:A-kernel-first}\\
 M(\eta_x)A&=q_xf_{\eta_x}sI_p+
 \frac{q_x^2f_{\eta_x}}{p+2}\{2A+(\tr A)I_p\}.
 \label{eq:A-kernel-second}
\end{align}
\end{subequations}
The trace of \eqref{eq:A-kernel-second}, followed by
\eqref{eq:A-kernel-mass}, gives
\[
 M(\eta_x)\tr A
 =pq_xf_{\eta_x}s+q_x^2f_{\eta_x}\tr A
 =pq_xf_{\eta_x}(s+q_x\tr A/p)=0.
\]
Thus \(\tr A=0\), then \(s=0\) by \eqref{eq:A-kernel-mass}.
Equation~\eqref{eq:A-kernel-second} reduces to \(H_xA=0\).
Since \(L_x,H_x>0\), \eqref{eq:A-kernel-first} also gives \(u=0\).
The square Jacobian with respect to \((T,S,r)\) is therefore invertible.

The implicit function theorem gives a unique \(C^2\) solution of
\eqref{eq:A-moments} near \((0,I_p,q_x)\).
The uniform limiting argument places every conditional minimizer in
this solution neighborhood after shrinking \(J\) and \(\varepsilon_0\).
The clean trimming is its ellipsoid indicator, up to \(F\)-null sets,
so the conditional minimizer is unique.
The equations also define a smooth extension to nearby parameter
values outside the feasible probability model.
Finally, every feasible trimming has some atomic weight \(w\in[0,1]\).
At that weight its determinant is at least the conditional minimum,
which is attainable. Minimizing the continuous conditional objective
over the compact weight interval gives exactly all full MCD solutions.
\end{proof}

\subsection{Expansion near the entry of contamination}
\label{sec:boundary-expansion}

Define
\begin{equation}
 a_x=\frac{q_x\eta_x}{M(\eta_x)}-1,\qquad
 b_x=\frac{\eta_x}{f_{\eta_x}}-q_xa_x,\qquad
 t=\frac{\gamma-\eta_x}{\varepsilon}.
 \label{eq:A-abt}
\end{equation}
The coefficient \(a_x>0\) is the first-order average increase in
scaled scatter at \(\gamma=\eta_x\). It does not imply an increase
in every eigenvalue.
The coefficient \(b_x\) is the radius derivative when \(w=0\) and
\(\gamma=\eta_x\) remains fixed.

\begin{lemma}
\label{lem:boundary-expansion}
Under Condition~\ref{cond:elliptical}, take \(J\) and
\(\varepsilon_0\) from Lemma~\ref{lem:conditional}.
For \(\gamma=\eta_x+\varepsilon t\), uniformly over bounded \(t\)
and \(w\in[0,1]\),
\begin{align}
 \rho_{\gamma,w}^2
 &=q_x+\varepsilon\{b_x+(t-w)/f_{\eta_x}\}
       +O(\varepsilon^2),\label{eq:A-radius-expansion}\\
 B_\gamma(w)
 &=\frac{\varepsilon}{f_{\eta_x}}
       \{\beta_xw-t-\eta_x\}+O(\varepsilon^2).
       \label{eq:A-gap-expansion}
\end{align}
Moreover, for every \(\gamma\in J\), \(0<\varepsilon<\varepsilon_0\),
and \(w\in[0,1]\),
\begin{equation}
 \partial_w\ell_\gamma(w)
   =\frac{\varepsilon c_\gamma}{\gamma}B_\gamma(w).
 \label{eq:A-envelope}
\end{equation}
At the weight endpoints the derivatives are those of the smooth
extension in Lemma~\ref{lem:conditional}.
\end{lemma}

\begin{proof}
First fix \(\gamma=\eta_x\) and write
\[
 T_{\gamma,w}=\varepsilon u+O(\varepsilon^2),\quad
 \Sigma_{\gamma,w}=I_p+\varepsilon A+O(\varepsilon^2),\quad
 \rho_{\gamma,w}^2=q_x+\varepsilon s+O(\varepsilon^2).
\]
For \(J_k=J_k(T,\Sigma,r)\) in \eqref{eq:A-moment-integrals},
\eqref{eq:A-moment-derivatives} gives
\[
 J_0=\eta_x+\varepsilon J'_0(0)+O(\varepsilon^2),\quad
 J_1=\varepsilon J'_1(0)+O(\varepsilon^2),\quad
 J_2=\frac{M(\eta_x)}pI_p+\varepsilon J'_2(0)+O(\varepsilon^2).
\]
Substitution in \eqref{eq:A-mass} yields
\[
 \eta_x+\varepsilon\{J'_0(0)-\eta_x+w\}
       +O(\varepsilon^2)=\eta_x,
\]
and therefore
\begin{equation}
 f_{\eta_x}(s+q_x\tr A/p)=\eta_x-w.
 \label{eq:A-linear-mass}
\end{equation}
Likewise, \eqref{eq:A-first-moment} becomes
\[
 \varepsilon\{J'_1(0)+wx\}+O(\varepsilon^2)
       =\eta_x\varepsilon u+O(\varepsilon^2),
\]
so that
\begin{equation}
 (\eta_x-2q_xf_{\eta_x}/p)u=L_xu=wx.
 \label{eq:A-linear-first}
\end{equation}
For \eqref{eq:A-second-moment}, use
\(\eta_x/c_{\eta_x}=M(\eta_x)/p\) and \(TT^\top=O(\varepsilon^2)\).
The coefficients of \(\varepsilon\) on its two sides give
\[
 J'_2(0)-\frac{M(\eta_x)}pI_p+wxx^\top
             =\frac{M(\eta_x)}p A.
\]
Multiplying by \(p\) and substituting \eqref{eq:A-moment-derivatives},
\begin{equation}
 \begin{split}
 M(\eta_x)A={}&-M(\eta_x)I_p+q_xf_{\eta_x}sI_p\\
 &+\frac{q_x^2f_{\eta_x}}{p+2}\{2A+(\tr A)I_p\}+pwxx^\top.
 \end{split}
 \label{eq:A-linear-second}
\end{equation}

Taking the trace of \eqref{eq:A-linear-second} and using
\eqref{eq:A-linear-mass} gives, in order,
\begin{align}
 M(\eta_x)\tr A
 &=-pM(\eta_x)+pq_xf_{\eta_x}s+q_x^2f_{\eta_x}\tr A+pwq_x
       \nonumber\\
 &=-pM(\eta_x)+pq_x(\eta_x-w)+pwq_x
   =p\{q_x\eta_x-M(\eta_x)\}.
 \label{eq:A-trace-solution}
\end{align}
Thus \(\tr A=pa_x\), and \eqref{eq:A-linear-mass} implies
\(s=(\eta_x-w)/f_{\eta_x}-q_xa_x=b_x-w/f_{\eta_x}\).

To obtain the remaining entries, put \(A_0=A-(\tr A)I_p/p\).
Subtract \(p^{-1}\) times the trace of
\eqref{eq:A-linear-second}, multiplied by \(I_p\), from that equation:
\[
 M(\eta_x)A_0
  =\frac{2q_x^2f_{\eta_x}}{p+2}A_0
        +pw\{xx^\top-q_xI_p/p\}.
\]
Since \(H_x>0\), this gives
\(A_0=w(pxx^\top-q_xI_p)/H_x\).
Together with \eqref{eq:A-linear-first},
\begin{equation}
 u=\frac{wx}{L_x},\qquad
 A=a_xI_p+\frac{w(pxx^\top-q_xI_p)}{H_x},\qquad
 s=b_x-\frac{w}{f_{\eta_x}}.
 \label{eq:A-first-variations}
\end{equation}

We now let \(\gamma\) vary.
Let \(G(\gamma,\varepsilon,\zeta)\) collect the entries of
\(T,\Sigma,r\) from Lemma~\ref{lem:conditional}, and set
\(a_0=(\eta_x,0,0)\).
Write \(\partial_iG\) and \(\partial_{ij}G\) for entrywise first
and second partial derivatives in these three coordinates.
For \(v=(\varepsilon t,\varepsilon,\varepsilon w)\), Taylor's formula is
\begin{equation}
 \begin{split}
 G(a_0+v)={}&G(a_0)+\sum_{i=1}^3v_i\partial_iG(a_0)\\
 &+\sum_{i,j=1}^3v_iv_j
       \int_0^1(1-u)\partial_{ij}G(a_0+uv)\dd u .
 \end{split}
 \label{eq:A-Taylor}
\end{equation}
Fix any bound \(|t|\le T_*\).
All the segments in \eqref{eq:A-Taylor} lie in one compact
\(C^2\) neighborhood for small \(\varepsilon\).
If \(K_*\) bounds the norms of its second derivatives, the remainder
norm is at most
\[
 \tfrac12K_*\Big(\sum_i|v_i|\Big)^2
 \le\tfrac32K_*\|v\|^2
 \le\tfrac32K_*\varepsilon^2(T_*^2+2).
\]
At \(\varepsilon=\zeta=0\), the identities
\(T_\gamma(F)=0\), \(\Sigma_\gamma(F)=I_p\), and
\(r(\gamma,0,0)=q_\gamma\) imply
\[
 \partial_\gamma T=0,\qquad
 \partial_\gamma\Sigma=0,\qquad
 \partial_\gamma r=1/f_{\eta_x}.
\]
Equation~\eqref{eq:A-first-variations}, evaluated at \(w=0\)
and then at \(w=1\), identifies
\[
 \begin{aligned}
 \partial_\varepsilon T&=0,&
 \partial_\zeta T&=x/L_x,\\
 \partial_\varepsilon\Sigma&=a_xI_p,&
 \partial_\zeta\Sigma&=(pxx^\top-q_xI_p)/H_x,\\
 \partial_\varepsilon r&=b_x,&
 \partial_\zeta r&=-1/f_{\eta_x}.
 \end{aligned}
\]
All derivatives in these two displays are evaluated at \(a_0\).
Thus \eqref{eq:A-Taylor} gives the same first-order center and scatter
as \eqref{eq:A-first-variations}, and
\[
 r(\eta_x+\varepsilon t,\varepsilon,\varepsilon w)
 =q_x+\varepsilon t/f_{\eta_x}
       +\varepsilon b_x-\varepsilon w/f_{\eta_x}
       +O(\varepsilon^2),
\]
proving \eqref{eq:A-radius-expansion}.

Using \(\Sigma^{-1}=I_p-\varepsilon A+O(\varepsilon^2)\),
\begin{align}
 D_{\gamma,w}(x)
 &=q_x-\varepsilon\{2x^\top u+x^\top Ax\}
                  +O(\varepsilon^2)\nonumber\\
 &=q_x-\varepsilon\left\{q_xa_x+
          w\left(\frac{2q_x}{L_x}
                  +\frac{(p-1)q_x^2}{H_x}\right)\right\}
                  +O(\varepsilon^2)\nonumber\\
 &=q_x-\varepsilon(q_xa_x+wV_x)+O(\varepsilon^2).
 \label{eq:A-atom-distance-expansion}
\end{align}
Here \(V_x\) is defined in \eqref{eq:A-beta}.
Subtracting \eqref{eq:A-radius-expansion} gives
\[
 \begin{split}
 B_\gamma(w)
 &=-\varepsilon\{q_xa_x+b_x+wV_x+(t-w)/f_{\eta_x}\}
           +O(\varepsilon^2)\\
 &=\frac{\varepsilon}{f_{\eta_x}}
       \{(1-f_{\eta_x}V_x)w-t-\eta_x\}+O(\varepsilon^2),
 \end{split}
\]
by \eqref{eq:A-abt}. Equation~\eqref{eq:A-beta} now proves
\eqref{eq:A-gap-expansion}.

To prove \eqref{eq:A-envelope}, differentiate a retained measure of
fixed total mass \(\gamma\) along the conditional weight \(w\).
Write a dot for this derivative. Since
\(\int(z-T)\dd\nu=0\), differentiation of its centred covariance gives
\begin{equation}
 \dot C=\gamma^{-1}\int(z-T)(z-T)^\top\dd\dot\nu,\qquad
 \partial_w\log\det C
 =\gamma^{-1}\int(z-T)^\top C^{-1}(z-T)\dd\dot\nu .
 \label{eq:A-logdet-derivative}
\end{equation}
Here \(\dot\nu=\varepsilon\delta_x+\dot\nu_F\), with
\(\dot\nu_F\) the signed boundary derivative of
\((1-\varepsilon)\1_{E_{\gamma,w}}F\).
For a continuous function \(a\), the formula for a moving domain is
\[
 \int a\dd\dot\nu_F
  =(1-\varepsilon)\int_{\partial E_{\gamma,w}}
          a(z)g(\|z\|^2)v_w(z)\dd S(z),
\]
where \(v_w\) is the outward normal velocity and \(\dd S\) is
surface measure. Smoothness and boundedness of the ellipsoids
justify the formula. For \(p=1\), it is the endpoint derivative.
Differentiating \eqref{eq:A-mass} gives
\(\dot\nu_F(\R^p)=-\varepsilon\).
On the support of \(\dot\nu_F\), contained in
\(\partial E_{\gamma,w}\), the raw squared Mahalanobis distance is
\(c_\gamma\rho_{\gamma,w}^2\).
Consequently \eqref{eq:A-logdet-derivative} becomes
\[
 \partial_w\ell_\gamma(w)
 =\gamma^{-1}\{\varepsilon c_\gamma D_{\gamma,w}(x)
             -\varepsilon c_\gamma\rho_{\gamma,w}^2\},
\]
which is \eqref{eq:A-envelope}.
\end{proof}

The following derivative bounds will be used to locate the peak.
Write \(r(\gamma,\varepsilon,\zeta)\) and
\(B(\gamma,\varepsilon,\zeta)\) for the squared-radius and gap
functions. At \(a_0\), \(r(\gamma,0,0)=q_\gamma\) and
\(B(\gamma,0,0)=q_x-q_\gamma\).
For the smooth location and scatter functions in
Lemma~\ref{lem:conditional}, define
\[
 D_{\gamma,\varepsilon,\zeta}(z)
 =\{z-T(\gamma,\varepsilon,\zeta)\}^{\top}
   \Sigma(\gamma,\varepsilon,\zeta)^{-1}
   \{z-T(\gamma,\varepsilon,\zeta)\}.
\]
Thus \(D_{\gamma,\varepsilon,\varepsilon w}(z)=D_{\gamma,w}(z)\).
Equations~\eqref{eq:A-first-variations} and
\eqref{eq:A-atom-distance-expansion} give
\(\left.\partial_\zeta D_{\gamma,\varepsilon,\zeta}(x)
  \right|_{(\gamma,\varepsilon,\zeta)=a_0}=-V_x\).
It follows that
\begin{equation}
 \begin{aligned}
 \partial_\gamma r&=f_{\eta_x}^{-1},&
 \partial_\zeta r&=-f_{\eta_x}^{-1},\\
 \partial_\gamma B&=-f_{\eta_x}^{-1},&
 \partial_\zeta B&=-V_x+f_{\eta_x}^{-1}=\beta_x/f_{\eta_x}
 \end{aligned}
 \qquad\text{at }a_0 .
 \label{eq:A-base-derivatives}
\end{equation}
For either \(G=r\) or \(G=B\), the mean-value formula is
\[
 \partial_iG(a_0+v)-\partial_iG(a_0)
   =\sum_{j=1}^3v_j\int_0^1\partial_{ij}G(a_0+uv)\dd u .
\]
The bound on second derivatives used in \eqref{eq:A-Taylor} makes this
\(O(\varepsilon)\) uniformly for bounded \(t\) and \(w\in[0,1]\).
Applying \(\partial_w=\varepsilon\partial_\zeta\) gives
\begin{align}
 \partial_\gamma\rho_{\gamma,w}^2
 &=f_{\eta_x}^{-1}+O(\varepsilon),&
 \partial_w\rho_{\gamma,w}^2
 &=-\varepsilon/f_{\eta_x}+O(\varepsilon^2),\nonumber\\
 \partial_\gamma B_\gamma(w)
 &=-f_{\eta_x}^{-1}+O(\varepsilon),&
 \partial_w B_\gamma(w)
 &=\varepsilon\beta_x/f_{\eta_x}+O(\varepsilon^2).
 \label{eq:A-derivatives}
\end{align}
On the fixed compact interval \(J\), uniform continuity of the
derivatives at \((\gamma,0,0)\) also gives
\begin{equation}
 0<m_0\le\partial_\gamma\rho_{\gamma,w}^2\le M_0
 \quad(w=0,1),\qquad
 \partial_\gamma B_\gamma(w)\le-b_0<0
 \quad(0\le w\le1),
 \label{eq:A-fixed-derivative-bounds}
\end{equation}
for constants \(m_0,M_0,b_0>0\), after reducing \(\varepsilon_0\).

Differentiating \eqref{eq:A-envelope} gives
\begin{equation}
 \partial_w^2\ell_\gamma(w)
   =\frac{\varepsilon^2c_\gamma}{\gamma}
            \partial_\zeta B(\gamma,\varepsilon,\varepsilon w).
 \label{eq:A-objective-curvature}
\end{equation}
By \eqref{eq:A-base-derivatives} and continuity, after shrinking \(J\)
the objective is strictly convex for \(\beta_x>0\), and strictly
concave for \(\beta_x<0\).

\subsection{Proof of Theorem~\ref{thm:population-peak}}
\label{sec:population-peak}

Let \(W_\gamma=\argmin_{0\le w\le1}\ell_\gamma(w)\).
The radius convention in Section~A.1 is equivalently
\begin{equation}
 \rho_\gamma^2=\max_{w\in W_\gamma}\rho_{\gamma,w}^2.
 \label{eq:A-radius-selection}
\end{equation}
At a minimizing weight, \eqref{eq:A-envelope} gives
\(B_\gamma(0)\ge0\) if \(w=0\),
\(B_\gamma(1)\le0\) if \(w=1\), and \(B_\gamma(w)=0\) if \(0<w<1\).
These are the one-sided first-order conditions on \([0,1]\).
At zero, increasing the weight cannot decrease the objective, so
\(\partial_w\ell_\gamma(0)\ge0\). At one, the derivative from the
left must be nonpositive. At an interior minimum it is zero.
The factor \(\varepsilon c_\gamma/\gamma\) in
\eqref{eq:A-envelope} is positive, so the same signs hold for \(B_\gamma\).
The atom is therefore on or outside the trimming ellipsoid boundary when \(w=0\),
on or inside it when \(w=1\), and on it when \(0<w<1\).
To verify the quantile interpretation, write
\(r=\rho_{\gamma,w}^2\) and \(D=D_{\gamma,w}\).
Equation~\eqref{eq:A-mass} reads
\[
 (1-\varepsilon)F(D\le r)+\varepsilon w=\gamma.
\]
If \(w=0\), every smaller threshold has probability below \(\gamma\).
At \(r\), the full distribution has probability at least \(\gamma\),
including any boundary atom. If \(w=1\), the full probability at
\(r\) is \(\gamma\), and reducing the threshold strictly reduces the
uncontaminated probability. If \(0<w<1\), the atom lies on the boundary and
\[
 P_{\varepsilon,x}(D<r)=\gamma-\varepsilon w<\gamma,\qquad
 P_{\varepsilon,x}(D\le r)=\gamma+\varepsilon(1-w)>\gamma.
\]
Thus each optimal conditional radius is the full
\(\gamma\)-quantile.

Continuity of \(\ell_\gamma\) on \([0,1]\) makes \(W_\gamma\)
nonempty and compact, so its largest conditional radius is attained.
If \(\gamma_j\to\gamma\), first take a subsequence realizing
\(\limsup_j\rho_{\gamma_j}^2\), and then a further subsequence of
maximizing weights with \(w_j\to w_*\).
For every \(v\in[0,1]\),
\(\ell_{\gamma_j}(w_j)\le\ell_{\gamma_j}(v)\).
Passing to the limit gives \(w_*\in W_\gamma\), and continuity of
the conditional radius yields
\[
 \limsup_j\rho_{\gamma_j}^2
   =\rho_{\gamma,w_*}^2\le\rho_\gamma^2.
\]
The selected radius function is therefore upper semicontinuous and attains
a maximum on every closed subinterval of \(J\).

\begin{proof}[Proof of Theorem~\ref{thm:population-peak}]
For \(w=0\), put \(\alpha=\gamma/(1-\varepsilon)\).
The conditional solution is the clean MCD at mass \(\alpha\), with
raw covariance \(M(\alpha)I_p/(p\alpha)\).
Its scaled covariance is \(s_0I_p\), where
\(s_0=c_\gamma M(\alpha)/(p\alpha)>0\). Consequently
\[
 D_{\gamma,0}(x)=q_x/s_0,\qquad
 \rho_{\gamma,0}^2=q_\alpha/s_0,\qquad
 B_\gamma(0)=(q_x-q_\alpha)/s_0.
\]
The gap vanishes exactly at
\(\gamma_a=(1-\varepsilon)\eta_x\), is positive to its left,
and is negative to its right.
Take \(m_0,M_0,b_0\) from \eqref{eq:A-fixed-derivative-bounds}.
In particular,
\(m_0\le\partial_\gamma\rho_{\gamma,w}^2\le M_0\) for \(w=0,1\).
Thus, for \(w=0,1\) and \(\gamma_1\le\gamma_2\) in \(J\),
\begin{equation}
 m_0(\gamma_2-\gamma_1)
 \le\rho_{\gamma_2,w}^2-\rho_{\gamma_1,w}^2
 \le M_0(\gamma_2-\gamma_1).
 \label{eq:A-endpoint-increments}
\end{equation}
The constants do not depend on \(\varepsilon\).
We will also use \(\partial_\gamma B_\gamma(w)<0\) on this same
neighborhood. These uniform signs allow the transition analysis to
extend to the fixed left-hand portion of \(I_\varepsilon\).
Choose \(\delta_->0\) small enough that the left endpoint of
\(I_\varepsilon\) remains in \(J\).

\emph{Case 1: \(\beta_x>0\).}
By \eqref{eq:A-objective-curvature}, the minimizing weight is unique.
For \(\gamma\le\gamma_a\), it is zero.
During partial inclusion, it is the unique root of \(B_\gamma(w)=0\)
in \((0,1)\).
It reaches one at the zero \(\gamma_z\) of \(B_\gamma(1)\).
To locate this zero, put
\(z_\varepsilon=\eta_x+\varepsilon(\beta_x-\eta_x)
=\gamma_a+\beta_x\varepsilon\).
Equation~\eqref{eq:A-gap-expansion} gives a constant \(K_z\) with
\[
 |B_{z_\varepsilon}(1)|\le K_z\varepsilon^2 .
\]
For \(K>K_z/b_0\), \eqref{eq:A-fixed-derivative-bounds} implies
\[
 \begin{split}
 B_{z_\varepsilon-K\varepsilon^2}(1)
   &\ge(-K_z+b_0K)\varepsilon^2>0,\\
 B_{z_\varepsilon+K\varepsilon^2}(1)
   &\le(K_z-b_0K)\varepsilon^2<0.
 \end{split}
\]
Continuity and strict decrease in \(\gamma\) give a unique root
between these points. Therefore
\begin{equation}
 \gamma_z=\gamma_a+\beta_x\varepsilon+O(\varepsilon^2).
 \label{eq:A-gamma-z}
\end{equation}
Since \(\gamma_z-\gamma_a=\beta_x\varepsilon+O(\varepsilon^2)>0\),
the partial-inclusion interval is nonempty.
On this interval \(B_\gamma(0)<0<B_\gamma(1)\).
Strict increase in \(w\) gives its unique interior zero, which is
the unique minimum of the strictly convex objective.
Throughout partial inclusion, implicit differentiation and
\eqref{eq:A-derivatives} give
\[
 \frac{\dd w}{\dd\gamma}
   =-\frac{\partial_\gamma B_\gamma(w)}{\partial_wB_\gamma(w)}
   =\frac1{\beta_x\varepsilon}+O(1).
\]
The denominator is
\(\varepsilon\beta_x/f_{\eta_x}+O(\varepsilon^2)\), so it is at
least \(\varepsilon\beta_x/(2f_{\eta_x})\) for small \(\varepsilon\).
Factoring this denominator as
\((\varepsilon\beta_x/f_{\eta_x})\{1+O(\varepsilon)\}\)
justifies the ratio and its uniform remainder.
The optimized radius must also account for the changing weight:
\[
 \frac{\dd\rho_\gamma^2}{\dd\gamma}
 =\partial_\gamma\rho_{\gamma,w}^2
     +\partial_w\rho_{\gamma,w}^2\frac{\dd w}{\dd\gamma}
 =\frac1{f_{\eta_x}}-\frac1{\beta_xf_{\eta_x}}+O(\varepsilon)
 =-\frac{V_x}{\beta_x}+O(\varepsilon)<0.
\]
The minimizing weight and radius are continuous at both endpoints.
Taking \(\delta_+<\beta_x/2\), the radius function on \(I_\varepsilon\) increases up
to \(\theta_\varepsilon=\gamma_a\) and decreases thereafter.
Indeed, the right endpoint then lies strictly before \(\gamma_z\)
for all sufficiently small \(\varepsilon\).
On the left of \(\gamma_a\), the derivative is at least \(m_0\).
On the right within \(I_\varepsilon\), the derivative is at most
\(-V_x/(2\beta_x)\). Integrating from a point at distance at least
\(\tau\varepsilon\) to \(\gamma_a\) therefore gives the desired
height loss on either side.
This proves the exact peak location, and integration of the derivative
bounds gives \eqref{eq:A-peak-gap} with
\[
 c_\tau=\tau\min\{m_0,V_x/(2\beta_x)\}.
\]

\emph{Case 2: \(\beta_x<0\).}
Strict concavity restricts the minimizing weights to \(0\) and \(1\).
For any \(0<w<1\), strict concavity gives
\(\ell_\gamma(w)>(1-w)\ell_\gamma(0)+w\ell_\gamma(1)\),
which is no smaller than the smaller endpoint value. No interior
weight can therefore minimize.
Define \(I(\gamma,\varepsilon)=\int_0^1B_\gamma(w)\dd w\).
Integrating \eqref{eq:A-envelope} over \(w\in[0,1]\) gives
\begin{equation}
 \ell_\gamma(1)-\ell_\gamma(0)
       =\frac{\varepsilon c_\gamma}{\gamma}I(\gamma,\varepsilon).
 \label{eq:A-endpoint-objective}
\end{equation}
The positive multiplier preserves the sign.
At \(\varepsilon=0\), \(I(\gamma,0)=q_x-q_\gamma\).
Also \(\partial_\gamma I<0\) on a sufficiently small \(J\).
Apply the implicit function theorem to \(I=0\) at \((\eta_x,0)\).
Integration of \eqref{eq:A-gap-expansion} gives
\begin{equation}
 I(\eta_x+\varepsilon t,\varepsilon)
  =\frac{\varepsilon}{f_{\eta_x}}
        \{\beta_x/2-t-\eta_x\}+O(\varepsilon^2).
 \label{eq:A-integrated-gap}
\end{equation}
Thus there is a unique switch \(\gamma_s\) satisfying
\[
 \ell_{\gamma_s}(0)=\ell_{\gamma_s}(1),\qquad
 \gamma_s=\theta_\varepsilon+O(\varepsilon^2).
\]
In detail, the implicit function theorem first gives a root with
\(\gamma_s-\eta_x=O(\varepsilon)\).
Substituting \(t=(\gamma_s-\eta_x)/\varepsilon\) into
\eqref{eq:A-integrated-gap} and dividing by \(\varepsilon/f_{\eta_x}\) yields
\(t=\beta_x/2-\eta_x+O(\varepsilon)\).
This is precisely the expansion for \(\gamma_s\).
Strict decrease of \(I\) makes its sign positive to the left of the
root and negative to the right. The sign of the endpoint objective
difference then identifies the selected weight.
The minimizing weight is zero for \(\gamma<\gamma_s\) and one for
\(\gamma>\gamma_s\). At the switch both endpoints minimize the
determinant.
Subtracting \eqref{eq:A-radius-expansion} at \(w=1\) from the
same formula at \(w=0\), with
\(t=(\gamma_s-\eta_x)/\varepsilon=O(1)\), gives
\begin{equation}
 \rho_{\gamma_s,0}^2-\rho_{\gamma_s,1}^2
       =\frac{\varepsilon}{f_{\eta_x}}+O(\varepsilon^2)>0.
 \label{eq:A-radius-drop}
\end{equation}
The convention \eqref{eq:A-radius-selection} takes the value at \(w=0\)
there. Choose \(\delta_+\) so that \(M_0\delta_+<1/(8f_{\eta_x})\).
For small \(\varepsilon\), the jump is at least
\(\varepsilon/(2f_{\eta_x})\), and subsequent growth of the \(w=1\)
branch before the right endpoint of \(I_\varepsilon\) is less than
\(\varepsilon/(4f_{\eta_x})\).
To verify the latter bound, choose \(C_s\) such that
\(|\gamma_s-\theta_\varepsilon|\le C_s\varepsilon^2\).
For \(\gamma_s<\gamma\le\theta_\varepsilon+\delta_+\varepsilon\),
\eqref{eq:A-endpoint-increments} bounds that growth by
\(M_0(\delta_+\varepsilon+C_s\varepsilon^2)\).
The choice \(M_0\delta_+<1/(8f_{\eta_x})\) leaves a fixed margin,
so the expression is below \(\varepsilon/(4f_{\eta_x})\) for small
\(\varepsilon\). Consequently every point to the right of the
switch is below the switch value by at least
\(\varepsilon/(4f_{\eta_x})\).
The switch is therefore the unique maximum on \(I_\varepsilon\).
Each left-hand point outside the \(\tau\varepsilon\)-neighborhood
of \(\theta_\varepsilon\) is at least
\(\tau\varepsilon/2\) from \(\gamma_s\).
Indeed, this distance is at least
\(\tau\varepsilon-C_s\varepsilon^2\), which is at least
\(\tau\varepsilon/2\) once \(C_s\varepsilon\le\tau/2\).
More explicitly, \eqref{eq:A-endpoint-increments} gives
\[
 \rho_{\max}^2-\rho_\gamma^2
 =\rho_{\gamma_s,0}^2-\rho_{\gamma,0}^2
 \ge m_0(\gamma_s-\gamma)\ge m_0\tau\varepsilon/2
 \quad(\gamma\le\theta_\varepsilon-\tau\varepsilon).
\]
For \(\gamma\ge\theta_\varepsilon+\tau\varepsilon\) in
\(I_\varepsilon\), the radius deficit is at least
\(\varepsilon/(4f_{\eta_x})\) by \eqref{eq:A-radius-drop} and
\eqref{eq:A-endpoint-increments}.
This gives \eqref{eq:A-peak-gap} with
\[
 c_\tau=\min\{m_0\tau/2,\,1/(4f_{\eta_x})\}.
\]

\emph{Case 3: \(\beta_x=0\).}
Now \(\theta_\varepsilon=\gamma_a\).
Fix the transition range \(|t+\eta_x|\le1\).
Since \(\beta_x=0\), denote the remainder as
\[
 R_B(\varepsilon,t,w)
 =B_{\eta_x+\varepsilon t}(w)
       +\varepsilon(t+\eta_x)/f_{\eta_x}.
\]
Equation~\eqref{eq:A-gap-expansion} supplies a fixed \(K_0<\infty\)
such that
\[
 \sup_{\substack{|t+\eta_x|\le1\\0\le w\le1}}
       |R_B(\varepsilon,t,w)|\le K_0\varepsilon^2 .
\]
Choose \(C_0>2f_{\eta_x}K_0\) and reduce \(\varepsilon\) so that
\(C_0\varepsilon\le1\).
At \(\gamma=\theta_\varepsilon\pm C_0\varepsilon^2\),
\(t=-\eta_x\pm C_0\varepsilon\) belongs to this range. Hence
\begin{equation}
 \begin{split}
 B_{\theta_\varepsilon-C_0\varepsilon^2}(w)
       &\ge(C_0/f_{\eta_x}-K_0)\varepsilon^2>0,\\
 B_{\theta_\varepsilon+C_0\varepsilon^2}(w)
       &\le(-C_0/f_{\eta_x}+K_0)\varepsilon^2<0 .
 \end{split}
 \label{eq:A-critical-signs}
\end{equation}
The inequality \(\partial_\gamma B_\gamma(w)\le-b_0<0\) in
\eqref{eq:A-fixed-derivative-bounds} extends the positive sign
to all smaller \(\gamma\) in \(J\), and the negative sign to all
larger \(\gamma\) in \(J\), uniformly in \(w\).
Equation~\eqref{eq:A-envelope} makes the minimizing weight uniquely
zero on the left and uniquely one on the right.
Put \(\gamma_b=\theta_\varepsilon-C_0\varepsilon^2\) and
\(\gamma_c=\theta_\varepsilon+C_0\varepsilon^2\).
For \(\gamma_c\le\gamma\le\theta_\varepsilon+\delta_+\varepsilon\),
first compare \(w=0\) and \(w=1\) at the same retained fraction
\(\gamma_b\), using \eqref{eq:A-radius-expansion}.
Then bound the increase of the \(w=1\) branch from \(\gamma_b\)
to \(\gamma\) by \eqref{eq:A-endpoint-increments}:
\begin{align}
 \rho_{\gamma_b}^2-\rho_\gamma^2
 &=\{\rho_{\gamma_b,0}^2-\rho_{\gamma_b,1}^2\}
     -\{\rho_{\gamma,1}^2-\rho_{\gamma_b,1}^2\}\nonumber\\
 &\ge\frac{\varepsilon}{f_{\eta_x}}
       -M_0(\gamma-\gamma_b)-O(\varepsilon^2)\nonumber\\
 &\ge\frac{\varepsilon}{f_{\eta_x}}
       -M_0\delta_+\varepsilon-O(\varepsilon^2).
 \label{eq:A-critical-loss}
\end{align}
Choosing \(M_0\delta_+<1/(8f_{\eta_x})\) makes this at least
\(\varepsilon/(2f_{\eta_x})\) for small \(\varepsilon\).
The left branch is strictly increasing, and the radius function attains a
maximum. Every maximizer must therefore lie in
\([\gamma_b,\gamma_c]\), proving \eqref{eq:A-peak-location}.
For \(\gamma\le\theta_\varepsilon-\tau\varepsilon\), comparison
with \(\gamma_b\) gives a gap of at least
\[
 m_0(\tau\varepsilon-C_0\varepsilon^2)
       \ge m_0\tau\varepsilon/2.
\]
Points with \(\gamma\ge\theta_\varepsilon+\tau\varepsilon\)
eventually lie beyond \(\gamma_c\). Equation~\eqref{eq:A-critical-loss}
then bounds their deficit below by \(\varepsilon/(2f_{\eta_x})\).
This proves \eqref{eq:A-peak-gap}.
For instance, one may take
\(c_\tau=\min\{m_0\tau/2,1/(2f_{\eta_x})\}\) in this case.
No convexity or uniqueness claim is needed inside the band
\([\gamma_b,\gamma_c]\): compactness and upper semicontinuity
already ensure that a maximum is attained there.
Reducing the upper bound on \(\varepsilon\) ensures that all chosen
intervals lie in \(J\).
In each of the three cases, \(c_\tau\) and the admissible upper
bound on \(\varepsilon\) may depend on the fixed \(F,p,x,\tau\),
but not on the varying \(\varepsilon\) or \(\gamma\).
\end{proof}

The proof distinguishes continuous partial inclusion when \(\beta_x>0\),
an endpoint switch when \(\beta_x<0\), and a transition confined to an
\(O(\varepsilon^2)\) band when \(\beta_x=0\).
At a switch, choosing the larger radius makes the maximum attained.
It does not alter the unique solution on either side.
For \(\beta_x=0\), uniqueness is not required inside the transition band.
All three cases yield a peak displaced from the clean radial rank by
an \(O(\varepsilon)\) amount, as given by \eqref{eq:A-theta}.

\subsection{Sample distances at varying retained fractions}
\label{sec:sample-grid}

For this section, extend \eqref{eq:sample-mcd}--\eqref{eq:sample-distance}
to any integer \(p+1\le h\le n-1\).
At \(\gamma=h/n\), choose any determinant-minimizing set of \(h\)
indices, counting repeated observations with their multiplicities.
For a fixed target fraction, use \(h_n(\gamma)=\lfloor n\gamma\rfloor\)
and the correction \(c_{h_n(\gamma)/n}\).
Write
\begin{equation}
 \widehat\rho_{n,\gamma}^2
   =\max_{i\in\mathcal I_{n,\gamma}}\widehat D_{n,\gamma}(X_i),
 \qquad
 \widehat d_{{\rm out},n}^2(\gamma)
   =\min_{i\notin\mathcal I_{n,\gamma}}\widehat D_{n,\gamma}(X_i).
 \label{eq:A-sample-radii}
\end{equation}

The next lemma is the separation property underlying C-steps
\citep{rousseeuw1999fast}. We include the argument to allow ties.
\begin{lemma}
\label{lem:sample-separation}
For an MCD subset with positive-definite covariance, selected distances
do not exceed unselected distances. Consequently
\[
 \widehat\rho_{n,\gamma}^2=\widehat D_{n,\gamma,(h_n(\gamma))},
 \qquad
 \widehat d_{{\rm out},n}^2(\gamma)
        =\widehat D_{n,\gamma,(h_n(\gamma)+1)} .
\]
\end{lemma}
\begin{proof}
Let \(h=h_n(\gamma)\), let \(T,C\) be the raw estimates, and put
\(d_i^2=(X_i-T)^\top C^{-1}(X_i-T)\).
Then \(h^{-1}\sum_{i\in\mathcal I_{n,\gamma}}d_i^2=p\).
If \(d_j^2<d_k^2\) for an unselected \(j\) and selected \(k\),
exchange these two indices.
The new mean and covariance satisfy
\[
 \begin{aligned}
 T'&=T+(X_j-X_k)/h,\\
 C'&=C+\frac{(X_j-T)(X_j-T)^\top-(X_k-T)(X_k-T)^\top}{h}
              -(T'-T)(T'-T)^\top .
 \end{aligned}
\]
Hence
\[
 \tr(C^{-1}C')\le p+(d_j^2-d_k^2)/h<p,\qquad
 \frac{\det C'}{\det C}
 \le\{\tr(C^{-1}C')/p\}^p<1,
\]
contradicting optimality.
The scaled distances are \(d_i^2/c_{h/n}\), so the ordering is
unchanged. The two order-statistic identities follow, including ties.
\end{proof}

The compactness argument below follows
\citet[Propositions~3.1--3.2 and Lemma~A.1]{CL12}.
Their condition (3.1) holds because each hyperplane has
\(P_{\varepsilon,x}\)-probability at most \(\varepsilon<\gamma\).
Their condition on the outer shell (4.2) holds because the continuous
component has positive density.
The subsequential optimality argument follows the framework of
\citet[Theorem~4.1]{CL12}.
We extend the argument to varying retained fractions and to the adjacent
ordered distance, allowing nonunique population minimizers.

Let \(\mathcal M_\gamma\) be the set of population MCD pairs
\((T,C)\) at retained fraction \(\gamma\).
For such a pair define
\[
 \begin{aligned}
 r_\gamma(T,C)&=\inf\{t>0:
 P_{\varepsilon,x}((Z-T)^\top(c_\gamma C)^{-1}(Z-T)\le t)\ge\gamma\},\\
 R_\varepsilon^+(\gamma)
   &=\max_{(T,C)\in\mathcal M_\gamma}r_\gamma(T,C).
 \end{aligned}
\]
The maximum exists by the compactness proved below.
On \(J\), \(R_\varepsilon^+(\gamma)=\rho_\gamma^2\) under
\eqref{eq:A-radius-selection}.

\begin{lemma}
\label{lem:varying-fraction}
Under Condition~\ref{cond:elliptical}, let \(X_i\) be independent
with distribution \(P_{\varepsilon,x}\), where
\(0\le\varepsilon<1/2\), \(p\), and \(x\) are fixed
(\(P_{0,x}=F\)).
Fix \(\mathcal K=[a,b]\subset(\varepsilon,1-\varepsilon)\).
On one probability-one event, the following assertions hold.
\begin{enumerate}[label=(\roman*),leftmargin=2em]
\item For every sequence \(n_j\to\infty\),
\(h_j/n_j\in\mathcal K\), \(h_j/n_j\to\gamma_*\), and every choice
of sample MCD subsets, their estimates have convergent subsequences.
Each limit is in \(\mathcal M_{\gamma_*}\).
Along a subsequence with limit \((T_*,C_*)\),
both the \(h_j\)th and \((h_j+1)\)st scaled distances converge to
\(r_{\gamma_*}(T_*,C_*)\).
\item If the population MCD pair is unique throughout a compact
\(\mathcal K_0\subset\mathcal K\), then
\begin{equation}
 \sup_{h/n\in\mathcal K_0}
 \left\{\|\widehat T_{n,h/n}-T_{h/n}\|
 +\|\widehat C_{n,h/n}-C_{h/n}\|
 +|Y_{n,h}-\rho_{h/n}^2|\right\}\longrightarrow0 .
 \label{eq:A-unique-uniform}
\end{equation}
\item For every closed \(A\subset\mathcal K\),
\begin{equation}
 \limsup_{n\to\infty}\sup_{h/n\in A}Y_{n,h}
       \le\sup_{\gamma\in A}R_\varepsilon^+(\gamma),
 \label{eq:A-envelope-upper-bound}
\end{equation}
where an empty supremum is \(-\infty\).
\end{enumerate}
All suprema over \(h\) are over integer subset sizes.
\end{lemma}

\begin{proof}
Put \(P=P_{\varepsilon,x}\),
\(P_n=n^{-1}\sum_i\delta_{X_i}\), and
\(N_{x,n}=\sum_i\1\{X_i=x\}\).
Work on an event of probability one where \(P_n\Rightarrow P\),
\(N_{x,n}/n\to\varepsilon\), and empirical probabilities of slabs,
balls, and ellipsoids converge uniformly.
These sets form Vapnik--Chervonenkis classes.
The positive measures
\[
 Q_n=P_n-(N_{x,n}/n)\delta_x
       \ \Rightarrow\ (1-\varepsilon)F
\]
also converge weakly.
This single event will be used for all subsequences and their
limiting ellipsoids.

\emph{Step 1: uniform compactness.}
Choose \(a_0\in(\varepsilon,a)\) and
\(\kappa=(a-a_0)/4\).
For a unit \(u\), center \(v\in\R\), and half-width \(d>0\), define
the slab
\[
 S(u,v,d)=\{z:|u^\top z-v|\le d\}.
\]
Spherical symmetry and continuity of the projected clean distribution give
\(\sup_{u,v}F(S(u,v,d))\to0\) as \(d\downarrow0\).
Choose \(d\) with \(\sup_{u,v}P(S(u,v,d))\le a_0\).
Uniform empirical convergence gives the bound \(a_0+\kappa\)
eventually.
For any retained empirical measure \(\nu_n\) of mass
\(\gamma_n=h/n\in[a,b]\), with mean \(T_n\) and raw covariance \(C_n\),
\[
 \begin{split}
 \nu_n(S(u,u^\top T_n,d)^c)
 &\ge\gamma_n-P_n(S(u,u^\top T_n,d))\\
 &\ge a-a_0-\kappa\ge(a-a_0)/2,\\
 u^\top C_nu
 &=\gamma_n^{-1}\int\{u^\top(z-T_n)\}^2\dd\nu_n\\
 &\ge d^2(a-a_0)/(2\gamma_n)
 \ge d^2(a-a_0)/2=:\lambda_0>0.
 \end{split}
\]
Choose \(R_0\) so that \(P(B_{R_0})>b\).
Eventually the ball contains more than \(bn\) observations.
Any subset in it has covariance trace at most \(R_0^2\), so
optimality gives
\[
 \det C_n\le D_0:=(R_0^2/p)^p,\qquad
 \lambda_{\max}(C_n)\le D_0\lambda_0^{-(p-1)}=:\Lambda_0.
\]
Next choose \(R_1\) with \(P(B_{R_1}^c)<a/4\).
Eventually \(P_n(B_{R_1}^c)<a/2\), whence
\(\nu_n(B_{R_1})\ge\gamma_n-P_n(B_{R_1}^c)\ge a/2\).
Consequently
\begin{equation}
 \begin{split}
 \gamma_n\tr C_n
 &=\int\|z-T_n\|^2\dd\nu_n\\
 &\ge\int_{B_{R_1}}\|z-T_n\|^2\dd\nu_n\\
 &\ge\nu_n(B_{R_1})\max\{\|T_n\|-R_1,0\}^2\\
 &\ge\tfrac a2\max\{\|T_n\|-R_1,0\}^2 .
 \end{split}
 \label{eq:A-sample-mean-bound}
\end{equation}
Since \(\gamma_n\tr C_n\le bp\Lambda_0\),
\(\|T_n\|\le R_1+(2bp\Lambda_0/a)^{1/2}=:T_0\).

Choose \(R_2>\max(R_0,T_0,\|x\|)\) with
\[
 (R_2-T_0)^2/\Lambda_0>(R_0+T_0)^2/\lambda_0 .
\]
Every point outside \(B_{R_2}\) then has a larger raw Mahalanobis distance
than every point in \(B_{R_0}\).
An optimal subset containing an outer point would leave an inner point
unselected, contrary to Lemma~\ref{lem:sample-separation}.
Thus all sample minimizers for all \(\gamma_n\in[a,b]\) eventually
have common bounded support.

The same bounds hold for population minimizers, using \(P\) in place
of \(P_n\) and the measure exchange
\eqref{eq:A-replacement-trace}.
Existence and ellipsoidal characterization follow from
\citet[Theorems~3.1--3.2]{CL12}.
Every minimizing retained measure has the form
\((1-\varepsilon)\1_EF+\varepsilon w\delta_x\), with \(E\)
bounded and \(0\le w\le1\).

\emph{Step 2: varying fractions and competitors with exactly \(h_j\) observations.}
Consider any sequence in part (i).
A subsequence of the retained measures converges weakly to \(\nu_*\).
For each nonnegative bounded continuous \(a\),
\(\int a\dd\nu_n\le\int a\dd P_n\). Passage to the limit gives
\(0\le\nu_*\le P\).
Its mass is \(\gamma_*\).
Common bounded support implies convergence of all first and second
moments, hence of the raw estimates to a feasible pair \((T_*,C_*)\).

Choose any population minimizer
\(\nu^\circ=(1-\varepsilon)\1_{E^\circ}F+
\varepsilon w^\circ\delta_x\) at mass \(\gamma_*\).
Select all non-atomic sample observations in \(E^\circ\) and
\(\lfloor w^\circ N_{x,n}\rfloor\) copies of \(x\).
Since \(F(\partial E^\circ)=0\), weak convergence of \(Q_n\) gives
\[
 \int_{E^\circ} a(z)\dd Q_n(z)
 \longrightarrow(1-\varepsilon)\int_{E^\circ}a(z)\dd F(z),
 \qquad a(z)=1,\ z,\ zz^\top .
\]
The functions are bounded on \(E^\circ\). Their only boundary
discontinuities are \(F\)-null.
Also \(\lfloor w^\circ N_{x,n}\rfloor/n\to\varepsilon w^\circ\).
Thus the temporary subset has limiting mass \(\gamma_*\) and the
moments of \(\nu^\circ\).
If its size is \(m_n\), then
\(|m_n-h_j|/n_j\to0\).
Choose one larger ball containing \(E^\circ\) and \(x\), with
probability greater than \(b\).
Remove excess points, or add unused points from that ball, to obtain
exactly \(h_j\) observations.
For a radius \(R\) bounding this ball, the changes in normalized
mass, first moment, and second moment are bounded by
\[
 \frac{|m_n-h_j|}{n_j},\qquad
 R\frac{|m_n-h_j|}{n_j},\qquad
 R^2\frac{|m_n-h_j|}{n_j},
\]
respectively, and all tend to zero.
This competitor has limiting covariance \(C^\circ\).
Therefore
\[
 \det C_*\ge\min_{\mathcal M_{\gamma_*}}\det C,\qquad
 \det C_*=\lim\det C_n
       \le\lim\det C_n^{\rm comp}=\det C^\circ .
\]
Both bounds coincide, proving \((T_*,C_*)\in\mathcal M_{\gamma_*}\).

No strong law indexed by a random limiting ellipsoid was used.
On the fixed event \(Q_n\Rightarrow(1-\varepsilon)F\), the
boundary-null convergence applies to every such limiting ellipsoid.
The argument therefore covers every subsequence on the same event.

The population optimizer correspondence has the same property that limits of minimizers remain optimal when \(\gamma_j\to\gamma_*\).
For completeness, to approximate a fixed competitor of mass
\(\gamma_*\), remove \(O(|\gamma_j-\gamma_*|)\) retained clean mass
if \(\gamma_j<\gamma_*\), or add that amount of available clean
mass from a fixed ball if \(\gamma_j>\gamma_*\).
There is positive retained clean mass because \(\gamma_*>\varepsilon\),
and positive available clean mass because \(\gamma_*<1-\varepsilon\).
The bounded moment changes tend to zero.
This gives the matching determinant bound exactly as above.
Together with Step 1 it proves compactness of the population
optimizer sets and their property that limits of minimizers remain optimal.

\emph{Step 3: adjacent distances and uniform consequences.}
Along a subsequence on which the location and covariance estimates converge,
\(\Sigma_n=c_{\gamma_n}C_n\to c_{\gamma_*}C_*=:\Sigma_*\).
Write
\[
 D_n(z)=(z-T_n)^\top\Sigma_n^{-1}(z-T_n)
       =\widehat D_{n,\gamma_n}(z).
\]
Put \(D_*(z)=(z-T_*)^\top\Sigma_*^{-1}(z-T_*)\) and
\(r_*=r_{\gamma_*}(T_*,C_*)\).
The quadratic forms converge uniformly on each bounded ball.
Their sublevel ellipsoids at bounded thresholds lie in a common ball
by Step 1.
Choose \(0<t_-<r_*<t_+\) with \(t_\pm\ne D_*(x)\).
The positive continuous component yields strict margins:
\[
 P(D_*\le t_-)<\gamma_*<P(D_*\le t_+).
\]
For either threshold, uniform convergence on the common ball and
the absence of boundary mass give
\(P(D_n\le t_\pm)\to P(D_*\le t_\pm)\).
Uniform empirical convergence for ellipsoids then implies, eventually,
\[
 P_n(D_n\le t_-)<h_j/n_j<(h_j+1)/n_j<P_n(D_n\le t_+).
\]
Indeed, each limiting margin is a fixed positive number, whereas
\(|h_j/n_j-\gamma_*|+1/n_j\to0\).
Both adjacent order statistics lie in \((t_-,t_+]\).
Letting the thresholds approach \(r_*\) proves part (i).
The same probability comparison proves continuity of population radii
along convergent population location and covariance pairs and retained fractions.
Thus \(R_\varepsilon^+\) attains its defining maximum and is upper
semicontinuous on \(\mathcal K\).

If the population pair is unique on \(\mathcal K_0\), the property that limits of minimizers remain optimal proves continuity of \(T_\gamma,C_\gamma,\rho_\gamma^2\)
there.
Failure of \eqref{eq:A-unique-uniform} would give an offending sequence
with \(\gamma_j\to\gamma_*\in\mathcal K_0\).
Part (i) and uniqueness force convergence to the same population location, covariance,
and radius, a contradiction.
Finally, failure of \eqref{eq:A-envelope-upper-bound} would give a
sequence in the closed set \(A\), with a limiting fraction
\(\gamma_*\in A\), and a limiting sample radius exceeding
\(\sup_A R_\varepsilon^+\).
Part (i) bounds that limit by
\(r_{\gamma_*}(T_*,C_*)\le R_\varepsilon^+(\gamma_*)\),
again a contradiction.
\end{proof}

\subsection{The first window maximum on the full integer path}
\label{sec:full-window-proof}

The local expansion must first be connected to all earlier candidate
fractions. The next lemma makes that step explicit.
\begin{lemma}
\label{lem:prepeak}
Under Condition~\ref{cond:elliptical}, fix \(x\) with \(\eta_x>1/2\).
After reducing \(\delta_-\) and the upper bound on \(\varepsilon\)
in Theorem~\ref{thm:population-peak}, there are constants
\(C_1,m>0\) such that the population MCD solution is unique, has
\(w=0\), and satisfies
\begin{equation}
 \frac{\dd}{\dd\gamma}\rho_{\gamma,0}^2\ge m
 \quad\text{for }\quad
 1/2\le\gamma\le\theta_\varepsilon-C_1\varepsilon^2 .
 \label{eq:A-prepeak-increase}
\end{equation}
Moreover, \eqref{eq:A-peak-gap} holds with
\(I_\varepsilon\) replaced by
\([1/2,\theta_\varepsilon+\delta_+\varepsilon]\).
\end{lemma}
\begin{proof}
Choose \(b>\eta_x\) with \(b<1\).
We first extend Lemma~\ref{lem:conditional} to \([1/2,b]\).

\emph{Step 1: bounds for the conditional minimizers.}
Choose \(0<\varepsilon_1<\min\{1/2,1-b\}\). For
\(\gamma\in[1/2,b]\), \(0\le\varepsilon\le\varepsilon_1\), and
\(0\le\zeta\le\varepsilon\), the required probability under \(F\) satisfies
\begin{equation}
 0<a_-:=\frac{1/2-\varepsilon_1}{1-\varepsilon_1}
 \le\frac{\gamma-\zeta}{1-\varepsilon}
 \le\frac{b}{1-\varepsilon_1}=:a_+<1.
 \label{eq:A-compact-mass}
\end{equation}
The compactness argument of Lemma~\ref{lem:conditional}, with
\(\gamma_-=1/2\), \(\gamma_+=b\), and the bounds in
\eqref{eq:A-compact-mass}, applies uniformly. Its slab bound,
bounded competitor and exchange inequality
\eqref{eq:A-replacement-trace} give constants
\(\lambda_0,\Lambda_0,T_0,R_2>0\) such that every conditional
minimizing retained measure \(\nu\) satisfies
\begin{equation}
 \supp(\nu)\subset B_{R_2},\qquad
 \lambda_0I_p\preceq C\preceq\Lambda_0I_p,\qquad
 \|T\|\le T_0.
 \label{eq:A-compact-conditional}
\end{equation}
Here \(A\preceq B\) means that \(B-A\) is positive semidefinite.
In particular, all retained first and second moments are bounded,
uniformly over the parameters in \eqref{eq:A-compact-mass}.
The same compactness argument as in Lemma~\ref{lem:conditional}
also gives existence and the moment equations \eqref{eq:A-moments}.

\emph{Step 2: convergence to the uncontaminated solutions.}
Let \(\varepsilon_j\downarrow0\), \(\gamma_j\in[1/2,b]\), and
\(0\le\zeta_j\le\varepsilon_j\). Choose any conditional minimizing
retained measure \(\nu_j\). By \eqref{eq:A-compact-conditional},
pass to a subsequence on which
\(\gamma_j\to\gamma_*\) and \(\nu_j\Rightarrow\nu_*\).
Then
\[
 0\le\nu_*\le F,\qquad \nu_*(\R^p)=\gamma_*,
 \qquad
 \int z\dd\nu_j\to\int z\dd\nu_*,
 \qquad
 \int zz^\top\dd\nu_j\to\int zz^\top\dd\nu_*.
\]
For the matching upper bound, put
\(\alpha_j=(\gamma_j-\zeta_j)/(1-\varepsilon_j)\).
The feasible competitor
\[
 \nu_j^\circ=(1-\varepsilon_j)
       \1_{\{\|z\|^2\le q_{\alpha_j}\}}F+\zeta_j\delta_x
\]
has covariance tending to \(M(\gamma_*)I_p/(p\gamma_*)\).
Feasibility of \(\nu_*\) and optimality of \(\nu_j\) therefore give
\[
 \left\{\frac{M(\gamma_*)}{p\gamma_*}\right\}^{p}
 \le \det C(\nu_*)=\lim_j\det C(\nu_j)
 \le \lim_j\det C(\nu_j^\circ)
 =\left\{\frac{M(\gamma_*)}{p\gamma_*}\right\}^{p}.
\]
The unique MCD solution under \(F\) yields
\(T_j\to0\) and \(\Sigma_j=c_{\gamma_j}C_j\to I_p\).
For \(0<s<q_{\gamma_*}\), these limits and the absence of
\(F\)-mass on spheres imply
\[
 F\{(z-T_j)^\top\Sigma_j^{-1}(z-T_j)\le q_{\gamma_*}\pm s\}
 \longrightarrow F_{R^2}(q_{\gamma_*}\pm s).
\]
Since \(F_{R^2}(q_{\gamma_*}-s)<\gamma_*<
F_{R^2}(q_{\gamma_*}+s)\) and \(\alpha_j\to\gamma_*\),
the mass equation \eqref{eq:A-mass} gives
\(q_{\gamma_*}-s<\rho_{\gamma_j,w_j}^2<q_{\gamma_*}+s\)
eventually, where \(w_j=\zeta_j/\varepsilon_j\).
Hence the conditional radii tend to \(q_{\gamma_*}\).
If convergence were not uniform, one could choose such a sequence
and a constant \(v>0\) with
\[
 \|T_j\|+\|\Sigma_j-I_p\|
       +|\rho_{\gamma_j,w_j}^2-q_{\gamma_j}|\ge v.
\]
The preceding subsequence argument contradicts this inequality.
Thus, uniformly over all conditional minimizers,
\begin{equation}
 \sup_{\substack{\gamma\in[1/2,b]\\0\le w\le1}}
 \{\|T_{\gamma,w}\|+\|\Sigma_{\gamma,w}-I_p\|
                  +|\rho_{\gamma,w}^2-q_\gamma|\}
 \longrightarrow0
 \quad(\varepsilon\downarrow0).
 \label{eq:A-compact-clean-limit}
\end{equation}

\emph{Step 3: a common interval of smoothness.}
At \((T,\Sigma,r)=(0,I_p,q_\gamma)\), replace
\(\eta_x,q_x\) by \(\gamma,q_\gamma\) in
\eqref{eq:A-moment-derivatives} and \eqref{eq:A-kernel}.
The moment equations \eqref{eq:A-moments} retain the same form.
Define
\[
 L_\gamma=\gamma-\frac{2q_\gamma f_\gamma}{p},\qquad
 H_\gamma=M(\gamma)-\frac{2q_\gamma^2f_\gamma}{p+2}.
\]
The integral identities \eqref{eq:A-LH-integrals}, with
\(q_x\) replaced by \(q_\gamma\), imply positivity. By continuity,
\begin{equation}
 \inf_{\gamma\in[1/2,b]}L_\gamma>0,\qquad
 \inf_{\gamma\in[1/2,b]}H_\gamma>0.
 \label{eq:A-compact-LH}
\end{equation}
The kernel calculation \eqref{eq:A-kernel-mass}--\eqref{eq:A-kernel-second},
with \(\eta_x,q_x\) replaced by \(\gamma,q_\gamma\), then makes
the Jacobian invertible at every \(\gamma\in[1/2,b]\).

For each \(\gamma_*\in[1/2,b]\), the implicit function theorem
gives a \(C^2\) solution on a parameter neighborhood of
\((\gamma_*,0,0)\), unique in a neighborhood of
\((0,I_p,q_{\gamma_*})\).
Choose an open interval \(U_{\gamma_*}\) about \(\gamma_*\)
whose closure lies inside that parameter neighborhood and whose
uncontaminated solutions lie inside the corresponding solution
neighborhood. Equation~\eqref{eq:A-compact-clean-limit} then puts
every feasible conditional minimizer with \(\gamma\in U_{\gamma_*}\)
in this solution neighborhood for sufficiently small \(\varepsilon\).
It must equal the implicit solution of \eqref{eq:A-moments}.
Compactness gives a finite cover
\[
 [1/2,b]\subset\bigcup_{j=1}^{N}U_{\gamma_j}.
\]
Take the minimum of the finitely many positive bounds on
\(\varepsilon\). At any feasible parameter in two overlapping
neighborhoods, both implicit solutions equal the same conditional
minimizer and are therefore equal. Their derivatives agree on the
interior \(0<\zeta<\varepsilon\) of the feasible region and, by
continuity, at its boundary. We obtain one unique conditional solution
on \([1/2,b]\), locally represented by these \(C^2\) functions.
Uniform continuity of their derivatives on the smaller compact
neighborhoods yields
\begin{equation}
 \begin{split}
 \sup_{\substack{\gamma\in[1/2,b]\\0\le w\le1}}
       |B_\gamma(w)-(q_x-q_\gamma)|&\longrightarrow0,\\
 \sup_{\gamma\in[1/2,b]}
       |\partial_\gamma\rho_{\gamma,0}^2-f_\gamma^{-1}|
       &\longrightarrow0.
 \end{split}
 \label{eq:A-compact-derivatives}
\end{equation}
The envelope identity \eqref{eq:A-envelope}, obtained from
\eqref{eq:A-moments}, holds throughout this interval.

\emph{Step 4: exclusion of the atom before the transition.}
Choose \(d>0\) with \(\eta_x-d>1/2\) and \(\eta_x-d\) in the
interior of \(J\).
Put \(v_d=q_x-q_{\eta_x-d}>0\). By
\eqref{eq:A-compact-derivatives}, for sufficiently small \(\varepsilon\),
\[
 B_\gamma(w)\ge q_x-q_\gamma-v_d/2\ge v_d/2>0
 \quad(1/2\le\gamma\le\eta_x-d,\ 0\le w\le1).
\]
Equation~\eqref{eq:A-envelope} gives
\(\partial_w\ell_\gamma(w)>0\), and hence
\(\argmin_{0\le w\le1}\ell_\gamma(w)=\{0\}\), on this interval.

On \([\eta_x-d,\theta_\varepsilon-C_1\varepsilon^2]\subset J\),
the minimizing weight is also uniquely zero:
\begin{enumerate}[label=(\roman*),leftmargin=2em]
 \item If \(\beta_x>0\), the objective is strictly convex by
 \eqref{eq:A-objective-curvature}, and
 \(B_\gamma(0)\ge0\) for \(\gamma\le
 \gamma_a=\theta_\varepsilon\), as shown at the start of the proof
 of Theorem~\ref{thm:population-peak}.
 Thus its minimum is attained only at \(w=0\).
 \item If \(\beta_x<0\), only \(w=0,1\) can minimize.
 Choose \(C_1\) greater than the constant bounding
 \(|\gamma_s-\theta_\varepsilon|/\varepsilon^2\).
 Then \(\gamma\le\theta_\varepsilon-C_1\varepsilon^2<\gamma_s\).
 Equations~\eqref{eq:A-endpoint-objective} and
 \eqref{eq:A-integrated-gap}, together with
 \(\partial_\gamma I<0\), give
 \(\ell_\gamma(1)-\ell_\gamma(0)>0\).
 Only \(w=0\) minimizes.
 \item If \(\beta_x=0\), choose \(C_1>C_0\), where \(C_0\)
 is from \eqref{eq:A-critical-signs}. For
 \(\gamma\le\theta_\varepsilon-C_1\varepsilon^2\) in \(J\),
 \eqref{eq:A-fixed-derivative-bounds} gives
 \begin{equation}
 \begin{split}
 B_\gamma(w)&\ge B_{\theta_\varepsilon-C_0\varepsilon^2}(w)
       +b_0(\theta_\varepsilon-C_0\varepsilon^2-\gamma)>0,
       \qquad 0\le w\le1.
 \end{split}
 \label{eq:A-left-critical-positive}
 \end{equation}
 By \eqref{eq:A-envelope}, only \(w=0\) minimizes.
\end{enumerate}
The conditional solution is unique by Step 3, so the full MCD
solution is unique whenever \(1/2\le\gamma\le
\theta_\varepsilon-C_1\varepsilon^2\).
Set \(m=\tfrac12\min_{\gamma\in[1/2,b]}f_\gamma^{-1}>0\).
For small \(\varepsilon\), \eqref{eq:A-compact-derivatives} gives
\[
 \partial_\gamma\rho_{\gamma,0}^2
 \ge f_\gamma^{-1}-m\ge m
 \quad(1/2\le\gamma\le\theta_\varepsilon-C_1\varepsilon^2),
\]
which is \eqref{eq:A-prepeak-increase}.

Choose \(\delta_->0\) small enough, and then reduce the upper bound
on \(\varepsilon\), so that
\(1/2<\theta_\varepsilon-\delta_-<
\theta_\varepsilon-C_1\varepsilon^2\).
The population MCD solution at \(\theta_\varepsilon-\delta_-\)
is therefore unique and retains no mass at \(x\).
For any earlier fraction,
\[
 \rho_\gamma^2\le\rho_{\theta_\varepsilon-\delta_-}^2
             \le\rho_{\max}^2-c_\tau\varepsilon
 \quad(1/2\le\gamma\le\theta_\varepsilon-\delta_-),
\]
where the second inequality is \eqref{eq:A-peak-gap} and
\(\delta_->\tau\varepsilon\) for small \(\varepsilon\).
For \(\gamma\in[\theta_\varepsilon-\delta_-,
\theta_\varepsilon+\delta_+\varepsilon]\) with
\(|\gamma-\theta_\varepsilon|\ge\tau\varepsilon\),
\eqref{eq:A-peak-gap} applies directly.
\end{proof}

\begin{proof}[Proof of Proposition~\ref{prop:population-first-window-entry}]
Fix \(F,p,x,b,\kappa\).
All subsequent upper bounds on \(\varepsilon\) are independent of
\(c\in(0,\kappa\varepsilon]\).

\emph{Step 1: derivative bounds and ordered weights.}
Steps 1--3 of the proof of Lemma~\ref{lem:prepeak} give a common
\(C^2\) family of unique conditional solutions on \(\Gamma_b\).
At \(\varepsilon=0\), the derivatives with respect to \(\gamma\)
of the radius and gap are \(f_\gamma^{-1}\) and \(-f_\gamma^{-1}\).
Uniform continuity on the finite cover therefore gives \(m,b_0>0\) with
\begin{equation}
 \partial_\gamma \rho_{\gamma,w}^2\ge m,\qquad
 \partial_\gamma B_\gamma(w)\le-b_0
 \quad(\gamma\in\Gamma_b,\ 0\le w\le1).
 \label{eq:pfw-uniform-signs}
\end{equation}
By \eqref{eq:A-envelope}, the normalized objective
\begin{equation}
 \Phi_\gamma(w)=\frac{\gamma}{\varepsilon c_\gamma}
       \{\ell_\gamma(w)-\ell_\gamma(0)\}
       =\int_0^w B_\gamma(v)\dd v
 \label{eq:pfw-normalized-objective}
\end{equation}
satisfies \(W_\gamma=\argmin_{[0,1]}\Phi_\gamma\).
Continuity makes \(W_\gamma\) nonempty and compact, with a closed graph.
For \(\gamma_1<\gamma_2\), \(w_i\in W_{\gamma_i}\) and \(w_1>w_2\),
optimality and \eqref{eq:pfw-uniform-signs} would give
\[
 0\le\int_{w_2}^{w_1}B_{\gamma_2}(v)\dd v
   <\int_{w_2}^{w_1}B_{\gamma_1}(v)\dd v\le0.
\]
Hence
\begin{equation}
 \gamma_1<\gamma_2,\quad w_i\in W_{\gamma_i}
       \quad\Longrightarrow\quad w_1\le w_2.
 \label{eq:pfw-ordered-weights}
\end{equation}
In particular, \(\underline w_\gamma=\min W_\gamma\) is nondecreasing
and left continuous: for \(\gamma_j\uparrow\gamma\), its limit
\(u\le\underline w_\gamma\) belongs to \(W_\gamma\) by the closed
graph property, so \(u=\underline w_\gamma\).

\emph{Step 2: every peak has optimal weight zero.}
Theorem~\ref{thm:population-peak} places every maximizer on
\(I_\varepsilon\) within \(C\varepsilon^2\) of \(\theta_\varepsilon\).
For \(\beta_x>0\), its proof gives the unique peak at
\((1-\varepsilon)\eta_x\), with optimal weight zero.
For \(\beta_x<0\), the peak is the unique switch \(\gamma_s\),
where \(W_{\gamma_s}=\{0,1\}\). Equation~\eqref{eq:A-radius-drop}
gives \(\rho_{\gamma_s,0}^2-\rho_{\gamma_s,1}^2>0\), so the selected weight is zero.

For \(\beta_x=0\), fix \(K>C\) and let
\(H_\varepsilon=[\theta_\varepsilon-K\varepsilon^2,
\theta_\varepsilon+K\varepsilon^2]\).
For sufficiently small \(\varepsilon\), every peak is interior to
\(H_\varepsilon\subset J\cap\operatorname{int}(I_\varepsilon)
\cap\operatorname{int}(\Gamma_b)\).
By \eqref{eq:A-abt}, \eqref{eq:A-theta} and the definition of
\(H_\varepsilon\), \(t=-\eta_x+O(\varepsilon)\) throughout this
interval. Hence \eqref{eq:A-derivatives} gives
constants \(K_B,C_R,c_R>0\) such that
\begin{equation}
 \begin{aligned}
 |\partial_w B_\gamma(w)|&\le K_B\varepsilon^2,&
 \partial_\gamma B_\gamma(w)&\le-b_0,\\
 |\partial_\gamma \rho_{\gamma,w}^2|&\le C_R,&
 \partial_w \rho_{\gamma,w}^2&\le-c_R\varepsilon
 \end{aligned}
 \quad(\gamma\in H_\varepsilon,\ 0\le w\le1).
 \label{eq:pfw-critical-bounds}
\end{equation}
Thus the largest radius is attained at \(\underline w_\gamma\).
Let \(\gamma_*\) be any peak, and set \(w_* =\underline w_{\gamma_*}\).
If \(w_*=1\), \eqref{eq:pfw-ordered-weights} gives \(W_u=\{1\}\)
for \(u>\gamma_*\). Choosing such a point in \(H_\varepsilon\) yields
\(\rho_u^2-\rho_{\gamma_*}^2\ge m(u-\gamma_*)>0\), a contradiction.

If \(0<w_*<1\), choose \(\gamma_j\uparrow\gamma_*\) in
\(H_\varepsilon\) and put \(w_j=\underline w_{\gamma_j}\).
Left continuity gives \(w_j\to w_*\), so eventually \(0<w_j<1\).
The first-order conditions are
\(B_{\gamma_j}(w_j)=B_{\gamma_*}(w_*)=0\).
For \(d_j=\gamma_*-\gamma_j>0\) and \(e_j=w_*-w_j\ge0\),
\eqref{eq:pfw-critical-bounds} yields
\begin{equation}
 \begin{split}
 0&=\int_{\gamma_j}^{\gamma_*}\partial_s B_s(w_j)\dd s
      +\int_{w_j}^{w_*}\partial_v B_{\gamma_*}(v)\dd v\\
  &\le-b_0d_j+K_B\varepsilon^2e_j,
 \qquad
 e_j>0,\quad d_j\le(K_B/b_0)\varepsilon^2e_j.
 \end{split}
 \label{eq:pfw-critical-gap-comparison}
\end{equation}
Applying the same two-segment integration to \(\rho_{\gamma,w}^2\), and then
\eqref{eq:pfw-critical-gap-comparison}, gives
\begin{equation}
 \begin{split}
 \rho_{\gamma_*}^2-\rho_{\gamma_j}^2
 &\le C_Rd_j-c_R\varepsilon e_j\\
 &\le\{(C_RK_B/b_0)\varepsilon^2-c_R\varepsilon\}e_j<0,
 \end{split}
 \label{eq:pfw-critical-radius-comparison}
\end{equation}
Equation~\eqref{eq:pfw-critical-radius-comparison} contradicts maximality.
Therefore, for every sign of \(\beta_x\),
\begin{equation}
 0\in W_{\gamma_*},\qquad
 \rho_{\gamma_*}^2=\rho_{\gamma_*,0}^2.
 \label{eq:pfw-zero-at-peak}
\end{equation}
There cannot be two peaks, since \eqref{eq:pfw-uniform-signs}
makes \(\rho_{\gamma,0}^2\) strictly increasing. Denote the unique peak
by \(\gamma_\varepsilon\).

\emph{Step 3: the endpoint at which exclusion remains optimal.}
For \(\gamma_1<\gamma_2\) and \(w>0\),
\eqref{eq:pfw-normalized-objective} yields
\begin{equation}
 \Phi_{\gamma_1}(w)-\Phi_{\gamma_2}(w)
 =-\int_{\gamma_1}^{\gamma_2}\int_0^w\partial_s B_s(v)\dd v\dd s
 \ge b_0(\gamma_2-\gamma_1)w>0.
 \label{eq:pfw-objective-order}
\end{equation}
Equations~\eqref{eq:pfw-zero-at-peak} and \eqref{eq:pfw-objective-order}
imply \(\Phi_\gamma(w)>\Phi_{\gamma_\varepsilon}(w)\ge0\)
for \(\gamma<\gamma_\varepsilon\), \(w>0\), hence \(W_\gamma=\{0\}\).
If \(0\in W_\gamma\) for some \(\gamma>\gamma_\varepsilon\),
choose \(u\in I_\varepsilon\cap\Gamma_b\) with
\(\gamma_\varepsilon<u<\gamma\).
Then \eqref{eq:pfw-objective-order} gives \(W_u=\{0\}\), whereas
\eqref{eq:pfw-uniform-signs} gives
\(\rho_u^2-\rho_{\gamma_\varepsilon}^2\ge
m(u-\gamma_\varepsilon)>0\), a contradiction.
Consequently,
\[
 W_\gamma=\{0\}\ (1/2\le\gamma<\gamma_\varepsilon),\qquad
 W_\gamma\subset(0,1]\ (\gamma_\varepsilon<\gamma\le b).
\]

\emph{Step 4: the first population window maximum.}
For \(\gamma<\gamma_\varepsilon\), take
\(u=\gamma+\tfrac12\min\{c,\gamma_\varepsilon-\gamma\}\).
Then \(u\in\Gamma_b\), \(|u-\gamma|<c\) and
\(\rho_u^2>\rho_\gamma^2\), so \eqref{eq:population-window-set} gives
\(\gamma\notin\mathcal A_{\varepsilon,c}^{(b)}\).
Since \(\kappa<\delta_+\) and
\(\gamma_\varepsilon\to\eta_x\in(1/2,b)\),
\eqref{eq:A-peak-location} gives, for sufficiently small \(\varepsilon\),
\begin{equation}
 C\varepsilon^2+\kappa\varepsilon<\min\{\delta_-,\delta_+\varepsilon\},
 \qquad
 [\gamma_\varepsilon-c,\gamma_\varepsilon+c]
       \subset I_\varepsilon\cap\Gamma_b
 \quad(0<c\le\kappa\varepsilon).
 \label{eq:pfw-window-inclusion}
\end{equation}
Maximality on \(I_\varepsilon\) and \eqref{eq:pfw-window-inclusion}
give \(\gamma_\varepsilon\in\mathcal A_{\varepsilon,c}^{(b)}\).
This proves both identities in \eqref{eq:pfw-identification}.
\end{proof}

\begin{proof}[Proof of Theorem~\ref{thm:contaminated-window}]
Fix \(\tau,\varepsilon,c\) as in the theorem.
Choose \(\varepsilon_\tau\) small enough for the bounds below to hold.
Put
\[
 U=\theta_\varepsilon+\delta_+\varepsilon<1-\varepsilon,
 \qquad \Delta=c_\tau\varepsilon>0 .
\]
Take the probability-one event of Lemma~\ref{lem:varying-fraction}
for a compact interval containing \([1/2,U]\).
All candidate covariances, including those with fractions above \(U\),
are eventually positive definite: choose a slab probability bound
strictly between \(\varepsilon\) and \(1/2\), and apply the lower
eigenvalue calculation in that lemma to every subset of size
\(h\ge h_{\min}\).
Thus \eqref{eq:mcd-window-selector} uses its ordinary definition
eventually.

Let \(M_0\) be the upper derivative bound in
\eqref{eq:A-fixed-derivative-bounds}, and take \(C,C_1\) from
Theorem~\ref{thm:population-peak} and Lemma~\ref{lem:prepeak}.
Equations~\eqref{eq:A-peak-location} and \eqref{eq:A-derivatives} give
\begin{equation}
 \rho_{\max}^2\le\rho_{\theta_\varepsilon+C\varepsilon^2,0}^2,
 \label{eq:A-peak-upper-zero}
\end{equation}
because the conditional radius decreases in \(w\) throughout the
transition interval. Define
\begin{equation}
 d_\varepsilon=\min\{\tau\varepsilon/2,
                    c_\tau\varepsilon/(8M_0)\},\qquad
 \gamma_0=\theta_\varepsilon-d_\varepsilon/2.
 \label{eq:A-d-epsilon}
\end{equation}
For sufficiently small \(\varepsilon\),
\begin{equation}
 \begin{gathered}
 C_1\varepsilon^2<d_\varepsilon/2,\qquad
 \theta_\varepsilon-\tau\varepsilon/2<\gamma_0
       <\theta_\varepsilon-C_1\varepsilon^2,\\
 \rho_{\max}^2-\rho_{\gamma_0}^2
 \le M_0(d_\varepsilon/2+C\varepsilon^2)
 \le\Delta/4.
 \end{gathered}
 \label{eq:A-comparison-point}
\end{equation}
Here Lemma~\ref{lem:prepeak} gives the unique solution with \(w=0\)
at \(\gamma_0\), and the last bound follows from
\eqref{eq:A-peak-upper-zero} and \eqref{eq:A-fixed-derivative-bounds}.
Let \(h_0=\lfloor n\gamma_0\rfloor\).
By Lemmas~\ref{lem:prepeak} and \ref{lem:varying-fraction}(ii),
\[
 e_n:=\sup_{\substack{h\in\mathcal H_n\\h/n\le\gamma_0}}
        |Y_{n,h}-\rho_{h/n,0}^2|\longrightarrow0.
\]

\emph{Exclude every early window maximum.}
For \(h/n\le\theta_\varepsilon-\tau\varepsilon\), let
\(k=\min(h+g_n,h_0)\).
For all large \(n\), \(k\in\mathcal W_n(h)\) and
\[
 \frac{k-h}{n}
  =\min\{g_n/n,h_0/n-h/n\}
  \ge\min\{c/2,\tau\varepsilon/4\}=:d_*>0.
\]
Since \(1/2\le h/n<k/n\le h_0/n\le\gamma_0\),
equation~\eqref{eq:A-prepeak-increase} yields
\[
 Y_{n,k}-Y_{n,h}
 \ge\rho_{k/n,0}^2-\rho_{h/n,0}^2-2e_n
 \ge md_*-2e_n>0
\]
eventually.
No index at or below \(\theta_\varepsilon-\tau\varepsilon\)
can be a window maximum.

\emph{Produce a window maximum near the peak.}
Let
\(\mathcal J_n=\{h\in\mathcal H_n:h/n\le U\}\).
Choose
\[
 j_n\in\argmax_{h\in\mathcal J_n}Y_{n,h}.
\]
On the closed set
\[
 A=\{\gamma\in[1/2,U]:
           |\gamma-\theta_\varepsilon|\ge\tau\varepsilon\},
\]
Lemma~\ref{lem:prepeak} gives
\(\sup_A R_\varepsilon^+\le\rho_{\max}^2-\Delta\).
By Lemma~\ref{lem:varying-fraction}(iii), eventually
\[
 \sup_{\substack{h\in\mathcal J_n\\h/n\in A}}Y_{n,h}
       <\rho_{\max}^2-3\Delta/4 .
\]
The population MCD solution at \(\gamma_0\) is unique.
Lemma~\ref{lem:varying-fraction}(i), applied to \(h_0/n\to\gamma_0\), gives
\[
 Y_{n,h_0}>\rho_{\gamma_0}^2-\Delta/4
                 \ge\rho_{\max}^2-\Delta/2 .
\]
Since \(Y_{n,j_n}\ge Y_{n,h_0}\), it follows that
\[
 \left|j_n/n-\theta_\varepsilon\right|<\tau\varepsilon
 \quad\text{for every }j_n\in\argmax_{h\in\mathcal J_n}Y_{n,h}.
\]
For all sufficiently large \(n\), \(g_n/n\le c\).
Using \(c<(\delta_+-\tau)\varepsilon\), every
\(k\in\mathcal W_n(j_n)\) therefore satisfies
\begin{equation}
 k/n\le j_n/n+g_n/n
    <\theta_\varepsilon+\tau\varepsilon+c
    <\theta_\varepsilon+\delta_+\varepsilon=U.
 \label{eq:A-window-in-prefix}
\end{equation}
Because \(\mathcal W_n(j_n)\subseteq\mathcal H_n\),
\eqref{eq:A-window-in-prefix} gives
\[
 \mathcal W_n(j_n)\subseteq
 \{h\in\mathcal H_n:h/n\le U\}=\mathcal J_n.
\]
Thus \(Y_{n,j_n}\ge Y_{n,k}\) for all \(k\in\mathcal W_n(j_n)\),
so \(j_n\in\mathcal A_n^{\MCD}\), including when maximizers are tied.
The earliest such index obeys
\[
 \theta_\varepsilon-\tau\varepsilon
       <\widehat h_{\MCD}/n
       \le j_n/n<\theta_\varepsilon+\tau\varepsilon .
\]
This proves \eqref{eq:contaminated-window-location}.
\end{proof}

\section{Proofs for sampling without contamination}
\label{sec:clean-window}

\subsection{Uniform convergence}

Here the observations are independent with distribution \(F\), and
the sample scatter uses \(c_{h/n}\) from \eqref{eq:A-radial}.
The population distance is
\begin{equation}
 d_{\rm out}^2(\gamma;F)=\rho_\gamma^2(F)=q_\gamma.
 \label{eq:B-clean-path}
\end{equation}
It is continuous and strictly increasing.

\begin{proof}[Proof of Lemma~\ref{thm:clean-uniform}]
Apply Lemma~\ref{lem:varying-fraction}(ii) with
\(\varepsilon=0\) and \(\mathcal K_0=\mathcal K=[a,b]\).
Under Condition~\ref{cond:elliptical}, the population MCD solution
is unique at every \(\gamma\in[a,b]\), with
\[
 T_\gamma=0,\qquad C_\gamma=\frac{M(\gamma)}{p\gamma}I_p,
 \qquad \rho_\gamma^2=q_\gamma.
\]
The lemma gives uniform convergence of the location, raw covariance
and outside distance. Since \(c_\gamma\) is continuous on \([a,b]\),
\[
 \sup_{\substack{h\in\mathcal H_n\\h/n\in[a,b]}}
 \|\widehat\Sigma_{n,h/n}-I_p\|
 \le\left(\sup_{\gamma\in[a,b]}c_\gamma\right)
 \sup_{\substack{h\in\mathcal H_n\\h/n\in[a,b]}}
 \|\widehat C_{n,h/n}-C_{h/n}\|\longrightarrow0
\]
almost surely. Together with the uniform convergence of the location
and \(Y_{n,h}\) supplied by Lemma~\ref{lem:varying-fraction}(ii),
this proves \eqref{eq:B-uniform-path}.
\end{proof}

\subsection{Window maxima and the selected fraction}

\begin{proof}[Proof of Proposition~\ref{prop:B-floor-control}]
Continuity and strict increase of \(q_\gamma\) on the compact
comparison interval imply \(\Delta_c>0\).
For \(n\ge2/c\), \(cn\ge2\), so \(g_n=\lfloor cn\rfloor\) and
\(cn-1\le g_n\le cn\). Therefore
\begin{equation}
 c/2\le c-1/n\le g_n/n\le c.
 \label{eq:B-floor-bound}
\end{equation}
Given \(h/n\in[\gamma_L,\gamma_U]\), let \(h^+=h+g_n\).
Since \(\gamma_U+c<1\), \eqref{eq:B-floor-bound} gives, for large \(n\),
\[
 h_{\min}\le h<h^+=h+g_n
 \le n(\gamma_U+c)\le n-1,\qquad |h^+-h|=g_n.
\]
Thus
\[
 h^+\in\mathcal H_n,\qquad
 h^+\in\mathcal W_n(h),\qquad
 \gamma_L\le h^+/n\le\gamma_U+c,
\]
so \(h^+/n\in\Gamma_c\). It follows that
\[
 q_{h^+/n}-q_{h/n}\ge q_{h/n+c/2}-q_{h/n}\ge\Delta_c.
\]
Whenever \(r_n<\Delta_c/3\),
\[
 Y_{n,h^+}-Y_{n,h}
 \ge q_{h^+/n}-q_{h/n}-2r_n>\Delta_c/3>0.
\]
This excludes all candidates in \([\gamma_L,\gamma_U]\) simultaneously.
We have proved that, for all sufficiently large \(n\),
\begin{equation}
 \mathcal F_n\subseteq\{r_n\ge\Delta_c/3\}.
 \label{eq:B-event-inclusion}
\end{equation}
Lemma~\ref{thm:clean-uniform} gives \(r_n\to0\) almost surely.
Hence, on an event of probability one, there is a finite
\(N\) such that \(r_n<\Delta_c/3\) for every \(n\ge N\).
Equation~\eqref{eq:B-event-inclusion} then gives
\(\1_{\mathcal F_n}=0\) for every \(n\ge N\).
Also,
\[
 0\le\1\{r_n\ge\Delta_c/3\}\le1,\qquad
 \1\{r_n\ge\Delta_c/3\}\longrightarrow0
 \quad\text{almost surely}.
\]
Taking probabilities in \eqref{eq:B-event-inclusion} and applying
bounded convergence yields
\[
 0\le\Pp_F(\mathcal F_n)
 \le\Pp_F(r_n\ge\Delta_c/3)
 =\Ee_F\1\{r_n\ge\Delta_c/3\}\longrightarrow0,
\]
where \(\Ee_F\) denotes expectation under sampling from \(F\).
\end{proof}

\begin{proof}[Proof of Corollary~\ref{cor:clean-retention}]
Fix \(0<\delta<1/2\), and let \(d=\min(c/2,\delta/4)>0\).
For each \(h\in\mathcal H_n\) with \(h/n\le1-\delta\), set
\(k=h+\lfloor nd\rfloor\).
Because \(d\le c/2<c\), we have
\(\lfloor nd\rfloor\le\lfloor nc\rfloor\le g_n\).
Also \(d\le\delta/4\), so
\(k/n\le1-\delta+d\le1-3\delta/4\).
For large \(n\), this upper bound is below \((n-1)/n\),
and \(\lfloor nd\rfloor/n\ge d-1/n\ge d/2>0\).
Finally, \(h_{\min}/n\ge1/2\) because \(p\ge1\).
Consequently, for all sufficiently large \(n\),
\[
 h_{\min}\le h<k\le n-1,\qquad
 |k-h|=\lfloor nd\rfloor\le g_n,
 \qquad k\in\mathcal H_n\cap\mathcal W_n(h).
\]
Moreover,
\[
 \frac d2\le\frac{k-h}{n}\le d,\qquad
 \frac12\le\frac hn\le\frac kn\le1-\frac{3\delta}{4}.
\]
The continuous, positive function
\(q_{\gamma+d/2}-q_\gamma\) has a positive minimum on
\([1/2,1-\delta]\).
Denote this minimum by \(m_\delta>0\), and put
\[
 r_{n,\delta}=\sup_{\substack{j\in\mathcal H_n\\
                         j/n\in[1/2,1-3\delta/4]}}
                       |Y_{n,j}-q_{j/n}|.
\]
For every eligible \(h\) and its comparison index \(k\),
\[
 Y_{n,k}-Y_{n,h}
 \ge q_{k/n}-q_{h/n}-2r_{n,\delta}
 \ge m_\delta-2r_{n,\delta}.
\]
Lemma~\ref{thm:clean-uniform}, applied on
\([1/2,1-3\delta/4]\), therefore gives \(Y_{n,k}>Y_{n,h}\)
for all these indices simultaneously, eventually.
No candidate with \(h/n\le1-\delta\) can be a window maximum.

Under the continuous uncontaminated distribution, all subsets with at least \(p+1\)
observations have positive definite covariance almost surely.
Fix any \(p+1\) distinct indices.
Their covariance is singular only if the corresponding observations
lie in an affine hyperplane.
This event has probability zero because the observations are independent
and have a density.
There are only countably many such groups in an infinite sample.
Thus they are all affinely independent on one probability-one event,
and every larger subset has positive definite covariance as well.
Eventually \(h_{\min}\ge p+1\), so all candidate distances are
finite and their maximum exists.
The global maximum is a window maximum, so the candidate set is nonempty.
Its earliest member must satisfy
\(\widehat h_{\MCD}/n>1-\delta\) eventually.
Intersecting these events over a countable sequence \(\delta\downarrow0\)
proves the result.
\end{proof}

\section{Additional numerical results and method details}

\subsection{Scale correction of the raw estimates and reweighting}

For a given subset size \(h\), all four FDB methods compute the sample
covariance matrix of the subset with denominator \(h-1\) and multiply the
matrix by the Gaussian MCD consistency factor and the finite-sample correction
factor used by \citet{maechler2025robustbase}.
For fixed \(n\) and \(p\), each method evaluates both factors at its own
retained proportion \(\gamma=h/n\).
The covariance matrix in \eqref{eq:sample-mcd} uses denominator \(h\), so the
factor \(a_{n,p,h}\) used in the numerical methods is the product of these two
correction factors times \(h/(h-1)\).

The population scale correction factor \(c_\gamma\) in \eqref{eq:A-radial} is
determined by the uncontaminated distribution.
Under the standardized Gaussian model, \(R^2\) has a chi-squared distribution
with \(p\) degrees of freedom, and
\[
c_\gamma=\frac{\gamma}
{F_{\chi^2_{p+2}}(\chi^2_{p,\gamma})},
\]
where \(F_{\chi^2_\nu}\) denotes the distribution function of a chi-squared
distribution with \(\nu\) degrees of freedom, and \(\chi^2_{p,\gamma}\) denotes
the \(\gamma\)-quantile of a chi-squared distribution with \(p\) degrees of
freedom.
Write \(f_{\chi_\nu^2}\) for the corresponding density. The formula for the
chi-squared density gives
\[
t f_{\chi_p^2}(t)=p f_{\chi_{p+2}^2}(t),\qquad t>0 .
\]
Integrating both sides from \(0\) to \(\chi^2_{p,\gamma}\) gives
\(M(\gamma)=pF_{\chi^2_{p+2}}(\chi^2_{p,\gamma})\).
Substitution into \eqref{eq:A-radial} gives the expression for \(c_\gamma\)
above.

All four FDB methods use this Gaussian correction in practice, including in the
experiments where the uncontaminated observations come from a multivariate
\(t_3\) distribution.
Section~\ref{sec:selector} gives the reweighting cutoff and the method for
computing the final estimates.

\subsection{Gaussian envelope for FDB-FS}

\fs{} uses a Gaussian reference threshold \(b_{n,p,h}\) to select the initial
subset size \(h\).
For a candidate size \(h\) satisfying \(p<h<n\), let
\(Z_1,\ldots,Z_h,Z_{\rm new}\) be independent observations from \(N_p(0,I_p)\).
Use the first \(h\) observations to compute the sample mean \(\bar Z_h\) and
sample covariance matrix \(S_h\), where \(S_h\) has denominator \(h-1\).
By equation~(21) of \citet{atkinson2010forward}, the squared Mahalanobis
distance of the new observation \(Z_{\rm new}\) satisfies
\begin{equation}
	(Z_{\rm new}-\bar Z_h)^\top S_h^{-1}(Z_{\rm new}-\bar Z_h)
	\ \overset{d}{=}\ 
	\frac{p(h+1)(h-1)}{h(h-p)}\,F_{p,h-p}.
	\label{eq:C-deletion-distance}
\end{equation}
Since \(Z_{\rm new}-\bar Z_h\) has covariance \((h+1)I_p/h\), setting
\(V=\sqrt{h/(h+1)}(Z_{\rm new}-\bar Z_h)\) gives \(V\sim N_p(0,I_p)\).
Also, \(W=(h-1)S_h\sim\operatorname{Wishart}_p(I_p,h-1)\), and \(V\) and \(W\)
are independent.
Thus \(V^\top W^{-1}V\) has the same distribution as \(pF_{p,h-p}/(h-p)\).
Multiplying this quadratic form by \((h-1)(h+1)/h\) gives
\eqref{eq:C-deletion-distance}.

To determine the threshold, \fs{} uses the order-statistic transformation in
equation~(8) of \citet{riani2009unknown}.
If \(n\) squared distances are independent and have the same continuous
distribution, applying their common distribution function to the squared
distance of rank \(h+1\) gives a random variable with distribution
\(\operatorname{Beta}(h+1,n-h)\).
The \(0.999\)-quantile of this beta distribution is \(u_h\) in
\eqref{eq:fs-beta-quantile}.
Taking the \(u_h\)-quantile of the distribution on the right of
\eqref{eq:C-deletion-distance} then gives \(b_{n,p,h}\) in
\eqref{eq:fs-envelope}.
In practice, the squared distances for all observations use location and
scatter estimates from the same subset, so \(b_{n,p,h}\) serves as a Gaussian
reference threshold for \(\widetilde Y_{n,h}\).

\fs{} differs from the full forward search of \citet{riani2009unknown},
implemented here in \fsm{}, in the following respects.
\begin{enumerate}[leftmargin=1.6em,itemsep=0.4em]
	\item \emph{Subset construction}\quad
	\fs{} orders the observations by projection depth once and takes the first
	\(h\) observations for each \(h\).
	At each step, the forward search estimates location and scatter from the
	current \(h\) observations, recomputes the Mahalanobis distances of all
	observations, and takes the \(h+1\) observations with the smallest distances
	as the next subset \citep[Section~2]{riani2009unknown}.
	
	\item \emph{Threshold crossings}\quad
	\fs{} considers \(h\) in increasing order and takes the first \(h\) satisfying
	\(\widetilde Y_{n,h}>b_{n,p,h}\) as the initial subset size.
	If no crossing occurs, it takes \(n-1\).
	The forward search monitors the smallest Mahalanobis distance among
	observations outside the current subset, with signalling rules that vary by
	stage of the search.
	A single crossing of a higher threshold or crossings at several consecutive
	steps can trigger an outlier signal \citep[Section~5.1]{riani2009unknown}.
	
	\item \emph{Outlier identification}\quad
	Once \fs{} selects \(h\), it uses the initial subset to compute weights for
	all observations and classifies those with zero weight as outliers.
	The forward search first confirms an outlier signal when necessary.
	To identify outliers, it can also compute threshold curves for different
	assumed numbers \(n^\ast\) of regular observations and compare the same
	minimum Mahalanobis distance curve from the search with each threshold curve
	in turn \citep[Sections~5.2--5.3]{riani2009unknown}.
\end{enumerate}

\Needspace{26\baselineskip}
\subsection{Additional comparisons at fixed subset sizes}

For the experiments at \(r=15\), Tables~\ref{tab:contam010}
and~\ref{tab:contam030} in the main text each compare three of the four FDB
methods.
Table~\ref{tab:supp-omitted010} gives the additional results for \hf{} at
\(\varepsilon=0.10\), and Table~\ref{tab:supp-omitted030} gives those for
\hseven{} at \(\varepsilon=0.30\).

\begin{table}[H]
\centering
\caption{Performance of \hf{} at \(r=15\) and \(\varepsilon=0.10\).}
\label{tab:supp-omitted010}
\small
\setlength{\tabcolsep}{3pt}
\resizebox{\linewidth}{!}{%
\begin{tabular}{llrrrr}
\toprule
Setting & Contamination & \(e_\mu\) & \(e_\Sigma\) &
\(\mathrm{MSE}_\Sigma\) & \(L_{\mathrm{Stein}}\)\\
\midrule
A & Point & \(0.231\) (\(0.056\)) & \(0.724\) (\(0.091\)) & \(0.048\) (\(0.005\)) & \(5.701\) (\(0.872\))\\
 & Point mass & \(0.234\) (\(0.055\)) & \(0.723\) (\(0.088\)) & \(0.048\) (\(0.005\)) & \(5.704\) (\(0.883\))\\
 & Random & \(0.337\) (\(0.075\)) & \(0.558\) (\(0.073\)) & \(0.013\) (\(0.003\)) & \(0.769\) (\(0.166\))\\
 & Cluster & \(0.334\) (\(0.078\)) & \(0.555\) (\(0.072\)) & \(0.013\) (\(0.003\)) & \(0.768\) (\(0.167\))\\
\addlinespace
B & Point & \(0.233\) (\(0.023\)) & \(0.831\) (\(0.032\)) & \(0.010\) (\(3.426\mathrm{e}{-4}\)) & \(31.938\) (\(1.652\))\\
 & Point mass & \(0.235\) (\(0.023\)) & \(0.833\) (\(0.034\)) & \(0.010\) (\(3.466\mathrm{e}{-4}\)) & \(32.035\) (\(1.662\))\\
 & Random & \(0.339\) (\(0.033\)) & \(0.597\) (\(0.022\)) & \(0.002\) (\(1.048\mathrm{e}{-4}\)) & \(3.195\) (\(0.141\))\\
 & Cluster & \(0.342\) (\(0.033\)) & \(0.597\) (\(0.023\)) & \(0.002\) (\(1.077\mathrm{e}{-4}\)) & \(3.192\) (\(0.145\))\\
\addlinespace
C & Point & \(0.235\) (\(0.012\)) & \(0.911\) (\(0.015\)) & \(0.003\) (\(3.735\mathrm{e}{-5}\)) & \(148.870\) (\(3.230\))\\
 & Point mass & \(0.235\) (\(0.012\)) & \(0.912\) (\(0.015\)) & \(0.003\) (\(3.865\mathrm{e}{-5}\)) & \(149.218\) (\(3.354\))\\
 & Random & \(0.345\) (\(0.017\)) & \(0.618\) (\(0.009\)) & \(5.939\mathrm{e}{-4}\) (\(7.591\mathrm{e}{-6}\)) & \(12.661\) (\(0.171\))\\
 & Cluster & \(0.346\) (\(0.017\)) & \(0.617\) (\(0.009\)) & \(5.935\mathrm{e}{-4}\) (\(7.602\mathrm{e}{-6}\)) & \(12.647\) (\(0.176\))\\
\bottomrule
\end{tabular}}
\end{table}

\begin{table}[H]
\centering
\caption{Performance of \hseven{} at \(r=15\) and \(\varepsilon=0.30\).}
\label{tab:supp-omitted030}
\small
\setlength{\tabcolsep}{3pt}
\resizebox{\linewidth}{!}{%
\begin{tabular}{llrrrr}
\toprule
Setting & Contamination & \(e_\mu\) & \(e_\Sigma\) &
\(\mathrm{MSE}_\Sigma\) & \(L_{\mathrm{Stein}}\)\\
\midrule
A & Point & \(14.849\) (\(0.243\)) & \(3.538\) (\(0.091\)) & \(2.386\mathrm{e}{3}\) (\(45.489\)) & \(486.683\) (\(4.955\))\\
 & Point mass & \(14.846\) (\(0.234\)) & \(3.547\) (\(0.091\)) & \(2.385\mathrm{e}{3}\) (\(43.763\)) & \(486.592\) (\(4.734\))\\
 & Random & \(0.367\) (\(0.080\)) & \(0.585\) (\(0.071\)) & \(0.016\) (\(0.004\)) & \(0.845\) (\(0.162\))\\
 & Cluster & \(5.554\) (\(0.804\)) & \(2.392\) (\(0.067\)) & \(138.215\) (\(30.575\)) & \(112.638\) (\(13.287\))\\
\addlinespace
B & Point & \(36.144\) (\(0.363\)) & \(4.451\) (\(0.034\)) & \(2.565\mathrm{e}{3}\) (\(25.386\)) & \(2.558\mathrm{e}{3}\) (\(13.430\))\\
 & Point mass & \(36.133\) (\(0.348\)) & \(4.467\) (\(0.035\)) & \(2.564\mathrm{e}{3}\) (\(23.983\)) & \(2.558\mathrm{e}{3}\) (\(12.586\))\\
 & Random & \(0.374\) (\(0.039\)) & \(0.656\) (\(0.025\)) & \(0.003\) (\(1.324\mathrm{e}{-4}\)) & \(3.847\) (\(0.151\))\\
 & Cluster & \(10.311\) (\(0.339\)) & \(2.821\) (\(0.018\)) & \(37.161\) (\(1.669\)) & \(301.798\) (\(6.796\))\\
\addlinespace
C & Point & \(76.912\) (\(0.521\)) & \(5.169\) (\(0.016\)) & \(2.707\mathrm{e}{3}\) (\(15.781\)) & \(1.056\mathrm{e}{4}\) (\(32.486\))\\
 & Point mass & \(76.806\) (\(0.521\)) & \(5.189\) (\(0.018\)) & \(2.704\mathrm{e}{3}\) (\(15.819\)) & \(1.055\mathrm{e}{4}\) (\(32.635\))\\
 & Random & \(0.377\) (\(0.020\)) & \(0.678\) (\(0.010\)) & \(7.184\mathrm{e}{-4}\) (\(8.305\mathrm{e}{-6}\)) & \(15.106\) (\(0.158\))\\
 & Cluster & \(14.585\) (\(0.237\)) & \(3.127\) (\(0.008\)) & \(9.281\) (\(0.202\)) & \(613.979\) (\(6.548\))\\
\bottomrule
\end{tabular}}
\end{table}

\FloatBarrier
\Needspace{24\baselineskip}
\subsection{Initial subset-size selection at \texorpdfstring{\(r=15\)}{r = 15}}

Table~\ref{tab:supp-hfrac} gives the mean initial subset proportion
\(\widehat h/n\) selected by \method{} before reweighting.

\begin{table}[H]
\centering
\caption{Initial subset proportion \(\widehat h/n\) selected by FDB-AH at
\(r=15\).}
\label{tab:supp-hfrac}
\footnotesize\renewcommand{\arraystretch}{0.96}
\begin{tabular}{llcc}
\toprule
Setting & Contamination & \(\varepsilon=0.100\) & \(\varepsilon=0.300\)\\
\midrule
A & Cluster & 0.892 (0.047) & 0.700 (0.007)\\
A & Random & 0.897 (0.052) & 0.698 (0.017)\\
A & Point & 0.878 (0.075) & 0.698 (0.010)\\
A & Point mass & 0.867 (0.089) & 0.698 (0.011)\\
\addlinespace
B & Cluster & 0.900 (0.000) & 0.700 (0.000)\\
B & Random & 0.901 (0.002) & 0.700 (0.001)\\
B & Point & 0.900 (\(7.421\mathrm{e}{-4}\)) & 0.698 (0.003)\\
B & Point mass & 0.900 (\(5.238\mathrm{e}{-4}\)) & 0.699 (0.002)\\
\addlinespace
C & Cluster & 0.900 (0.000) & 0.700 (0.000)\\
C & Random & 0.901 (\(8.007\mathrm{e}{-4}\)) & 0.701 (0.002)\\
C & Point & 0.900 (\(3.935\mathrm{e}{-4}\)) & 0.698 (0.002)\\
C & Point mass & 0.900 (\(2.962\mathrm{e}{-4}\)) & 0.699 (0.001)\\
\bottomrule
\end{tabular}
\end{table}

\Needspace{16\baselineskip}
\subsection{Results across signal strengths}

\begin{figure}[H]
\centering
\includegraphics[width=0.98\textwidth]{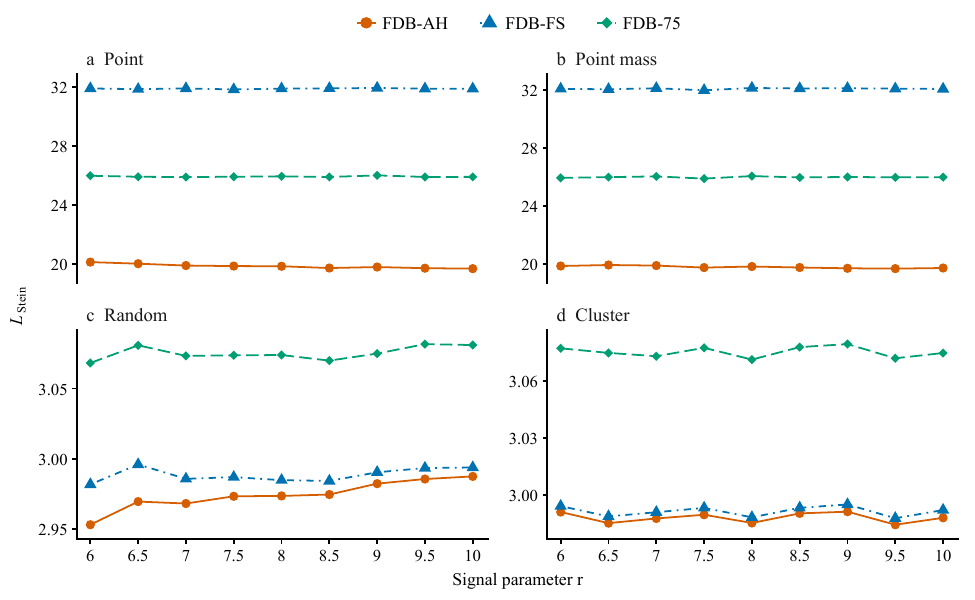}
\caption{Mean \(L_{\mathrm{Stein}}\) after reweighting across signal strengths \(r\)
(Setting B, \(\varepsilon=0.10\)).}
\label{fig:eps010}

\end{figure}

Figure~\ref{fig:eps010} shows how mean \(L_{\mathrm{Stein}}\) after reweighting varies with
signal strength \(r\) at \(\varepsilon=0.10\) in Setting B.
In the point contamination experiment at \(r=6\), the mean values of \(L_{\mathrm{Stein}}\)
for \method{}, \fs{}, and \hseven{} are \(20.132\), \(31.890\), and \(25.980\),
respectively.
In the same simulations, the mean for \hf{} is \(31.894\).

\begin{figure}[!htbp]
\centering
\includegraphics[width=0.98\textwidth]{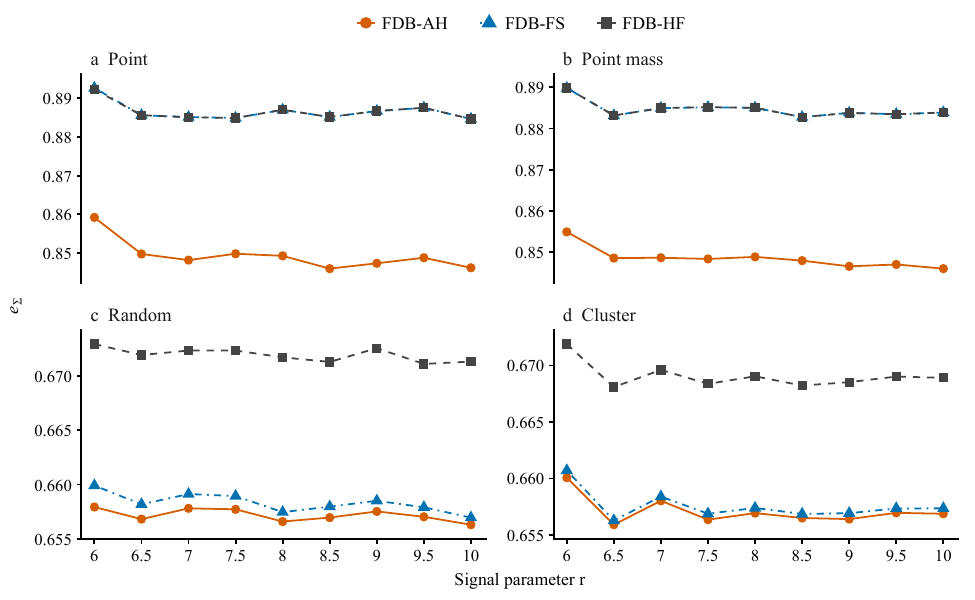}
\caption{Mean \(e_\Sigma\) after reweighting across signal
strengths \(r\) (Setting B, \(\varepsilon=0.30\)).}
\label{fig:eps030}

\end{figure}

Figure~\ref{fig:eps030} shows how the mean \(e_\Sigma\) after
reweighting varies with \(r\) at \(\varepsilon=0.30\).
In the cluster contamination experiment at \(r=6\), the means for
\method{} and \fs{} are similar, at \(0.660\) and \(0.661\), respectively, and
the corresponding mean for \hf{} is \(0.672\).
In this experiment, the mean values of \(L_{\mathrm{Stein}}\) for \method{} and \hseven{} are
\(3.859\) and \(46.246\), respectively.

In Figure~\ref{fig:eps030}, the curves for \fs{} and \hf{} coincide from
\(r=6.5\) onwards under point contamination and at all nine values of \(r\)
under point-mass contamination.
The curves coincide because, at these values of \(r\), \fs{} selects
\(h_{\min}=275\) in every run, the fixed initial subset size used by \hf{}.

\FloatBarrier
\subsection{Comparison with FSDA}
\label{sec:supp-fs-comparison}

We compare \method{} with the full forward search procedure
\citep{riani2009unknown} implemented in \fsm{}.
To assess the accuracy of location and scatter estimation, we use the
scale-corrected and reweighted estimates from \method{} and the estimates
returned directly by \texttt{fsdac::FSM(X,msg=FALSE)} for \fsm{}.

The comparison includes all experiments in Section~\ref{sec:simulation}.
We repeat each experiment \(1000\) times in Settings A and B and \(100\) times
in Setting C.
Both methods use the same \(90{,}900\) simulated datasets, and \method{}
returns estimates in every run.
\fsm{} fails in \(3\) runs because of numerical problems, all under point-mass
contamination in Setting B (\(\varepsilon=0.30\), \(r=7.5\) or \(9\)).
For each method, mean losses are computed over successful runs and mean running
times over all runs.

\begin{figure}[!htbp]
	\centering
	\includegraphics[width=0.98\textwidth]{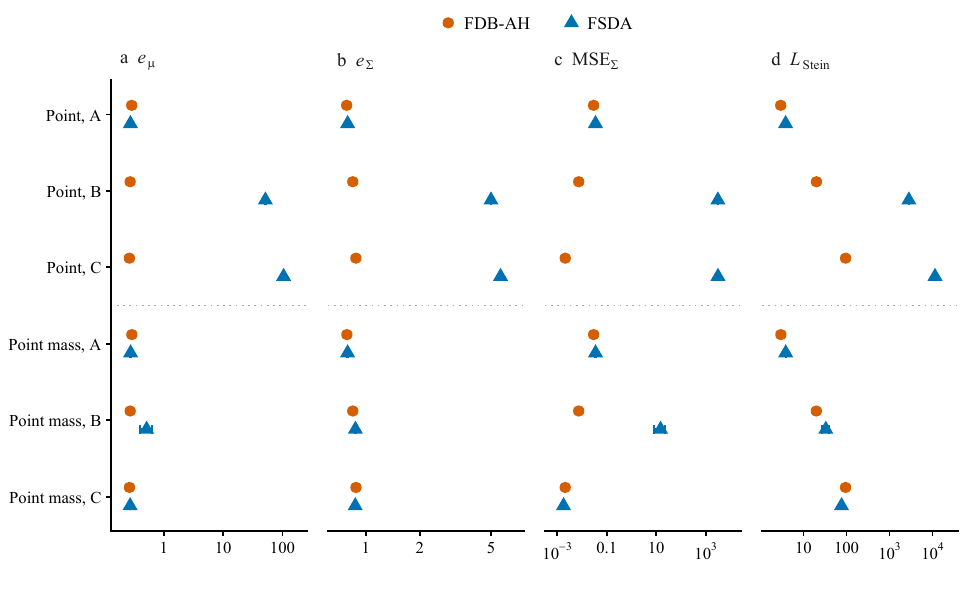}
	\caption{Four losses for \method{} and \fsm{} under point and point-mass
	contamination (Settings A--C, \(\varepsilon=0.30\), \(r=15\)).
		The horizontal axis is on a logarithmic scale. Points show mean losses, and
		horizontal error bars extend one standard error on either side of the mean.}
	\label{fig:supp-fs-losses}
\end{figure}

At \(\varepsilon=0.30\) and \(r=15\), the mean \(L_{\mathrm{Stein}}\) for
\method{} is much lower than for \fsm{} under point contamination in both
Settings B and C (Figure~\ref{fig:supp-fs-losses}).
The means for the two methods are \(20.017\) and \(2.814\mathrm{e}{3}\),
respectively, in Setting B, and \(95.312\) and \(1.141\mathrm{e}{4}\) in
Setting C.
In both experiments, \method{} also has lower mean \(e_\Sigma\) and
\(\mathrm{MSE}_\Sigma\) than \fsm{}.

To examine this difference, we take one sample from each of the two point
contamination experiments and inspect the search history for \fsm{}.
The default \fsm{} method, \texttt{unibiv}, selects the initial subset using
the coordinates after the transformation \(X=YG\).
In these two samples from Settings B and C, the initial subsets contain \(51\)
and \(201\) observations, respectively, all of them contaminated.
After signal confirmation, \fsm{} still retains contaminated observations in
both samples.
The uncontaminated observations in these two experiments come from a
multivariate \(t_3\) distribution, while the default \fsm{} thresholds are
computed under a Gaussian model.

In the point-mass contamination experiment for Setting C shown in
Figure~\ref{fig:supp-fs-losses}, \fsm{} has lower means for \(e_\Sigma\),
\(\mathrm{MSE}_\Sigma\), and \(L_{\mathrm{Stein}}\) than \method{}.
In the uncontaminated experiments, all four mean losses for \fsm{} are slightly
lower than for \method{} in Settings A and B, and the four means are similar
for the two methods in Setting C.

\begin{table}[htbp]
	\centering
	\caption{Mean running time (seconds) for \method{} and \fsm{}.}
	\label{tab:supp-fs-time}
	\small
	\begin{tabular}{lrrrr}
		\toprule
		Setting & Experiments & Repetitions & \method{} & \fsm{}\\
		\midrule
		A & 9 & 1000 & \(0.018\) & \(0.013\)\\
		B & 81 & 1000 & \(0.124\) & \(0.862\)\\
		C & 9 & 100 & \(6.146\) & \(167.362\)\\
		\bottomrule
	\end{tabular}
\end{table}

Each row of Table~\ref{tab:supp-fs-time} summarizes running times across all
experiments in one setting.
The timings for Setting B also include experiments at the nine signal strengths
\(r=6,6.5,\ldots,10\).
The experiments used an Intel Core i7-13700 processor and R 4.5.2.

\FloatBarrier
\subsection{Comparison with Stable MCD}
\label{sec:supp-stable-comparison}

We compare \method{} with Stable MCD \citep{heng2026stability}.
Stable MCD uses \(50\) pairs of bootstrap samples to compute a combined
instability measure for \(20\) candidate retained proportions,
\(0.5,0.525,\ldots,0.975\), with parameters \texttt{classifier="MD"} and
\texttt{sd\_ratio=3}.
It selects the retained proportion with the smallest instability. In case of
ties, it takes the smallest such proportion.
The subset updates and stopping rule follow the implementation in the
\texttt{StableMCD} package. All settings satisfy \(n>p\), and Stable MCD uses
all \(p\) variables directly.
To compute projection depth, Stable MCD and \method{} use \(\max(1000,100p)\)
and \(\max(1000,10p)\) directions, respectively.
To assess the accuracy of location and scatter estimation, we use the
scale-corrected and reweighted estimates for \method{} and, for Stable MCD, the
estimates returned directly by \texttt{mcd} at the selected proportion with
\texttt{reweighting=FALSE}.

\begin{figure}[!htbp]
	\centering
	\includegraphics[width=0.98\textwidth]{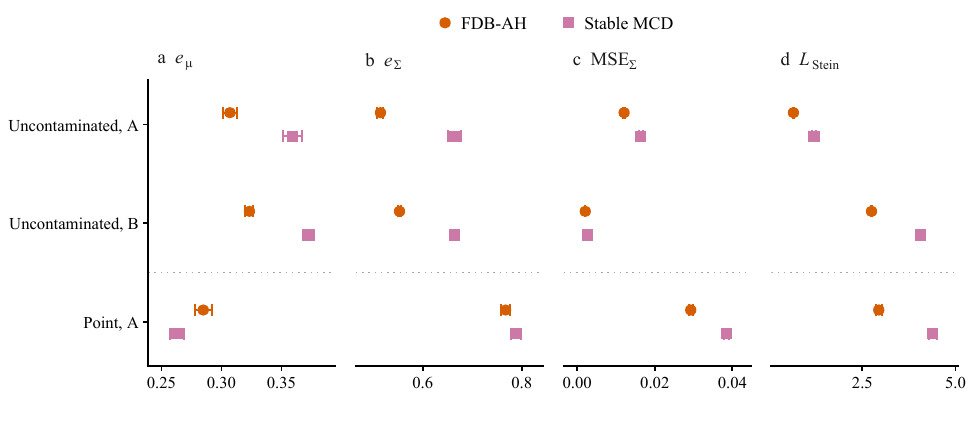}
	\caption{Four losses for \method{} and Stable MCD. The figure includes the
	uncontaminated experiments in Settings A and B and the point contamination
	experiment in Setting A at \(\varepsilon=0.30\) and \(r=15\).
		Points show mean losses, and horizontal error bars extend one standard error
		on either side of the mean.}
	\label{fig:supp-stable-losses}
\end{figure}

In Settings A and B, we repeat each experiment \(100\) times, using the same
\(9{,}000\) simulated datasets for both methods.
\method{} returns estimates in every run. Stable MCD fails in \(38\) runs
because of singular matrices, all under point-mass contamination at
\(\varepsilon=0.30\).
Of these failures, \(37\) occur in Setting A at \(r=15\) and \(1\) in Setting B
at \(r=6\).
Mean losses for each method are computed over its successful runs.
Both methods complete all \(100\) runs successfully in each experiment shown in
Figure~\ref{fig:supp-stable-losses}.

In the uncontaminated experiments in Settings A and B, \method{} selects larger
initial subsets on average, and all four mean losses of its final estimates are
lower than for Stable MCD (Figure~\ref{fig:supp-stable-losses}).
In Setting A, the mean initial subset proportions selected by \method{} and
Stable MCD are \(0.978\) and \(0.800\), respectively, and their mean values of
\(L_{\mathrm{Stein}}\) are \(0.650\) and \(1.199\).
In Setting B, the mean selected proportions are \(0.998\) and \(0.784\),
respectively, and the mean values of \(L_{\mathrm{Stein}}\) are \(2.746\) and \(4.067\).

In the point contamination experiment in Setting A (\(\varepsilon=0.30\),
\(r=15\)), \method{} has a lower mean \(L_{\mathrm{Stein}}\) (\(2.940\)) than Stable MCD
(\(4.390\)), while Stable MCD has the smaller mean \(e_\mu\).
In Setting B, Stable MCD has lower mean values for both
\(\mathrm{MSE}_\Sigma\) and \(L_{\mathrm{Stein}}\) than \method{} in some
point contamination experiments and in all point-mass contamination experiments.

\Needspace{8\baselineskip}
Table~\ref{tab:supp-stable-time} summarizes the running times of the two
methods.
In Settings A and B, the two methods are timed in batches on the same simulated
samples, and the table reports the mean time per run, including the time spent
on failed runs. In Setting C, only uncontaminated samples are timed.

\begin{table}[htbp]
	\centering
	\caption{Mean running time (seconds) for \method{} and Stable MCD.}
	\label{tab:supp-stable-time}
	\small
	\begin{tabular}{lrrrr}
		\toprule
		Setting & Experiments & Repetitions & \method{} & Stable MCD\\
		\midrule
		A & 9 & 100 & \(0.020\) & \(3.598\)\\
		B & 81 & 100 & \(0.132\) & \(24.725\)\\
		C & 1 & 100 & \(6.221\) & \(845.636\)\\
		\bottomrule
	\end{tabular}
\end{table}

\FloatBarrier

\section{SKAB data analysis}
\label{sec:supp-skab}

We analyse one normal-operation record and one rotor-imbalance record from
SKAB, both collected in 2020 on a water-circulation testbed
\citep{katser2020skab}.
The normal-operation record contains \(9405\) observations, from which we
select \(1500\) at approximately equally spaced row indices.
In the rotor-imbalance experiment, the operators introduced a step change in
rotor imbalance. We use the full record of \(1090\) observations.
The data providers labelled \(743\) observations as normal and \(347\) as
anomalous, and we refer to the latter as fault observations below.
We retain all eight sensor variables in both records, including the root mean
square vibration accelerations from two accelerometers, motor current, pressure
in the pipe downstream of the pump, motor housing temperature, circulating
fluid temperature measured by a thermocouple, motor voltage, and flow rate.
We standardize each variable within each record by subtracting its median and
dividing by \(Q_n\), the robust scale estimator of \citet{rousseeuw1993qn}.
If \(Q_n\) is zero or non-finite, we use the sample standard deviation. In both
records, only the pressure variable requires this substitution.

We compare the six methods from Section~\ref{sec:real-data} on each record.
Once the records have been selected, we use the providers' labels only to count
normal and fault observations in the initial subsets and to compute the final
error rates.
The FPR is the proportion of normal observations
classified as anomalous, and the FNR is the proportion of
fault observations classified as normal.
The normal-operation record contains no fault observations, so its FNR is
undefined.
Each method completes \(100\) runs successfully on each fixed record.
Table~\ref{tab:realdata-h} reports subset sizes, final error rates, and
running times averaged over these \(100\) runs, with error rates expressed as
percentages.

\begin{table}[H]
	\centering
	\begin{threeparttable}
		\caption{Results for the six methods on the two SKAB records.}
		\label{tab:realdata-h}
		\small
		\setlength{\tabcolsep}{3pt}
		\renewcommand{\arraystretch}{1.15}
		\begin{tabular*}{\linewidth}{@{\extracolsep{\fill}}lrrrr@{}}
			\toprule
			Method & Initial subset size \(h\) & FPR (\%) & FNR (\%) & Time (s)\\
			\midrule
			\multicolumn{5}{l}{\textit{Normal-operation record (\(n=1500,\ p=8\))}}\\
			\method & 1499.000 & 4.061 & --- & 0.314\\
			\fs & 758.130 & 15.149 & --- & 0.074\\
			\hf & 754.000 & 15.173 & --- & 0.048\\
			\hseven & 1125.000 & 10.239 & --- & 0.048\\
			\fsm & --- & 25.600 & --- & 1.611\\
			Stable MCD & 1454.520 & 3.032 & --- & 15.631\\
			\addlinespace
			\multicolumn{5}{l}{\textit{Rotor-imbalance record (\(n=1090,\ p=8\))}}\\
			\method & 744.000 & 0.269 & 0.288 & 0.261\\
			\fs & 549.000 & 2.560 & 0.288 & 0.057\\
			\hf & 549.000 & 2.560 & 0.288 & 0.034\\
			\hseven & 817.000 & 0.135 & 13.499 & 0.034\\
			\fsm & --- & 0.269 & 0.288 & 0.160\\
			Stable MCD & 735.000 & 1.211 & 0.288 & 7.787\\
			\bottomrule
		\end{tabular*}
	\end{threeparttable}
\end{table}

In every run on the normal-operation record, \method{} includes \(1499\) of the
\(1500\) observations in the initial subset.
\hf{} and \hseven{} use smaller fixed initial subsets, and under the same
reweighting rule both have higher mean FPRs than \method{}
(Figure~\ref{fig:motivation}c).

On the rotor-imbalance record, \method{} selects \(h=744\) in all \(100\) runs.
In the run corresponding to the distance curve in Figure~\ref{fig:motivation}b,
the initial subset for \method{} contains all \(743\) normal observations and
\(1\) fault observation. Reweighting retains \(741\) normal observations and
the same fault observation.
\hf{} and \fs{} select the same \(549\) observations, of which \(548\) are
normal and \(1\) is a fault observation. After reweighting, both methods flag
\(23\) normal observations as anomalous.
The initial subset for \hseven{} contains \(817\) observations, including
\(74\) fault observations.
Reweighting retains \(48\) fault observations, of which \(46\) were in the
initial subset and \(2\) were outside it.

In each of the \(100\) runs on this record, \method{} gives \(2\) false
positives and \(1\) false negative.
\hf{} and \fs{} give identical final classifications in each run, with 9--31
false positives and always \(1\) false negative.
\hseven{} gives only \(1\) false positive per run but misses 44--51 fault
observations.
Figure~\ref{fig:motivation}d,e gives the means and ranges of both error rates
after reweighting for \method{}, \hf{}, and \hseven{}.
\fsm{} retains the same observations as \method{} in every run. Stable MCD
gives \(9\) false positives and \(1\) false negative in every run.

\FloatBarrier
\renewcommand{\bibfont}{\small}
\setlength{\bibsep}{3pt plus 1pt minus 1pt}

\end{document}